%% file: main.tex
\documentclass[a4paper,USenglish,numberwithinsect,cleveref, autoref, thm-restate]{lipics-v2021}

\usepackage{mathtools, amssymb, nicefrac}
\usepackage{stmaryrd}   
\usepackage{float}
\usepackage{xspace}
\usepackage[dvipsnames]{xcolor}
\usepackage{multicol}
\usepackage{tabularx, booktabs}
\usepackage{infer}          
\usepackage{tikz}
\usetikzlibrary{arrows,automata,positioning,shapes.geometric}

\let\originalleft\left
\let\originalright\right
\renewcommand{\left}{\mathopen{}\mathclose\bgroup\originalleft}
\renewcommand{\right}{\aftergroup\egroup\originalright}

\makeatletter
\@ifpackageloaded{stix}{%
}{%
  \DeclareFontEncoding{LS2}{}{\noaccents@}
  \DeclareFontSubstitution{LS2}{stix}{m}{n}
  \DeclareSymbolFont{stix@largesymbols}{LS2}{stixex}{m}{n}
  \SetSymbolFont{stix@largesymbols}{bold}{LS2}{stixex}{b}{n}
  \DeclareMathDelimiter{\lBrace}{\mathopen} {stix@largesymbols}{"E8}%
                                            {stix@largesymbols}{"0E}
  \DeclareMathDelimiter{\rBrace}{\mathclose}{stix@largesymbols}{"E9}%
                                            {stix@largesymbols}{"0F}
}
\makeatother

\newcommand\scalemath[2]{\scalebox{#1}{\mbox{\ensuremath{\displaystyle #2}}}}

\allowdisplaybreaks

\input{macros}

\title{Slicing of Probabilistic Programs based on Specifications}

\author{Marcelo Navarro}{Computer Science Department (DCC), University of Chile, Chile}{mnavarro@dcc.uchile.cl}{}{}

\author{Federico Olmedo\footnote{Corresponding author.}}{Computer Science Department (DCC), University of Chile, Chile}{folmedo@dcc.uchile.cl}{https://orcid.org/0000-0003-0217-6483}{}

\authorrunning{M. Navarro and F. Olmedo} 

\Copyright{Marcelo Navarro and Federico Olmedo} 



\category{} 

\relatedversion{} 

\funding{This research has been supported by the FONDECYT  Grant No.~11181208.} 

\acknowledgements{} 

\nolinenumbers 

\EventEditors{Karim Ali and Jan Vitek}
\EventNoEds{36}
\EventLongTitle{36th European Conference on Object-Oriented
  Programming (ECOOP 2022)}
\EventShortTitle{ECOOP 2022}
\EventAcronym{ECOOP}
\EventYear{2022}
\EventDate{June 6--10, 2022}
\EventLocation{Berlin, Germany}
\EventLogo{}
\ArticleNo{1}

\begin{document}

\maketitle

\begin{abstract}
\input{abstract}
\end{abstract}

\input{intro}

\input{foundations}

\input{vcgen}

\input{slicing}
\input{extension}

\input{applications}

\input{graphs}

\input{discussion}

\input{related}

\input{concl}

\pagebreak
\bibliography{bibfile}
\vfill

\pagebreak
\appendix
\renewcommand{\thefigure}{\arabic{figure}} 
\section*{Appendix A}
\input{appendixA}

\pagebreak
\section*{Appendix B}
\input{appendixB}

\end{document}

%% file: macros.tex
\newcommand{\lcfg}[2]{\ensuremath{\mathsf{LCFG} \left[ #1 \right]%
    \left( #2\right)}\xspace}
\newcommand{\slg}[3]{\ensuremath{\mathsf{SLG} \left[ #1 \right]%
    \left( #2, #3\right)}\xspace}

\newcommand{\glabel}[1]{\ensuremath{\mathit{label}\left(#1\right)}\xspace}

\newcommand{\ing}[1]{\ensuremath{\mathit{in}\left(#1\right)}\xspace}
\newcommand{\outg}[1]{\ensuremath{\mathit{out}\left(#1\right)}\xspace}
\newcommand{\bif}[1]{\ensuremath{{\sf bif}\left( #1 \right)}\xspace}
\newcommand{\fib}{{\sf fib}\xspace}
\newcommand{\pif}[1]{\ensuremath{{\sf pif}\left( #1 \right)}\xspace}
\newcommand{\fip}{{\sf fip}\xspace}
\newcommand{\don}[1]{\ensuremath{\Do\left( #1 \right)}\xspace}
\newcommand{\odn}{{\sf od}\xspace}
\newcommand{\start}{{\sf start}\xspace}
\newcommand{\dne}{{\sf end}\xspace}

\newcommand{\ie}{i.e.\@\xspace}
\newcommand{\wrt}{w.r.t.\@\xspace}
\newcommand{\eg}{e.g.\@\xspace}

\newcommand{\etal}{et al.\@\xspace}

\newcommand{\ZO}{\ensuremath{[0,\,\! 1]}\xspace}                             

\newcommand{\Lang}{\ensuremath{\mathsf{pWhile}}\xspace} 
\newcommand{\Prog}{\ensuremath{\mathcal{C}}\xspace}   
\newcommand{\prog}{\ensuremath{c}\xspace}   
\newcommand{\Inst}{\ensuremath{\mathcal{I}}\xspace}   
\newcommand{\inst}{\ensuremath{i}\xspace} 

\newcommand{\Var}{\ensuremath{\mathcal{V}}\xspace}
\newcommand{\Expr}{\ensuremath{\mathcal{E}}\xspace}
\newcommand{\Asser}{\ensuremath{\mathcal{A}}\xspace}

\newcommand{\Skip}{{\sf skip}\xspace}

\newcommand{\Ass}[2]{\ensuremath{{#1} \mathrel{\coloneqq} {#2}}}

\newcommand{\If}{{\sf if}\xspace}
\newcommand{\Then}{{\sf then}\xspace}
\newcommand{\Else}{{\sf else}\xspace}
\newcommand{\Cond}[3]{\If \, (#1) \,\Then \, \{#2\} \,\Else \, \{#3\}}
\newcommand{\While}{{\sf while}\xspace}
\newcommand{\Do}{{\sf do}\xspace}
\newcommand{\WhileDo}[2]{\While \, (#1) \, \Do \, \{#2\}}
\newcommand{\IWhileDo}[3]{\While \, (#1) \, \left[#2\right] \, \Do \, \{#3\}}

\newcommand{\seqn}[2]{\inst_{#1};\dots;\inst_{#2}}
\newcommand{\codeComment}[1]{\textcolor{brown}{\textnormal{\small
    \textbackslash\textbackslash\ \textsl{#1}}}}
\newcommand{\codeGraph}[1]{\textcolor{brown}{\textnormal{\small
    \ \textsl{#1}}}}

\newcommand{\SliceAway}[1]{\textcolor{BrickRed}{#1}}

\newcommand{\State}{\ensuremath{\mathcal{S}}\xspace}
\newcommand{\state}{\ensuremath{s}\xspace}

\newcommand{\sem}[1]{\llbracket #1 \rrbracket}

\newcommand{\Exp}{\ensuremath{\mathbb{E}}\xspace} 
\newcommand{\charFun}[1]{\left[#1\right]} 
\newcommand{\pimplies}{\ensuremath{\Rrightarrow}\xspace}

\newcommand{\spec}[4][\;]{ \lBrace #2 \rBrace #1 #3 #1 \lBrace #4 \rBrace}   
\newcommand{\pspec}[4][\;]{ \lBrace #2 \rBrace #1 #3 #1 \lBrace #4 \rBrace^{\circlearrowleft}}   
\newcommand{\tspec}[4][\;]{ \lBrace #2 \rBrace #1 #3 #1 \lBrace #4 \rBrace^{\downarrow}}   

\newcommand{\pre}{\ensuremath{P}\xspace}  
\newcommand{\post}{\ensuremath{Q}\xspace}  
\newcommand{\pree}{\ensuremath{f}\xspace}  
\newcommand{\poste}{\ensuremath{g}\xspace} 
\newcommand{\inv}{\ensuremath{\mathit{inv}}\xspace} 

\newcommand{\wlpsymbol}{\textnormal{\textsf{wlp}}\xspace}
\newcommand{\wlpd}[1]{\ensuremath{\wlpsymbol \left[ #1 \right] }\xspace}
\newcommand{\wlp}[2]{\ensuremath{\wlpsymbol \left[ #1 \right] \left( #2 \right)} \xspace}

\newcommand{\wpsymbol}{\textnormal{\textsf{wp}}\xspace}
\newcommand{\wpd}[1]{\ensuremath{\wpsymbol \left[ #1 \right] }\xspace}
\renewcommand{\wp}[2]{\ensuremath{\wpsymbol \left[ #1 \right] \left( #2 \right)}\xspace}

\newcommand{\wllpsymbol}{\textnormal{\textsf{w}(\textsf{l})\textsf{p}}\xspace}
\newcommand{\wllpd}[1]{\ensuremath{\wllpsymbol \left[ #1 \right] }\xspace}
\newcommand{\wllp}[2]{\ensuremath{\wllpsymbol \left[ #1 \right] \left( #2 \right)} \xspace}

\newcommand{\wprecsymbol}{\textnormal{\textsf{wpre}}\xspace}
\newcommand{\wprecd}[1]{\ensuremath{\wprecsymbol \left[ #1 \right] }\xspace}
\newcommand{\wprec}[2]{\wprecsymbol \left[ #1 \right] \left( #2 \right) \xspace}
\newcommand{\wprecid}[1]{\wprecsymbol_{\geq #1}}
\newcommand{\wpreci}[3]{\wprecid{#1}\left[ #2\right]\left( #3\right)\xspace}

\newcommand{\wprectsymbol}{\textnormal{\textsf{wpre}}^{\downarrow}\xspace}

\newcommand{\wprect}[2]{\wprectsymbol \left[ #1 \right] \left( #2 \right) \xspace}
\newcommand{\wprectid}[1]{\wprectsymbol_{\geq #1}}
\newcommand{\wprecti}[3]{\wprectid{#1}\left[ #2\right]\left( #3\right)\xspace}

\newcommand{\vcsymbol}{\textnormal{\textsf{vc}}\xspace}
\newcommand{\vcd}[1]{\vcsymbol \left[ #1 \right]\xspace}
\newcommand{\vc}[2]{\vcsymbol \left[ #1 \right] \left( #2 \right)\xspace}
\newcommand{\vci}[3]{\vcsymbol_{\geq #1} \left[ #2 \right] \left( #3
  \right)\xspace}

\newcommand{\vcgsymbol}{\textnormal{VCG}\xspace}
\newcommand{\vcg}[3]{\vcgsymbol\left[#2\right] \left( #1, #3 \right)\xspace}
\newcommand{\vcgd}[1]{\vcgsymbol\left[#1\right]\xspace}
\newcommand{\vcgi}[4]{\vcgsymbol_{\leq #1}\left[#3\right] \left( #2, #4 \right)\xspace}
\newcommand{\vcgid}[1]{\vcgsymbol_{\leq #1}\xspace}
\newcommand{\vcgii}[4]{\vcgsymbol_{\geq #1}\left[#3\right] \left( #2, #4 \right)\xspace}

\newcommand{\vctsymbol}{\textnormal{\textsf{vc}}^{\downarrow}\xspace}

\newcommand{\vct}[2]{\vctsymbol \left[ #1 \right] \left( #2 \right)\xspace}
\newcommand{\vcti}[3]{\vctsymbol_{\geq #1} \left[ #2 \right] \left( #3 \right)\xspace}

\newcommand{\vcgtsymbol}{\textnormal{VCG}^{\downarrow} \xspace}
\newcommand{\vcgt}[3]{\vcgtsymbol\left[#2\right] \left( #1, #3 \right)\xspace}

\newcommand{\vcgti}[4]{\vcgtsymbol_{\leq #1}\left[#3\right] \left( #2, #4 \right)\xspace}

\newcommand{\vcgtii}[4]{\vcgtsymbol_{\geq #1}\left[#3\right] \left( #2, #4 \right)\xspace}

\newcommand{\lrule}[1]{\textnormal{\small \textsf{[#1]}}\xspace}

\newcommand{\specSlice}[4]{\spec{#2}{#1 \preccurlyeq #4}{ #3}}
\newcommand{\pspecSlice}[4]{\pspec{#2}{#1 \preccurlyeq #4}{ #3}}
\newcommand{\tspecSlice}[4]{\tspec{#2}{#1 \preccurlyeq #4}{ #3}}

\newcommand{\removesymbol}{\ensuremath{\mathsf{remove}\xspace}}
\newcommand{\remove}[3]{\ensuremath{\removesymbol\left( #1, #2, #3 \right)\xspace}}

\newcommand{\sspec}[2]{ \left\langle #1, #2\right\rangle}
\newcommand{\sspecp}[2]{ \left\langle #1, #2\right\rangle^\circlearrowleft}
\newcommand{\sspect}[2]{ \left\langle #1, #2\right\rangle^\downarrow}

\newcommand{\localSpecSymbol}{\ensuremath{\leadsto}\xspace}

\newcommand{\localSpec}[5]{#2 \vdash \sspec{#1}{#3} \:\localSpecSymbol\:
  #4 \vdash \left\{ #5 \right\}}  
\newcommand{\localSpecp}[6]{#2 \vdash \sspecp{#1}{#3} \:\localSpecSymbol\:
  #5 \vdash \sspecp{#4}{#6}}  
\newcommand{\localSpect}[5]{%
#2 \vdash \sspect{#1}{#3}  \:\localSpecSymbol\:  #4 \vdash \left\{ #5
\right\}} 
\newcommand{\PChoiceSym}[1]{\ensuremath{[#1]}}
\newcommand{\PChoice}[3]{ \{ #1 \} \: \PChoiceSym{#2} \: \{#3\}}

\newcommand{\eval}[1]{[ #1]}

\providecommand{\eqdef}{\circeq}
\newcommand{\To}{\rightarrow}                          

\newcommand{\subst}[2]{\left[{#1}/{#2}\right]}

\newcommand{\by}[1]{\text{\small \{#1{}\}}}

\newcommand{\prob}[1]{\mathbb{P}({#1})}

%% file: abstract.tex
This paper presents the first slicing approach for probabilistic programs based on specifications. We show that when probabilistic programs are accompanied by their specifications in the form of pre- and post-condition, we can exploit this semantic information to produce specification-preserving slices strictly more precise than slices yielded by conventional techniques based on data/control dependency.

To achieve this goal, our technique is based on the backward propagation of post-conditions via the greatest pre-expectation transformer---the probabilistic counterpart of Dijkstra weakest pre-condition transformer. The technique is termination-sensitive, allowing to preserve the partial as well as the total correctness of probabilistic programs \wrt their specifications.  It is modular, featuring a local reasoning principle, and is formally proved correct.

As fundamental technical ingredients of our technique, we design and prove sound verification condition generators for establishing the partial and total correctness of probabilistic programs, which are of interest on their own and can be exploited elsewhere for other purposes.

On the practical side, we demonstrate the applicability of our approach by means of a few illustrative examples and a case study from the probabilistic modelling field. We also describe an algorithm for computing least slices among the space of slices derived by our technique.

%% file: intro.tex
\section{Introduction}\label{sec:intro}

Since its introduction by Weiser~\cite{Weiser:ICSE81}, \emph{program
  slicing} has been recognized for its wide range of applications in
the process of software development. The basic idea behind program
slicing is, given a program and a set of variables of interest (the
slicing criterion), to identify the program fragments that can be
safely removed without affecting the program behavior, with respect to
the said set of variables; the “subset” program so obtained is known
as a \emph{slice} of the original program. 
Among others, primary applications of slicing include testing, program
understanding, program debugging and extraction of reusable
components~\cite{Xu:2005}.

Different approaches have been proposed to compute program
slices~\cite{Tip:JFP95}. However, two shared---and sometimes conflicting---
requirements of these approaches are efficiency and precision. On the one
hand, one is interested in computing slices as fast as possible, and
on the other hand, in computing the least slices. Besides efficiency
and precision, another fundamental aspect of slice approaches lies in the
subset of \emph{language features} that they support. Current
approaches can be applied, for instance, to programs with procedural
abstractions, unstructured control flow, composite data types and
pointers, and concurrency primitives~\cite{Tip:JFP95}. However, a fundamental language
feature that is only partially supported is \emph{randomization}.

At the programming language level, randomization is typically supported by some form of
\emph{probabilistic choice}
. For instance, a
probabilistic program can flip a (fair or biased) coin and depending
on the observed outcome, continue its execution in one way or another.

Probabilistic programs have proved useful in a wealth of different domains. 
They are central in the field of machine learning due to their
compelling properties for representing probabilistic
models~\cite{Ghahramani:2015,DBLP:conf/sigsoft/ClaretRNGB13}.
They are the cornerstone of modern cryptography---modern public-key encryption
schemes\footnote{By ``public-key encryption schemes'' we here mean public-key encryption schemes understood as a whole, comprising all the public key generation, encryption and decryption phases.}  are by nature probabilistic~\cite{Goldwasser:1984}. 
They lie at the heart of
quantum computing---quantum programs are inherently probabilistic due
to the random outcomes of quantum
measurements~\cite{Sanders:2000}. Finally, they are the key ingredient
for implementing
randomized algorithms~\cite{Motwani:1995}.

In the past years, the field of probabilistic programming has seen a
resurgence, in particular, due to the emergence of new 
probabilistic modeling applications~\cite{Van:18}. A wealth of new probabilistic programming
systems have been developed, which  conveniently  allow representing
probabilistic models as programs, and querying them, \eg,  to determine
the probability of a given event or the expected value of a given
random variable. To enable this, probabilistic programming systems
implement some form of \emph{inference}, building an
explicit representation of the probability distribution implicitly
encoded by a program. 

Notoriously, three distinguished features of probabilistic programs makes the
problem of slicing even more crucial for this class of
programs. First, despite usually consisting in a few lines of code,
probabilistic programs may present a complex and intricate behaviour,
which is hard to grasp for an average programmer, even when
knowledgeable in probability theory. For example, the termination
analysis of probabilistic programs is full of
subtleties~\cite{Olmedo:JACM:2018}. Second, the process of inference
is known to be computationally highly expensive~\cite{Darwiche:2009},
which turns the problem of computing slices as small as possible even
more critical for this class of programs. Third, the development of
probabilistic programming systems is a daunting task, and several bugs
have been recently discovered in many of them~\cite{PPsBugs:FSE18}. Any tool aiding
program understanding is thus vital.

Hur \etal gave a first step toward supporting slicing for
probabilistic programs, extended with conditioning~\cite{Hur:2014}. In
a later work, Amtoft and Banerjee introduced probabilistic control-flow
graphs, which allows a direct adaptation of standard machinery
from the slicing literature to the case of probabilistic
programs~\cite{Amtoft:2016,Amtoft:TOPLAS:20}. Both works adopt the
classical slicing criteria where slicing is performed with respect to
a set of program variables of interest, typically the variables
influencing the program output. Said otherwise, these works aims at
identifying those program fragments that do not have a true influence
on the value of the program output variables (or other set of
variables, at other execution point).

There exist, however, more precise slicing techniques based on program
assertions instead of program variables. The idea here is to identify
those program fragments that contribute to establishing a given
program assertion, instead of fixing the variables’
value~\cite{Comuzzi:1996,Barros:2012}. The main benefit of this
approach is that it produces smaller slices, provided there exists a (functional)
specification of the program in the form of a pre- and
post-condition. As advocated by the \emph{design-by-contract}
methodology to software development \cite{Meyer:92}, pre- and post-conditions specify
program behaviour by constraining the set of final states
(post-condition) that are reachable
from a given set of initial states (pre-condition). However, slicing
techniques based on specifications have so far been restricted to
deterministic programs, and it is an open problem whether they can
 be applied to probabilistic programs as well.

The main contribution of this article is to give a positive answer to
the above problem. Concretely, given a probabilistic program together
with its specification, we show how it can be sliced in order to
\emph{preserve the specification}. To illustrate this, consider the
 program below, accompanied by its specification
\[
  \tspec[\quad]%
  {\mathit{if} \; y^2 \!\leq\! 0.5 \; \mathit{then} \; \tfrac{1}{2} \; \mathit{else} \; 0}
  {\Ass{x}{1.5 {-} y^2};\; \PChoice{\Ass{x}{x{-}1}}{\nicefrac{1}{2}}{\Ass{x}{x{-}2}}}%
  {x \geq 0}
\]
The program starts by assigning the value of $1.5 - y^2$ to variable $x$, then flips a fair
coin and depending on the observed outcome, it decrements $x$ by
either 1 or 2. The specification says that, upon termination, the program establishes
post-condition $x \geq 0$ with probability at least $\nicefrac{1}{2}$
provided that initially $y^2 \leq 0.5$, and with probability at least 0,
otherwise. In the same way that for deterministic programs pre-conditions provide only
\emph{sufficient conditions} for establishing post-conditions, for
probabilistic programs pre-conditions provide only \emph{lower bounds
for the probability} of establishing post-conditions.

To slice this program, we can apply existing techniques for probabilistic
programs, by selecting $x$ as the output variable of interest whose
value we want to preserve (since $x$ is the only program variable mentioned in the
post-condition).  However, the conventional dataflow analysis carried out
by these techniques will say that the only valid slice of the program
is the very same program.

On the contrary, our slicing technique implements a more precise
analysis that captures quantitative relations between program
variables, concluding that the proper subprogram
\[
  \Ass{x}{1.5 - y^2}
\]
is a valid slice that preserves the original program
specification.\footnote{Formally, we also rely on the assumption that
  $y$ is a real-valued variable, and therefore it always holds that
  $y^2 \geq 0$.} In effect, it is the \emph{least} slice preserving 
the specification. 

Besides yielding more precise slices, specification-based slicing
opens the windows to further applications~\cite{Lee:2001}. A prominent example is
software reuse by \emph{specification specialization} (weakening). Suppose a
probabilistic program is known to establish a given post-condition \eg,
with a minimal probability $\nicefrac{1}{2}$. Now, if we are to
use the program in a new context where it suffices to establish the
post-condition with probability only $\nicefrac{1}{4}$, we can slice
the original program \wrt this weakened specification to yield to a
potentially simpler and more efficient program which can safely be
used in this context.

At the technical level, our slicing technique works by propagating post-conditions backward
using (a variant of) the greatest pre-expectation
transformer~\cite{McIver:2004}---the probabilistic counterpart of Dijkstra's
weakest pre-condition transformer~\cite{Dijkstra:90}. This endows programs
with an axiomatic semantics, expressed in terms of a verification
condition generator (VCGen) that yields quantitative proof
obligations.

In particular, we design (and prove sound) VCGens for
both the partial (allowing divergence) and the total (requiring
termination)  correctness of probabilistic programs, making
our slicing technique \emph{termination-sensitive}. To handle
iteration, we assume that program loops are annotated with
invariants. To reason about (probabilistic) termination, we assume that loop annotations also include
(probabilistic) variants. 

Another appealing property of our slicing technique is its
\emph{modularity}: It yields valid slices of a program from valid
slices of its subprograms. Most importantly, this involves only 
\emph{local reasoning}. This is crucial for keeping the slice
computation tractable.


In this regard, besides developing the theoretical foundations of our slicing
technique, we also exhibit an algorithm for computing program
slices. Interestingly, the algorithm computes the \emph{least}
slice that can be derived from our slicing technique, according to a
proper notion of slice size, using, as main ingredient, a shortest-path algorithm.

Finally, we illustrate the application of
our technique through some examples, showing that it yields strictly
more precise slices than existing techniques.


\subparagraph*{Contributions of the article.} To summarize, the main contributions of this
article are as follows: 
\begin{itemize}
\item We develop the first slicing technique based on specifications
  for probabilistic (imperative) programs  (Sections~\ref{sec:slicing} and \ref{sec:extension}). The slicing technique is
  termination-sensitive, modular featuring local reasoning principles,
  and formally proved correct. 
\item We demonstrate the applicability of the technique by means of
  illustrative examples. These comprise a set of small ---yet
  instructive--- programs while developing
  the theory (Sections~\ref{sec:slicepc} and \ref{sec:slicetc}) as
  well as a case study from the probabilistic modelling field
  (Section~\ref{sec:applications}). All these examples confirm that our
  technique yields strictly more precise slices than other existing techniques,
  provided programs are accompanied by their specifications and
  annotated with loop invariants. 
\item We show that an adaption of the slicing algorithm introduced
  by Barros~\etal~\cite{Barros:2012} can be used for computing minimal slices of
  probabilistic programs, among the space of slices derived by our
  technique (Section~\ref{sec:graphs}).  

\item As a fundamental ingredient of our technique, we design and
  prove sound VCGens for establishing both the partial and the total
  correctness of probabilistic programs (Sections~\ref{sec:VCGen} and
  \ref{sec:VCGentc}). These VCGens are of self-interest and can be
  exploited elsewhere for other purposes, such as program
  verification.
\end{itemize}

\subparagraph*{Organization of the article.} The remainder of the article
is organized as follows. Section~\ref{sec:foundations} introduces
the probabilistic imperative language used for describing
programs and lays out their specification
model. Section~\ref{sec:VCGen} presents the VCGen
for characterizing the partial correctness of
programs. Section~\ref{sec:slicing} develops the specification-based
slicing technique for preserving the partial correctness of programs
and Section~\ref{sec:extension} extends this technique 
to the case of total correctness. Section~\ref{sec:applications}
applies the slicing technique to a probabilistic model from the
literature. Section~\ref{sec:graphs} presents an
algorithm for computing slices. Section~\ref{sec:discussion}
discusses some design decisions and limitations behind the 
slicing technique. Finally, Section~\ref{sec:related} overviews the
related work and  Section~\ref{sec:conclusion} concludes.

%% file: foundations.tex
\section{Programming and Specification Model}
\label{sec:foundations}

In this section we introduce the programming language used for
describing probabilistic programs and lay out their functional
specification model. While not new, this provides the basic background for
understanding the problem we address and fixes the programming model
we adopt for our development.

\subsection{Programming Language}
To describe probabilistic programs we adopt a simple imperative
language extended with probabilistic choices, dubbed \Lang.
A \emph{program} is a non-empty sequence of instructions, where an
\emph{instruction} is either
a no-op, an assignment, a conditional branching, a
probabilistic choice or a guarded loop, annotated with its invariant. Formally, it is given by
grammar:
%

    $$
        \begin{array}{r@{\ \,}c@{\ \,}l@{\qquad}l}
          \Inst
             & ::=  & \Skip                    & \mbox{no--op}                 \\
             & \mid & \Ass{\Var}{\Expr}        & \mbox{assignment}             \\
             & \mid & \Cond{\Expr}{\Prog}{\Prog} & \mbox{conditional branching}  \\
             & \mid & \PChoice{\Prog}{p}{\Prog}  & \mbox{probabilisitic choice}  \\
            & \mid & \IWhileDo{\Expr}{\Asser}{\Prog}    & \mbox{guarded loop}           \\           
          \\
          \Prog
             & ::=  & \Inst & \mbox{single instruction}\\
             & \mid & \Inst; \, \Prog            & \mbox{sequential composition} \\
       \end{array}
    $$

\noindent Note that the set of programs, denoted by \Prog, and the set of
instructions, denoted by \Inst,
are defined mutually recursively. In the definition, we assume a set
\Var of \emph{variables} and a set \Expr of \emph{expressions} over program
variables. Finally, we use \Asser to denote the set of program
\emph{assertions}, in particular, loop invariants. 

No-op's, assignments, conditional branchings and guarded loops are
standard. However, we assume that guarded loops are annotated with invariants
so that they can be given an axiomatic semantics based on
VCGens. Finally, instruction $\PChoice{\prog_1}{p}{\prog_2}$ represents a
\emph{probabilistic choice}: it behaves like $\prog_1$ with probability $p$
and like $\prog_2$ with the complementary probability $1-p$.

As usual, a program \emph{state} is mapping from variables to
values; we use \State to denote the set of program states. Given a
state $\state \in \State$ and a variable $x \in \Var$, we write
$\state\subst{x}{v}$ for the state that is obtained from $\state$, by
updating the value of $x$ to $v$. Finally,
we assume the presence of an \emph{interpretation function} 
$\sem{\Expr}$ for expressions, mapping program states to values. 

\noindent\emph{Notational convention.} Since sequential composition is associative, we omit parentheses in
programs consisting of three or more instructions. In general, we
write $\prog = \inst_1; \inst_2; \ldots ; \inst_n$ to denote a program
that consists in a sequence of $n$ instructions.

\medskip
\subsection{Program Specifications}
The (functional) specification of programs is given by a pair  of
\emph{pre-} and \emph{post-condition}, which are interpreted on the
program initial and final states, respectively. Intuitively, the pre-condition
provides a sufficient condition for establishing the post-condition. However,
the precise interpretation of this varies depending on the program
nature. If the program at stake is
deterministic, each initial state either establishes or not the
post-condition upon program termination. Therefore, \textit{i}) pre-conditions are
\emph{qualitative}, that is, predicates over (initial) program states, and
\textit{ii}) an initial state satisfying the pre-condition is
guaranteed to establish the post-condition, whereas nothing is
guaranteed about an initial state violating the pre-condition. On the
other hand, if the
program at stake is probabilistic, each initial state establishes the
post-condition \emph{with a certain probability}. Thus,  \textit{i}) pre-conditions become \emph{qualitative}, mapping each (initial) state to
a probability in the interval $[0,1]$, and  \textit{ii}) the
probabilities reported by such pre-conditions represent \emph{lower bounds} for the probability that the
program establishes the post-condition. For example,
specification
\[
\tspec[\quad]{ \mathit{if} \: y \geq
    0 \: \mathit{then} \: \tfrac{1}{2} \: \mathit{else} \: 0}{\PChoice{\Ass{x}{y}}{\nicefrac{1}{2}}{\Ass{x}{x+1}}}{x
  \geq 0}
 \]
 says that program $\PChoice{\Ass{x}{y}}{\nicefrac{1}{2}}{\Ass{x}{x+1}}$
    establishes post-condition $x \geq 0$ with probability
    at least $\tfrac{1}{2}$ from an initial state where $y \geq
    0$, and with probability
    at least $0$ from an initial state where $y <
    0$. (Note that the pre-condition is not tight, as it dismiss the
    case where the right branch of the probabilistic choice
    establishes the post-condition.) 
    
    In fact, both pre-conditions as well as post-conditions become
    quantitative in the probabilistic case: pre-conditions for the
    reason argued above and post-conditions because in the presence of
    sequential composition, say $c_1; c_2$, the established
    pre-condition of $c_2$ behaves as the post-condition of $c_1$,
    requiring thus a uniform treatment between pre- and
    post-conditions. Thus, pre- and post-conditions are both functions of type
    $\Exp \eqdef \State \To [0,1]$, known as \emph{expectations},
    mapping program states to probabilities. Therefore, in
    the rest of the presentation we usually refer to the pre- and
    post-condition of a probabilisitic program as its \emph{pre-} and
    \emph{post-expectation}, respectively.

    To accomodate this generalization, we lift predicates (in
    particular, post-conditions) to expectations in a standard manner,
    taking their \emph{characteristic function}, which maps states
    satisfying the predicate to 1, and states violating the predicate
    to 0. In terms of notation, if $G$ is a Boolean expression over
    program variables encoding a predicate, we use $\charFun{G}$ to
    denote its characteristic function. For example, the above
    (informal) specification is formally written~as
\[
\tspec[\quad]{\tfrac{1}{2} \charFun{y \geq 0}}{\PChoice{\Ass{x}{y}}{\nicefrac{1}{2}}{\Ass{x}{x+1}}}{\charFun{x
  \geq 0}}~.
 \]
 %

As already hinted, this pre-expectation is not ``tight'' or the most
precise, as it says that from an initial state where $y <0$, the
program terminates in a final state where $x \geq 0$ with probability
at least 0. Even though being (trivially) valid, there is room for
significant improvement on this bound. In general, if $\pree$ and
$\pree'$ are two valid pre-expectations for a probabilistic program
specification, and $\pree \leq \pree'$ (where the ``$\leq$'' should be
understood pointwise), we usually prefer $\pree'$ over $\pree$. Said
otherwise, we are typically interested in the \emph{greatest} pre-expectation.  
In fact,  \emph{greatest} pre-expectations are the probabilistic
counterpart of \emph{weakest} pre-conditions. That is, while predicates
are ordered by relation ``$\Rightarrow$'', expectations are ordered by
relation ``$\leq$''. To better highlight this analogy at the
notation level, in the rest
of the presentation we use symbol $\pimplies$ to denote the pointwise
relation ``$\leq$'' over expectations:

\begin{definition}[Entailment relation $\pimplies$ between expectations]\label{def:exp-entailment}
For a pair of expectations $\pree, \pree' \colon \Exp$, we let
\[
\pree \pimplies \pree' \quad=\quad \forall s \in \State.\; \;  \pree(s)
\leq \pree'(s)~.
\]  
\end{definition}

Importantly, this induces a \emph{consistent} extension from the
deterministic to the probabilistic case: if $\pre, \pre'$ are predicates and $\charFun{\pre},
\charFun{\pre'}$ denote their respective characteristic functions, then
$\pre \Rightarrow \pre'$ if and only if $\charFun{\pre} \pimplies
\charFun{\pre'}$.  

Now that we have presented an intuitive approximation to the notion
of specification for probabilistic programs, we proceed to
define it formally. Like the specification of deterministic
programs, that of probabilistic programs comes also in two flavors,
differentiating on whether they account for the possibility of
divergence, or not. The kind of specifications that we have presented
so far corresponds to \emph{total correctness} specifications, since
the reported probabilities refer to the probability of
\emph{terminating and} establishing the post-condition. On the other
hand, \emph{partial correctness} specifications refer to the
probability of \emph{either} terminating and establishing the
post-condition \emph{or} diverging.

Formally, total and partial correctness are defined in
terms of the respective expectation transformers
\[
 \wpd{\cdot} \colon \Exp \To \Exp%
 \qquad \text{and}  \qquad%
 \wlpd{\cdot}  \colon \Exp \To \Exp~,
\]
which 
generalize Dijkstra's \emph{weakest pre-condition} and
\emph{weakest liberal pre-condition} transformers~\cite{Dijkstra} from
the deterministic to the probabilistic
case~\cite{McIver:2004,Kozen:1985,DBLP:journals/jcss/Kozen81}.
\begin{definition}[Program specification]\label{def:spec}
      We say that a \Lang program $\prog$ satisfies the \emph{total
        correctness specification} given by
      pre-expectation $\pree$ and post-expectation $\poste$, written
      $\models \tspec{\pree}{\prog}{\poste}$, iff
      \[
        \pree ~\pimplies~ \wp{\prog}{\poste}~.
      \]  
      \noindent Likewise, we say that a \Lang program $\prog$
      satisfies the \emph{partial correctness specification} given by
      pre-expectation $\pree$ and post-expectation $\poste$, written
      $\models \pspec{\pree}{\prog}{\poste}$, iff
      \[
        \pree ~\pimplies~ \wlp{\prog}{\poste}~.
      \]  
\end{definition}

Transformers $\wpsymbol$ and $\wlpsymbol$ were originally introduced
by Kozen~\cite{Kozen:1985,DBLP:journals/jcss/Kozen81}
and then further extended by
McIver and Morgan~\cite{McIver:2004}. They are defined by
induction on the program structure, as shown in
Figure~\ref{fig:wllp}. For all language constructs other than loops,
both transformers follow the same rules. Let us briefly explain them.
$\wllpd{\Skip}$ behaves as the identity since $\Skip$ has no
effect. The pre-expectation of an assignment is obtained by updating
the program state and then applying the post--expectation, \ie
$\wllpd{\Ass{x}{E}}$ takes post--expectation $\poste$ to
pre--expectation
$\poste\subst{x}{E}=\lambda s.\, \poste
(s\subst{x}{\sem{E}(s)})$. 
$\wllpd{\Cond{G}{\prog_1}{\prog_2}}$ behaves
either as $\wllpd{\prog_1}$ or $\wllpd{\prog_2}$ according to the evaluation
of $G$. $\wllpd{\PChoice{\prog_1}{p}{\prog_2}}$ is obtained as a convex combination
of $\wllpd{\prog_1}$ and $\wllpd{\prog_2}$, weighted according to
$p$. $\wllpd{\prog_1;\prog_2}$ is obtained as the functional composition of
$\wllpd{\prog_1}$ and $\wllpd{\prog_2}$. Finally,
$\wllpd{\WhileDo{G}{\prog}}$ is defined using standard fixed
point techniques, the only difference being the limit fixed point
considered: $\wpsymbol$ takes the least and $\wlpsymbol$ takes the
greatest (according to the $\pimplies$ order between
expectations). Observe that, as expected, the definition of
$\wllpsymbol$ over loops dismiss annotated loop invariants (and we
thus omit them in Figure~\ref{fig:wllp}). 
\begin{figure}[t]
    $$
        \begin{array}{lcl}
            \wllp{\Skip}{\poste}                 & = & \poste                                                              \\[1.5pt]
            \wllp{\Ass{x}{E}}{\poste}            & = & \poste\subst{x}{E}                                                  \\[1.5pt]
            \wllp{\Cond{G}{\prog_1}{\prog_2}}{\poste}    & = & \eval{G} \cdot\wllp{\prog_1}{\poste} +  \eval{ \lnot G} \cdot \wllp{\prog_2}{\poste} \\[1.5pt]
            \wllp{\PChoice{\prog_1}{p}{\prog_2}}{\poste} & = & p\cdot \wllp{\prog_1}{\poste} + (1-p)\cdot \wllp{\prog_2}{\poste}                    \\[1.5pt]
            \wllp{\prog_1;\prog_2}{\poste}               & = & \wllp{\prog_1}{\wllp{\prog_2}{\poste}}                                          \\[10pt]
             \wp{\WhileDo{G}{\prog}}{\poste}  & = & \mu f.\: 
                                                    \charFun{\lnot
                          G} \cdot g + \charFun{G} \cdot \wp{\prog}{f} \\[1.5pt]
            \wlp{\WhileDo{G}{\prog}}{\poste}  & = & \nu f.\: 
                                                    \charFun{\lnot
                          G} \cdot g + \charFun{G} \cdot \wlp{\prog}{f}\\[1.5pt]
          
        \end{array}
    $$
    \caption{Expectation transformer \wlpsymbol and
      \wpsymbol.  \newline $\mu f.\: F(f)$ (resp.~$\nu f.\: F(f)$) represents
      the least (resp. greatest) fixed
      point of expectation transformer $F$ \wrt the entailment order $\pimplies$. }
    \label{fig:wllp}
  \end{figure}


We now illustrate the application of $\wpsymbol$ by means of an example.
\begin{example}\label{ex:ej1}
Consider the program $\prog_1$ below that starts by assigning
   $1.5 - y^2$ to $x$, and then randomly
   decrements $x$, by either 1 or 2.
   \[
   \prog_1 \colon \quad \Ass{x}{1.5 - y^2};\; \PChoice{\Ass{x}{x-1}}{\nicefrac{1}{2}}{\Ass{x}{x-2}}
 \]
 To obtain the probability that the program establishes post-condition $x \geq
 0$, we proceed as follows:
 \begin{align*}
\MoveEqLeft[1]
  \wp{ \prog_1 }{\charFun{x \geq 0}} \\
& =~ \wp{\Ass{x}{1.5 - y^2}}{\wp{\PChoice{\Ass{x}{x-1}}{\nicefrac{1}{2}}{\Ass{x}{x-2}}}{\charFun{x \geq 0}}} \\
& =~ \wp{\Ass{x}{1.5 - y^2}}{\tfrac{1}{2} \,
    \wp{\Ass{x}{x-1}}{\charFun{x \geq 0}} + \tfrac{1}{2} \, 
    \wp{\Ass{x}{x-2}}{\charFun{x \geq 0}} } \\
& =~ \wp{\Ass{x}{1.5 - y^2}}{\tfrac{1}{2} \,
   \charFun{x-1 \geq 0} + \tfrac{1}{2} \, 
    \charFun{x-2 \geq 0}} \\  
& =~ \tfrac{1}{2} \,
   \charFun{(1.5 - y^2)-1 \geq 0} + \tfrac{1}{2} \, 
    \charFun{(1.5 - y^2)-2 \geq 0} \\    
& =~ \tfrac{1}{2} \, \underbrace{\charFun{y^2\leq -0.5}}_{=\:0} +\: \tfrac{1}{2} \, \charFun{y^2 \leq 0.5}  \\  
& =~ \tfrac{1}{2} \,
   \charFun{y^2 \leq 0.5} 
 \end{align*}

\noindent We can then conclude that the program establishes the
post-condition $x \geq 0$ with (exact) probability $\tfrac{1}{2}$, when
executed from an initial state where $y^2 \leq 0.5$, and with (exact)
probability $0$, otherwise. \hfill $\triangle$
\end{example}

%% file: vcgen.tex
\section{Verification Condition Generator}
\label{sec:VCGen}
In this section we present the VCGen (Definition~\ref{def:VCGen}) that
will serve as the axiomatic semantic of programs for slicing
purposes. We prove it sound (Lemma~\ref{thm:vcgpartial-sound}) and
establish other subsidiary properties (Lemmas~\ref{thm:vc-monot}
and~\ref{thm:alt-vcg}) required for proving the correctness of our
slicing approach. While the soundness of the VCGen is not
``explicitly'' used in our development, it legitimates the notion of
slicing based on specifications (Definition~\ref{def:slice}) that we
adopt.

Our ultimate goal here is to design a slicing technique that is
specification-preserving: Given a program with its purported
specification, we would like to synthesize a ``subset'' of the program
that still complies with the specification. A fundamental
requirement for the practical adoption of this---and any
other---slicing technique is that it is amenable to
automation. However, determining whether a program complies with a
given specification is known to be an undecidable problem (primarily
because of the undecidability of entailment in first-order logic).

To address this limitation, we draw on a well-known tool from
the program verification community: Verification Condition Generators
\mbox{(VCGens)}. A \mbox{VCGen} is a tool that given a program \emph{annotated with
  loop invariants}, together with its purported specification, 
generates a set of proof obligations, also known as \emph{verification
conditions}, such that their validity entails the program correctness \wrt the
specification. The key point here is that these so-generated
verification conditions can be typically discharged by automated theorem
provers such as SMT Solvers.

The classical approach for designing VCGens leverages
predicate transformers, or in the case of probabilistic programs,
expectation transformers. %
The transformer $\wprecsymbol$ that we use for designing our VCGen
(see Figure~\ref{fig:wprec}) is an adaptation of the transformers
$\wllpsymbol$ from Figure~\ref{fig:wllp}, deviating from them in the case of loops to support the automatization enabled by annotated invariants. More specifically, for any instruction different from a loop, $\wprecsymbol$ behaves like $\wllpsymbol$ transforming a post-expectation into the greatest (\ie the most precise) pre-expectation establishing the post-expectation. For a loop, it simply returns the annotated loop invariant. 
The intuition behind this latter rule is that the VCGen
will generate the necessary proof obligations to ensure that the
annotated invariant ($\inv$ in Figure~\ref{fig:wprec}) is
a valid pre-expectation---though possibly not the greatest---\wrt the
given post-expectation  ($\poste$ in Figure~\ref{fig:wprec}).

\begin{figure}[t]
    $$
        \begin{array}{lcl}
            \wprec{\Skip}{\poste}                 & = & \poste                                                              \\[1.5pt]
            \wprec{\Ass{x}{E}}{\poste}            & = & \poste\subst{x}{E}                                                  \\[1.5pt]
            \wprec{\Cond{G}{\prog_1}{\prog_2}}{\poste}    & = & \eval{G} \cdot\wprec{\prog_1}{\poste} +  \eval{ \lnot G} \cdot \wprec{\prog_2}{\poste} \\[1.5pt]
            \wprec{\PChoice{\prog_1}{p}{\prog_2}}{\poste} & = & p\cdot \wprec{\prog_1}{\poste} + (1-p)\cdot \wprec{\prog_2}{\poste}                    \\[1.5pt]
            \wprec{\prog_1;\prog_2}{\poste}               & = & \wprec{\prog_1}{\wprec{\prog_2}{\poste}}                                          \\[1.5pt]
             \wprec{\IWhileDo{G}{\inv}{\prog}}{\poste}  & = & \inv
        \end{array}
    $$
    \caption{Expectation transformer \wprecsymbol used for defining
      the VCGen.}
    \label{fig:wprec}
\end{figure}

The transformer $\wprecsymbol$ satisfies appealing algebraic properties,
which include monotonicity and linearity:

\begin{lemma}[Basic properties of transformer $\wprecsymbol$]\label{thm:wprec-props}
 For any \Lang program $\prog$, any two expectations $\pree, \pree'
 \colon \Exp$
 and any probability $p \in \ZO$, it holds:
 \begin{flushleft}
\begin{tabular}{@{}l@{\hspace{3em}}l}
  	\textnormal{Monotonicity:}			& $\pree
                                                   \pimplies \pree'
                                                   ~~\implies~~
                                                   \wprec{\prog}{\pree} \pimplies \wprec{\prog}{\pree'}$\\[1ex]

	\textnormal{Linearity:} 			& $\wprec{\prog}{p
                                                  \pree + (1-p) 
                                                  \pree'} ~=~ p \cdot
\wprec{\prog}{\pree} + (1-p) \cdot \wprec{\prog}{\pree'}$\\[1ex]
	%
\end{tabular}
\end{flushleft}
\end{lemma}

\begin{proof}
 Both proofs proceed by induction on the program structure. For the
 case of loops, the results are immediate since the transformer is
 constant (always yielding the annotated loop invariant). For the
 remaining cases, the proofs follow the same arguments as for
 transformer \wpsymbol; see, \eg, \cite{kaminski2019advanced}.
\end{proof}

Having introduced the expectation transformer $\wprecsymbol$, we are now
in a position to define the VCGen for probabilistic programs. For
making the presentation more incremental, in this section we introduce the VCGen for establishing
partial correctness specifications only, and defer the treatment of total correctness to
Section~\ref{sec:extension}.

The VCGen takes a \Lang program $\prog$, a pre-expectation $\pree$ and
a post-expectation $\poste$, and returns a set  $\vcg{\pree}{\prog}{\poste}$ of verification
conditions such that
their validity entails that $\prog$ adheres to the specification
given by $\pree$ and $\poste$. The returned verification conditions are
entailment claims between expectations, \ie claims of the form $\pree'
\pimplies \poste'$ (which generalize the entailment between predicates
returned by VCGens for deterministic programs).


\begin{definition}[VCGen for partial correctness]\label{def:VCGen}
  The set of verification conditions $\vcg{\pree}{\prog}{\poste}$ for
  the partial correctness of a \Lang program $\prog$ \wrt pre-expectation $\pree$ and
  post-expectation $\poste$ is defined as:
  \[
  \vcg{\pree}{\prog}{\poste} ~=~ \{ \pree \pimplies
  \wprec{\prog}{\poste} \} \, \cup \, \vc{\prog}{\poste}~,
\]
where $\vc{\prog}{\poste}$ is defined in Figure~\ref{fig:VCGen}, by induction on the structure of
$\prog$.
\end{definition}

To extract the verification conditions, $\vcg{\pree}{\prog}{\poste}$
proceeds roughly as follows. First, it leverages transformer
$\wprecsymbol$ to compute a valid pre-expectation $\wprec{\prog}{\poste}$ that establishes the
declared post-expectation $\poste$, and then verifies that the declared
pre-expectation $\pree$ entails the so-computed pre-expectation. This
generates verification condition $\pree \pimplies
\wprec{\prog}{\poste}$.

However, while computing the pre-expectation
$\wprec{\prog}{\poste}$, the VCGen makes two assumptions about $\prog$ loops
that must be accounted for: On the one hand, that the annotated
invariants are indeed invariants, that is, that they are preserved
by the body of the respective loops. On the other hand, that the
invariants are strong enough as to establish the expectations that
should hold upon exit of the loops (as computed by transformer
$\wprecsymbol$). The verification conditions accounting for these
assumptions are captured by $\vc{\prog}{\poste}$ (see Figure~\ref{fig:VCGen}).

\begin{figure}[t]
    $$
        \begin{array}{lcl}
            \vc{\Skip}{\poste}              & = & \emptyset                                                                                                   \\[1.5pt]
            \vc{\Ass{x}{E}}{\poste}            & = & \emptyset                                                                                                   \\[1.5pt]
            \vc{\Cond{G}{\prog_1}{\prog_2}}{\poste}    & = & \vc{\prog_1}{\poste}
                                                 \:\cup\:  \vc{\prog_2}{\poste}                                                                                \\[1.5pt]
            \vc{\PChoice{\prog_1}{p}{\prog_2}}{\poste} & = & \vc{\prog_1}{\poste} \:\cup\: \vc{\prog_2}{\poste}                                                                              \\[1.5pt]
            \vc{\prog_1;\prog_2}{\poste}           & = & \vc{\prog_1}{\wprec{\prog_2}{\poste}} \:\cup\: \vc{\prog_2}{\poste}                                                                      \\[1.5pt]
            \vc{\IWhileDo{G}{\inv}{\prog}}{\poste}  & = & \left\{
                                                          \charFun{G}
                                                          \cdot \inv
                                                  \Rrightarrow
                                                  \wprec{\prog}{\inv},
                                                  \right.\\ & &\left.~ %
                                                              \charFun{\lnot
                                                                G}
                                                                \cdot \inv
                                                                \Rrightarrow
                                                                \poste
                                                                \right\}
                                                                \:\cup\: \vc{\prog}{\inv}
        \end{array}
    $$
    \caption{Verification condition generator for partial correctness
      of \Lang programs.}
    \label{fig:VCGen}
  \end{figure}

  Rules defining $\vc{\prog}{\poste}$ are mostly
  self-explanatory. The most important rule is the one for
  loops, as these are the only instructions that generate verification
  conditions. Concretely, $\vc{\IWhileDo{G}{\inv}{\prog}}{\poste}$
  extends the potential set of verification conditions induced by (the
  loops in) $\prog$ with two additional verification conditions: \emph{i}) $
  \eval{G} \cdot \inv \Rrightarrow \wprec{\prog}{\inv}$, which ensures that
  $\inv$ is indeed a loop invariant, and \emph{ii}) $\eval{\lnot
    G} \cdot \inv \pimplies \poste$, which ensures that upon loop exit, the invariant is strong
    enough as to establish post-expectation $\poste$. The
    remaining rules simply collect the verification conditions
    generated by loops. The only subtlety appears in the rule for
    sequential composition, where $\vc{\prog_1;\prog_2}{\poste}$ collects the
    verification conditions generated by $\prog_1$ applying  $\vcd{\prog_1}$ to
   $\wprec{\prog_2}{\poste}$ as this is the post-expectation of
   $\prog_1$ (yielded by \wprecsymbol)
   when $\poste$ is the post-expectation of $\prog_1; \prog _2$.


   

We next establish three relevant properties of the VCGen. First, the VCGen is \emph{sound} meaning that the validity of the verification conditions $\vcg{\pree}{\prog}{\poste}$ entail the validity of partial correctness specification $\pspec{\pree}{\prog}{\poste}$.

\begin{lemma}[Soundness of $\vcgsymbol$]\label{thm:vcgpartial-sound}
For any $\Lang$ program $\prog$ and any two expectations $\pree, \poste \colon \Exp$,
\[
 \models \vcg{\pree}{\prog}{\poste}  \quad \implies \quad \models \pspec{\pree}{\prog}{\poste}~.
\]
\end{lemma}
\begin{proof}
  The result follows as an immediate corollary of the following property:
  \[
    \models \vc{\prog}{\poste} %
    \quad \implies \quad %
    \wprec{\prog}{\poste} \pimplies \wlp{\prog}{\poste}~,
  \]
  which can be established by induction on the structure of $\prog$. See Appendix~A for details.
\end{proof}

The remaining two properties are required to prove the correctness of
the slicing techniques from Section~\ref{sec:slicing}
(Theorems~\ref{thm:slice-top} and \ref{thm:slice-nested}). One
property is the monotonicity of $\vcd{\prog}$ and $\vcgd{\prog}$:
   
\begin{lemma}[Monotonicity of  $\vcsymbol$/$\vcgsymbol$]\label{thm:vc-monot}
For any \Lang program $\prog$ and any four expectations $\pree,
\pree', \poste,
\poste' : \Exp$,
\[
  \begin{array}{c @{\qquad}c@{\qquad} c}
  \poste \pimplies \poste'   & \implies & \models
 \vc{\prog}{\poste} ~\Rightarrow~ \models \vc{\prog}{\poste'}~,\\[1ex]
\pree' \pimplies \pree \: \land \: \poste \pimplies \poste'   & \implies & \models
 \vcg{\pree}{\prog}{\poste} ~\Rightarrow~ \models \vcg{\pree'}{\prog}{\poste'}~.\\
  \end{array}
\]
\end{lemma}
\begin{proof}
Both monotonicity proofs rely on the monotonicity of $\wprecd{\prog}$
(Lemma~\ref{thm:wprec-props}). The monotonicity proof of $\vcd{\prog}$
proceeds by routine induction on the structure of $\prog$ (see
Appendix~A for details). The monotonicity proof of $\vcgd{\prog}$
follows as an immediate corollary. 
\end{proof}

The last property is an alternative characterization of
$\models \vcg{\pree}{\prog}{\poste}$ for the case where $\prog$
contains compound instructions featuring subprograms, that is,
conditional branches, probabilistic choices or loops. To state the
result we need variants of $\wprec{\prog}{\poste}$ and
$\vc{\prog}{\poste}$ that act on
$\prog$ suffixes, and variants of $\vcg{\pree}{\prog}{\poste}$
that act on $\prog$ suffixes and prefixes. Assuming that
$\prog = \seqn{1}{n}$, we then define:
\begin{align*}
\wpreci{j}{\prog}{\poste} &~=~
\begin{cases}
  \wprec{\inst_j;\seqn{j+1}{n}}{\poste} &\text{if $1 \leq j \leq n$}\\
  \poste &\text{if $j = n + 1$}
\end{cases}\\[1ex]
\vci{j}{\prog}{\poste} &~=~
\begin{cases}
  \vc{\inst_j;\seqn{j+1}{n}}{\poste} &\text{if $1 \leq j \leq n$}\\
  \emptyset &\text{if $j = n + 1$}
\end{cases}\\[1ex]
  \vcgii{j}{\pree}{\prog}{\poste} &~=~
\begin{cases}
  \vcg{\pree}{\inst_j;\seqn{j+1}{n}}{\poste} &\text{if $1 \leq j \leq n$}\\
  \emptyset  &\text{if $j = n+1$}
\end{cases}\\[1ex]
  \vcgi{j}{\pree}{\prog}{\poste} &~=~
\begin{cases}
  \vcg{\pree}{\seqn{1}{j}}{\poste} &\text{if $1 \leq j \leq n$}\\
  \{ \pree \pimplies \poste \}  &\text{if $j = 0$}
\end{cases}
\end{align*}

\begin{lemma}[Alt.~characterization of $\models \vcgsymbol$]\label{thm:alt-vcg}
For any \Lang program $\prog = \seqn{1}{n}$ and any two expectations $\pree,
\poste : \Exp$,
\begin{enumerate}
  \item If $\inst_j = \Cond{G}{\prog_1}{\prog_2}$ or  $\inst_j = \PChoice{\prog_1}{p}{\prog_2}$ for some $1 \leq j
    \leq n$, then 
    \[
      \models \vcg{\pree}{\prog}{\poste}
      \qquad \text{iff} \qquad
      \begin{array}[t]{l}
        \models \vci{j+1}{\prog}{\poste}~~\land\\[1ex]
        \models \vc{\prog_1}{\wpreci{j+1}{\prog}{\poste}}~~\land\\[1ex]
        \models \vc{\prog_2}{\wpreci{j+1}{\prog}{\poste}}~~\land\\[1ex]
         \models \vcgi{j-1}{\pree}{\prog}{\wpreci{j}{\prog}{\poste}}~.
      \end{array}
    \]%
\item If $\inst_j = \IWhileDo{G}{\inv}{\prog'}$ for some $1 \leq j
    \leq n$, then
    \[
      \models \vcg{\pree}{\prog}{\poste}
      \qquad \text{iff} \qquad
      \begin{array}[t]{l}
        \models \vcgii{j+1}{\charFun{\lnot G} \cdot \inv}{\prog}{\poste}~~\land\\[1ex]
        %
        %
        \models \vcg{\charFun{G} \cdot \inv}{\prog'}{\inv}~~\land\\[1ex]
         \models \vcgi{j-1}{\pree}{\prog}{\inv}~.
      \end{array}
    \]
  \end{enumerate}
\end{lemma}

\begin{proof}
It follows from the definition of $\models
\vcg{\pree}{\prog}{\poste}$, by splitting $\prog = \seqn{1}{n}$ into the prefix
before $\inst_j$, $\inst_j$ and the suffix after $\inst_j$.  See
Appendix~A for details.
\end{proof}

%% file: slicing.tex
\section{The Slice Transformation}
\label{sec:slicing}
In this section we present the two fundamental results
(Theorems~\ref{thm:slice-top} and \ref{thm:slice-nested}) that allow
identifying removable program fragments and underlie our slicing
approach. We prove the theorems correct and show application
examples (Examples~\ref{ex:pc1} and \ref{ex:randint}). This pair of theorems
form the cornerstone of our theoretical contribution. 

\medskip
\subsection{Specification-based Slice}

Roughly speaking, given a \Lang program $\prog$ together with its
specification,
a \emph{specification-based slice} is obtained by removing from $\prog$
those fragments that do not contribute to establishing the
specification. Thus, the notion of specification-based slice involves
a syntactic and a semantic component that we formally define next.

The syntactic component is captured by the relation
``\emph{being-portion-of}''  over programs, denoted by
``$\preccurlyeq$''. Informally, $\prog' \preccurlyeq \prog$ if
$\prog'$ is obtained from $\prog$ by removing some instructions. The 
relation is formally defined by the set of rules in
Figure~\ref{fig:is-portion-of}. The first two rules say
that we can obtain a portion
of a program consisting in a sequence of $n$ instructions by removing
either all its instructions (resulting in \Skip) or a proper
subsequence of contiguous instructions. The following four rules
represent congruence rules  for the sequential composition,
conditional branching, probabilistic choice and loops. Finally, the
last two rules encode the reflexivity and transitivity of the
relation.\footnote{In view of the congruence rule for sequential
  composition and the rule stating that $\Skip$ is a portion of any
  program, we could have discarded the rule that allows removing a
  proper subsequence of instructions of a program to obtain a portion
  thereof. However, we preferred to keep it because it yields cleaner
  program slices, \eg,  $\inst_1; \inst_5$ instead of $\inst_1; \Skip;
  \Skip; \Skip; \inst_5$, and also simplifies, to some degree, the
  proofs.}

\begin{figure}[t]
    \[
\begin{array}{c}
\infrule{}{\Skip \preccurlyeq \seqn{1}{n}}  \qquad %
\infrule{1<j \leq k \leq n \text { or } 1 \leq j \leq k<n}{\seqn{1}{j-1} ; \seqn{k+1}{n} \preccurlyeq
          \seqn{1}{j}; \ldots ; \seqn{k}{n}} \\[3.5ex]
\infrule{\inst_{j}^{\prime} \preccurlyeq
        \inst_{j} \qquad 1\leq j \leq n}{\inst_{1};\dots ;\inst_{j}^{\prime};\dots ;\inst_{n} \preccurlyeq
        \inst_{1};\dots ;\inst_{j};\dots ;\inst_{n}} \\[3.5ex]
\infrule{\prog_{1}^{\prime} \preccurlyeq
          \prog_{1} \qquad  \prog_{2}^{\prime} \preccurlyeq
          \prog_{2} }{ \Cond{G}{\prog_{1}^{\prime}}{\prog_{2}^{\prime}} \preccurlyeq
          \Cond{G}{\prog_{1}}{\prog_{2}} }   \\[3.5ex]
\infrule{\prog_{1}^{\prime} \preccurlyeq
          \prog_{1} \qquad \prog_{2}^{\prime} \preccurlyeq
          \prog_{2} }{ \PChoice{\prog_{1}^{\prime}}{p}{\prog_{2}^{\prime}} \preccurlyeq
          \PChoice{\prog_{1}}{p}{\prog_{2}} }   \\[3.5ex]
\infrule{\prog^{\prime} \preccurlyeq
          \prog}{\IWhileDo{G}{\inv}{\prog^{\prime}} \preccurlyeq
          \IWhileDo{G}{\inv}{\prog}} \\[3.5ex] 
\infrule{}{\prog \preccurlyeq \prog} \qquad \qquad
\infrule{\prog_1 \preccurlyeq \prog_2 \qquad \prog_2 \preccurlyeq \prog_3}{\prog_1 \preccurlyeq \prog_3} \\[3.5ex]  
\end{array}
\]
    \caption{Relation ``\emph{is-portion-of}'' over programs.}
    \label{fig:is-portion-of}
  \end{figure}

  As for the semantic component, we assume that the semantics of a
  program $\prog$ is given by the verification condition generator
  $\vcgd{\prog}$, or said otherwise, that a program $\prog$ satisfies
  a specification given by, say pre-expectation $\pree$ and
  post-expectation $\poste$, if and only if
  \mbox{$\models \vcg{\pree}{\prog}{\poste}$}. The ``if'' direction
  refers to the \emph{soundness} of the VCGen and was already
  established in Lemma~\ref{thm:vcgpartial-sound}. The ``only if''
  direction refers to the \emph{completeness} of the VCGen, that is,
  if a program satisfies a specification, then it is always possible
  to annotate the program with appropriate loop invariants such that
  the VCGen can establish the specification. In fact, one can prove
  that the exact semantics of a loop \wrt a post-expectation as given
  by transformer $\wlpsymbol$ is always a valid invariant, strong
  enough as to establish the post-expectation. For the rest of our
  development, we thus assume that programs are annotated with
  appropriate loop invariants as to establish the purported
  specification. (This is also a natural assumption for any other
  automated program verification task.)



\begin{definition}[Program slicing based on partial correctness specification~\cite{Barros:2012}]\label{def:slice}
  We say that \Lang program $\prog'$ is a \emph{specification-based
    slice} of \Lang program $\prog$ with respect to the partial
  correctness specification given by pre-expectation $\pree$ and
  post-expectation $\poste$, written
  $\pspecSlice{\prog'}{\pree}{\poste}{\prog}$, iff
    \begin{enumerate}
        \item $\prog' \preccurlyeq  \prog$, and
        \item $\models \vcg{\pree}{\prog}{\poste} ~\Rightarrow~ \models \vcg{\pree}{\prog'}{\poste}$
    \end{enumerate}
\end{definition}

Observe that it only makes sense to compute specification-based slices
of programs that adhere to their specifications: If a program violates
its declared specification, then any portion of the program (including,
\eg, \Skip) is a vacuously valid specification-based slice.

\medskip
\subsection{Removing Instructions}
\label{sec:slicepc}
We next present our two fundamental results for deriving 
specification-based slices of probabilistic programs. The first result
allows removing top-level instructions of a program, and the second
result, nested instructions.  


\subsubsection{Removing top-level instructions}
Given a program $\prog = \seqn{1}{n}$, the first result allows
slicing away a contiguous subsequence of instructions. We thus begin 
introducing the function $\removesymbol \colon \Lang \To \Lang$ that
captures this program transformation:
\[
\remove{j}{k}{\prog}~=~\begin{cases}
    \Skip                       & \quad\text{if}~ j=1 ~ \text{and}~ k=n \\
    \seqn{1}{j-1};\seqn{k+1}{n} & \quad\text{otherwise}
  \end{cases}
\]
In words, $\remove{i}{j}{\prog}$ slices away from $\prog$ from the 
$j$-th to $k$-th  instructions, inclusive.

The slicing criteria requires propagating the program
post-expectation, say $\poste$, backward, along all its
instructions, that is,
calculating $\wpreci{j}{\prog}{\poste}$ for all $j=1,\ldots,n$. Then,
if for some $1 \leq j \leq k \leq n$,  $\wpreci{j}{\prog}{\poste}$
happens to entail  $\wpreci{k+1}{\prog}{\poste}$, we can remove the
subsequence of instructions from $j$-th to $k$-th.

\begin{theorem}[Removing top-level instructions for partial correctness]\label{thm:slice-top}
Let $\prog = \seqn{1}{n}$ be a \Lang program together with its respective pre- and
post-expectation $\pree$ and $\poste$. Moreover, let $1\leq j \leq k
\leq n$. If \[\wpreci{j}{\prog}{\poste} ~\pimplies~ 
\wpreci{k+1}{\prog}{\poste}\] then, \[\pspecSlice{\remove{j}{k}{\prog}}{\pree}{\poste}{\prog}~.\]
\end{theorem}

\begin{proof}
We show that $\models \vcg{\pree}{\prog}{\poste}$ entails $\models
\vcg{\pree}{\remove{j}{k}{\prog}}{\poste}$:
\begin{align*}
                  & \models \vcg{\pree}{\prog}{\poste}                                                                                                   \\
  \Leftrightarrow & \qquad \by{def of $\vcgsymbol$}\displaybreak[0]                                                                                      \\
                  & \models  \{\pree \pimplies \wprec{\prog}{\poste} \} \, \cup \, \vc{\prog}{\poste}                                                    \\
  \Leftrightarrow & \qquad \by{$\prog=\seqn{1}{n}$  }\displaybreak[0]                                                                                    \\
                  & \models \{\pree \pimplies \wprec{\seqn{1}{n}}{\poste} \} \, \cup \, \vc{\seqn{1}{n}}{\poste}                                         \\
  \Leftrightarrow & \qquad \by{def of \wprecsymbol and \vcsymbol for sequential composition}\displaybreak[0]                                             \\
                  & \models \{\pree \pimplies \wprec{\seqn{1}{j-1}}{\wpreci{j}{\prog}{\poste}} \}                                                        \\
                  & \phantom{\models} \, \cup \, \vc{\seqn{1}{j-1}}{\wpreci{j}{\prog}{\poste}} \, \cup \, \vc{\seqn{j}{n}}{\poste}                       \\
  \Rightarrow     & \qquad \by{hypothesis, monotonicity of $\vcsymbol$ and $\wprecsymbol$}\displaybreak[0]                                               \\
                  & \models \{\pree \pimplies \wprec{\seqn{1}{j-1}}{\wpreci{k+1}{\prog}{\poste}} \}                                                      \\
                  & \phantom{\models} \, \cup \, \vc{\seqn{1}{j-1}}{\wpreci{k+1}{\prog}{\poste}}\, \cup \, \vc{\seqn{j}{n}}{\poste}                      \\
  \Leftrightarrow & \qquad \by{def of \vcsymbol for sequential composition}\displaybreak[0]                                                              \\
                  & \models \{\pree \pimplies \wprec{\seqn{1}{j-1}}{\wpreci{k+1}{\prog}{\poste}} \}                                                      \\
                  & \phantom{\models} \, \cup \, \vc{\seqn{1}{j-1}}{\wpreci{k+1}{\prog}{\poste}}\, \cup \, \vc{\seqn{j}{k}}{\wpreci{k+1}{\prog}{\poste}} \\
                  & \phantom{\models} \, \cup \, \vc{\seqn{k}{n}}{\poste}                                                                                \\
  \Leftrightarrow & \qquad \by{associativity and def of \vcsymbol for sequential composition}\displaybreak[0]                                                                                              \\
                  & \models \{\pree \pimplies \wprec{\remove{j}{k}{\prog}}{\poste} \}\, \cup \, \vc{\remove{j}{k}{\prog}}{\poste}                        \\
                  & \phantom{\models} \, \cup \, \vc{\seqn{j}{k}}{\wpreci{k+1}{\prog}{\poste}}                                                           \\
  \Leftrightarrow & \qquad \by{def of $\vcgsymbol$}\displaybreak[0]                                                                                      \\
                  & \models \vcg{\pree}{\remove{j}{k}{\prog}}{\poste}\, \cup \, \vc{\seqn{j}{k}}{\wpreci{k+1}{\prog}{\poste}}                            \\
  \Rightarrow     & \qquad \by{weakening}\displaybreak[0]                                                                    \\
                  & \models \vcg{\pree}{\remove{j}{k}{\prog}}{\poste} \qedhere                                                                                   
\end{align*}
\end{proof}

We next illustrate the application of Theorem~\ref{thm:slice-top} to
slice the program from  Example~\ref{ex:ej1}.

\begin{example}\label{ex:pc1}
 Consider the program $\prog_1$ from Example~\ref{ex:ej1},
  with pre-expectation $\pree=\tfrac{1}{2} \, \charFun{y^2 \leq 0.5}$ and
  post-expectation $\poste=\charFun{x \geq 0}$. Below we display the program,
  along with the expectations $\wpreci{j}{\prog_1}{\poste}$
  (abbreviated $\wprecid{j}$) for $j=1,2,3$, that are obtained by
  propagating $\poste$ backward.
\[
  \begin{array}{r@{\qquad}l}
   \scalemath{0.8}{f:}     & \codeComment{$\tfrac{1}{2} \, \charFun{y^2 \leq 0.5}$} \\
    \scalemath{0.8}{\wprecid{1}:} & \codeComment{$\tfrac{1}{2} \, \charFun{1.5 - y^2
                          \geq 1}+ \tfrac{1}{2} \, \charFun{1.5 - y^2 \geq 2}$} \\
    \scalemath{0.8}{i_1:}  & \Ass{x}{1.5 - y^2}; \\
    \scalemath{0.8}{\wprecid{2}:} & \codeComment{$\tfrac{1}{2} \, \charFun{x
                          \geq 1}+ \tfrac{1}{2} \, \charFun{x \geq 2}$} \\
    \scalemath{0.8}{i_2:} & \PChoice{\Ass{x}{x-1}}{\nicefrac{1}{2}}{\Ass{x}{x-2}} \\
    \scalemath{0.8}{\wprecid{3} = g:}    & \codeComment{$\charFun{x \geq 0}$}
  \end{array}
\]
By doing a case analysis on the value that variable $x$ can have in an
arbitrary state $\state$, taking $x \in (-\infty, 0)$, $x \in [0,1)$, $x \in
[1,2)$ or $x \in [2,\infty)$, it is not hard to see that in all four
cases, $\wpreci{2}{\prog_1}{\poste}(s) \leq
\wpreci{3}{\prog_1}{\poste}(s)$. In other words, \[\wpreci{2}{\prog_1}{\poste} ~\pimplies~
\wpreci{3}{\prog_1}{\poste}~,\] which in view of
Theorem~\ref{thm:slice-top} allows us to slice away the probabilistic
choice $\inst_2$ from $\prog_1$, while preserving its specification.\hfill $\triangle$
\end{example}

\subsubsection{Removing nested instructions}
Given a program $\prog = \seqn{1}{n}$, Theorem~\ref{thm:slice-top}
allows slicing away ``top-level'' instructions of $\prog$. For
example, if for some $1 \leq j \leq n$, instruction $\inst_j$ is a conditional
branching, Theorem~\ref{thm:slice-top} allows slicing away the entire
conditional branches. However, in some circumstances, we may obtain a
valid slice by removing instructions from either of its branches,
only. In general, this may be the case for any other \emph{compound}
instruction $\inst_j$ of $\prog$ such as a probabilistic choice or
a loop. Next, we present a complementary result to
Theorem~\ref{thm:slice-top} that enables this kind of slice.

To state the result, we need the notion of \emph{local
  specification}. Intuitively, if $\prog$ is a program \eg with a
conditional branching, then any specification of $\prog$ induces a ``local
specification'' on each of the two branches. In turn, if one of the
banches contains \eg a loop, the local specification of the branch induces
a (deeper) local specification on the loop body. Notationwise, we
write  
\[\localSpecp{\pree}{\prog}{\poste}{\pree'}{\prog'}{\poste'}\]  
to denote that specification $\pspec{\pree}{\prog}{\poste}$ induces local specification
$\pspec{\pree'}{\prog'}{\poste'}$ on the subprogram $\prog'$ of
$\prog$. The relation \localSpecSymbol is formally defined by the set
of rules in Figure~\ref{fig:local-spec}.

Let us briefly explain the rules. Assume that the specification of the
program at hand $\prog = \seqn{1}{n}$ is given by pre-expectation~$\pree$ and post-expectation~$\poste$. Furthermore, assume that its
instruction $\inst_j$ is compound. If $\inst_j$ is a conditional
branching, then the local specification induced on either of its
branches is as follows: the post-expectation is obtained by
propagating $\poste$ backward along $\prog = \seqn{1}{n}$, until reaching
$\inst_j$; the pre-expectation is obtained by further propagating the
so-calculated post-expectation along the branch, and restricting the
result to the branch respective guard  (rules \lrule{\localSpecSymbol\texttt{ift}}  and \lrule{\localSpecSymbol\texttt{iff}}). If $\inst_j$ is a probabilistic
choice, the local specification induced on either of its branches 
is defined similarly, except that pre-expectations are not guarded (rules \lrule{\localSpecSymbol\texttt{pl}}  and \lrule{\localSpecSymbol\texttt{pr}}). 
If $\inst_j$ is a loop, the local specification induced on its
body is as follows: the post-expectation is the loop invariant, and
the pre-expectation is the loop invariant, restricted to the loop
guard (rule \lrule{\localSpecSymbol\texttt{while}}). Finally, these
definitions can be applied recursively, to yield the local
specification of a subprogram at any depth level of the original program (rule \lrule{\localSpecSymbol\texttt{trans}}).

\begin{figure}[t]
    \[
\begin{array}{c}
  \infrule{\inst_j = \Cond{G}{\prog_1}{\prog_2}}%
  {\localSpecp{\pree}{\prog}{\poste}{\charFun{G} \cdot
  \wprec{\prog_1}{\wpreci{j+1}{\prog}{\poste}}
  }{\prog_1}{\wpreci{j+1}{\prog}{\poste}}}~\lrule{\localSpecSymbol\texttt{ift}} \\[4.5ex]
  %
  \infrule{\inst_j = \Cond{G}{\prog_1}{\prog_2}}%
  {\localSpecp{\pree}{\prog}{\poste}{\charFun{\lnot G} \cdot \wprec{\prog_2}{\wpreci{j+1}{\prog}{\poste}} }{\prog_2}{\wpreci{j+1}{\prog}{\poste}}}~\lrule{\localSpecSymbol\texttt{iff}}  \\[4.5ex]
  %
  \infrule{\inst_j = \PChoice{\prog_1}{p}{\prog_2}}%
  {\localSpecp{\pree}{\prog}{\poste}{\wprec{\prog_1}{\wpreci{j+1}{\prog}{\poste}} }{\prog_1}{\wpreci{j+1}{\prog}{\poste}}}~\lrule{\localSpecSymbol\texttt{pl}}  \\[4.5ex]
  %
  \infrule{\inst_j = \PChoice{\prog_1}{p}{\prog_2}}%
  {\localSpecp{\pree}{\prog}{\poste}{\wprec{\prog_2}{\wpreci{j+1}{\prog}{\poste}} }{\prog_2}{\wpreci{j+1}{\prog}{\poste}}}~\lrule{\localSpecSymbol\texttt{pr}}  \\[4.5ex]
  %
  \infrule{\inst_j = \IWhileDo{G}{\inv}{\prog^{\prime} }}%
  {\localSpecp{\pree}{\prog}{\poste}{\charFun{G} \cdot \inv}{\prog^{\prime}}{\inv}}~\lrule{\localSpecSymbol\texttt{while}}  \\[4.5ex] 
        %
%
\infrule{}{\localSpecp{\pree}{\prog}{\poste}{\pree}{\prog}{\poste}}~\lrule{\localSpecSymbol\texttt{refl}}  \\[4.5ex] 
%
  \infrule{\localSpecp{\pree_1}{\prog_1}{\poste_1}{\pree_2}{\prog_2}{\poste_2}%
  \qquad \localSpecp{\pree_2}{\prog_2}{\poste_2}{\pree_3}{\prog_3}{\poste_3}}%
  {\localSpecp{\pree_1}{\prog_1}{\poste_1}{\pree_3}{\prog_3}{\poste_3}}~\lrule{\localSpecSymbol\texttt{trans}}
\end{array}
\]
    \caption{Relation of local specification inducement. \newline For the first
    five rules we assume that $\prog = \seqn{1}{n}$ and $1 \leq j \leq n$.}
    \label{fig:local-spec}
  \end{figure}

  The value of local specifications resides in that they allow a
  \emph{modular} approach to slicing: 
  If a program with its specification induces a local specification on
  a given subprogram, then slicing the subprogram \wrt the local
  specification yields a valid slice of the original program (\wrt to
  its original specification).
  
\begin{theorem}[Removing nested instructions for partial correctnes]\label{thm:slice-nested}
 Let $\prog$ be a \Lang program together with its respective pre- and
post-expectation $\pree$ and $\poste$, and let $\prog'$ be a
subprogram of $\prog$ such that
$\localSpecp{\pree}{\prog}{\poste}{\pree'}{\prog'}{\poste'}$. If
\[
  \pspecSlice{\prog''}{\pree'}{\poste'}{\prog'}~,
\]
then
\[
  \pspecSlice{\prog\subst{\prog'}{\prog''}}{\pree}{\poste}{\prog}~,
\]
where $\prog\subst{\prog'}{\prog''}$ denotes the program that is
obtained from $\prog$ by replacing $\prog'$ with~$\prog''$. 
\end{theorem}
\begin{proof}
By induction on the derivation of
$\localSpecp{\pree}{\prog}{\poste}{\pree'}{\prog'}{\poste'}$. See
Appendix~A for details.
\end{proof}

Theorem~\ref{thm:slice-nested} embodies a local reasoning principle,
which is crucial for the simplicity (and elegance) of the technique, and
for keeping the computation of slices tractable.  

We now illustrate the application of Theorem~\ref{thm:slice-nested},
and more broadly, the application of specification-based slicing for
software reuse.

\begin{example}\label{ex:randint}
Consider the program below, that assigns to variable $r$ a random integer uniformly
distributed in the interval $[0,16)$:
\[
  \begin{array}{l@{\qquad}l}
    & \codeComment{$\tfrac{1}{16} \charFun{0 \leq K < 16}$} \\
    & \PChoice{\Ass{b_0}{0}}{\nicefrac{1}{2}}{\Ass{b_0}{1}}; \\[0.5ex]
               & \PChoice{\Ass{b_1}{0}}{\nicefrac{1}{2}}{\Ass{b_1}{1}};
    \\[0.5ex]
               & \PChoice{\Ass{b_2}{0}}{\nicefrac{1}{2}}{\Ass{b_2}{1}};
    \\[0.5ex]
               & \PChoice{\Ass{b_3}{0}}{\nicefrac{1}{2}}{\Ass{b_3}{1}};
    \\[0.5ex]
    & \Ass{r}{b_0 + 2 b_1 + 4 b_2 + 8 b_3}\\
    & \codeComment{$\charFun{r=K}$} \\
  \end{array}
\]
Intuitively, it encodes $r$ as a four-bit binary number (we assume that $b_0,\ldots,b_3$
are $\{0,1\}$-valued variables) and randomly
assigns a value to each bit. Since all four bits and uniformly and 
independently distributed, $r$ takes each of the values $0,\ldots,15$
with probability
$\left(\nicefrac{1}{2}\right)^4=\nicefrac{1}{16}$. Formally, the program satisfies the specification
given by pre-expectation $\tfrac{1}{16} \charFun{0 \leq K < 16}$ and
post-expectation $\charFun{r=K}$. 

Now assume we would like to reuse the program in another
context where a random integer is required, but instead of requiring
that the integer be uniformly distributed in the interval $[0,16)$,
the context only requires that the integer be at least 8 with probability
(at least) $\nicefrac{1}{2}$. We can then slice the program with
respect to this weaker specification, given by pre-expectation $\pree
= \tfrac{1}{2}$ and post-expectation $\poste = \charFun{r \geq 8}$. Propagating the
post-expectation backward along the program and calculating the local
specification induced over the left branch of the last 
probabilistic choice yields the result below. Therein, for
convenience, we use $s_n$ as a shorthand for the partial sum $\sum_{i=0}^n
2^i b_i$.

\[
  \begin{array}{r@{\quad}l}
   & \codeComment{$f =  \tfrac{1}{2}$} \\
    & \codeComment{$\wprecid{1} = \sum_{j \in \{0,\ldots,15\}} \tfrac{1}{16}
                    \charFun{j \geq 8}$} \\[0.5ex]
    \scalemath{0.8}{i_1:}  & \PChoice{\Ass{b_0}{0}}{\nicefrac{1}{2}}{\Ass{b_0}{1}}; \\[0.5ex]
    & \codeComment{$\wprecid{2} = \sum_{j \in \{0,2,4,\ldots,12,14\}} \tfrac{1}{8}
                    \charFun{s_0 + j \geq 8}$} \\[0.5ex]
    \scalemath{0.8}{i_2:}  & \PChoice{\Ass{b_1}{0}}{\nicefrac{1}{2}}{\Ass{b_1}{1}}; \\[0.5ex]
    & \codeComment{$\wprecid{3} = \sum_{j \in \{0,4,8,12\}} \tfrac{1}{4}
                    \charFun{s_1 + j \geq 8}$} \\[0.5ex]
    \scalemath{0.8}{i_3:}  & \PChoice{\Ass{b_2}{0}}{\nicefrac{1}{2}}{\Ass{b_2}{1}}; \\[0.5ex]
    & \codeComment{$\wprecid{4} = \sum_{j \in \{0,8\}} \tfrac{1}{2}
                    \charFun{s_2 + j \geq 8}$} \\[0.5ex]
    \scalemath{0.8}{i_4:}  & \PChoice{\codeComment{$\pree_3$} \quad
                             \Ass{b_3}{0} \quad \codeComment{$\poste_3$} }{\nicefrac{1}{2}}{\Ass{b_3}{1}}; \\[0.5ex]
    & \codeComment{$\wprecid{5} = \charFun{s_3 \geq 8}$} \\[0.5ex]
    \scalemath{0.8}{i_5:}  &  \Ass{r}{b_0 + 2 b_1 + 4 b_2 + 8 b_3} \\[0.5ex]
   & \codeComment{$\wprecid{6} = g = \charFun{r \geq 8}$} \\[1.5ex]
  \end{array}
\]
where  $\pree_3 = \charFun{s_2 \geq 8}$ and $\poste_3 = \charFun{s_3
  \geq 8}$. 
%
%
%
Observe that since $s_3 \geq s_2$, it holds that
$\pree_3 \pimplies \poste_3$ 
and in view of Theorem~\ref{thm:slice-nested}, we can remove
the left branch ($\Ass{b_3}{0}$) of the probabilistic choice
initializing $b_3$.

Furthermore, appealing to Theorem~\ref{thm:slice-top} we can remove
instructions $i_1$ through $i_3$ because $ \wprecid{1} \pimplies
\wprecid{4}$. To see why, observe that 
\begin{align*}
  & \wprecid{1} ~= \sum_{j \in \{0,\ldots,15\}} \tfrac{1}{16}
                    \charFun{j \geq 8} ~=~ \tfrac{1}{16} \cdot 8 ~=~
    \tfrac{1}{2}\\
   & \wprecid{4} ~=~ \sum_{j \in \{0,8\}} \tfrac{1}{2}
                    \charFun{s_2 + j \geq 8} ~=~ \tfrac{1}{2}
    \underbrace{\charFun{s_2 + 0 \geq 8}}_{=\:0}  + \tfrac{1}{2}
     \underbrace{\charFun{s_2 + 8 \geq 8}}_{= \: 1} ~=~ \tfrac{1}{2}
\end{align*}

In summary, we obtain the following program slice:
  \begin{align*}
     %
               & \PChoice{\Skip}{\nicefrac{1}{2}}{\Ass{b_3}{1}};  \\[0.5ex]
    & \Ass{r}{b_0 + 2 b_1 + 4 b_2 + 8 b_3} 
  \end{align*}

\noindent Observe that slicing the program to preserve the value of
variable of $r$ (as allowed by existing techniques) would be futile
because there does not exist any proper such slice. \hfill $\triangle$
\end{example}

%% file: extension.tex
\section{Total Correctness}
\label{sec:extension}
In this section we adapt the slicing approach developed in the
previous section to preserve the \emph{total}---rather than
\emph{partial}---correctness of programs. In
Sections~\ref{sec:probterm}-\ref{sec:VCGentc} we develop the
prerequisites for the adaptation and in Section~\ref{sec:slicetc} we
present the two fundamental results for program slicing based on total
correctness specifications (Theorems~\ref{thm:slice-top-total} and
\ref{thm:slice-nested-total}). 

The slicing techniques developed in the previous section concern the
\emph{partial} correctness of programs: they guarantee that if a
program satisfies a partial correctness specification, then so do the
slices provided by Theorems~\ref{thm:slice-top} and
\ref{thm:slice-nested}. In other words, they aim at preserving (lower
bounds for) the probability that the resulting program slices either
terminate establishing the post-condition, or diverge. However, if the
program at hand satisfies a given \emph{total} correctness
specification where preconditions refer to (lower bounds for) the
probability of terminating \emph{and} establishing the
post-condition, we will certainly be interested in preserving the
total correctness for program slices, too. 

This is particularly desirable because when considering partial
correctness, loopy programs admit trivial slices where loop bodies
are simply removed. To see why, let us consider a program
containing \eg loop $\IWhileDo{G}{\inv}{\prog}$, together with its
specification. From Figure~\ref{fig:local-spec}, the
local specification induced on the loop body $\prog$ has
pre-expectation $\eval{G} \cdot inv$ and post-expectation
$\inv$. Thus, a trivial portion of the loop body $\prog$ that preserves the local
specification is $\Skip$, that is,  $\pspecSlice{\Skip}{\eval{G} \cdot
  inv}{\inv}{\prog}$. Therefore, in view of
Theorem~\ref{thm:slice-nested}, removing the loop body $\prog$ from
the original program yields a valid slice thereof.

This may raise doubts about the value of slicing based on partial
correctness specifications, as developed in the previous
section. However, this type of slicing turns out very useful at the practical
level for two reasons. First, it allows concluding that a program
slice never —or only with \emph{low probability}— terminates with an
\emph{incorrect} result (which is different from always —or with
\emph{high probability}— terminating with a \emph{correct}
result). Second, it allows for better ``separation of concerns" and
``tool synergy": one could slice a program \wrt its partial
correctness specification using the results from the previous section,
and exploit any other approach at hand to prove its termination.

\medskip
\subsection{Probabilistic Termination}\label{sec:probterm}
The \emph{de facto} notion of termination for probabilistic programs
is that of \emph{almost-sure termination} (AST), that is, 
termination with probability $1$. Roughly speaking, we say that a
$\Lang$ program is \emph{almost-sure terminating} from an initial
state $\state$ if the probabilities of all its finite executions sum
up to $1$. Note that this does not prohibit the presence of infinite
executions, but instead requires them to have an overall null probability. For
example, the program
\[
\Ass{c}{1};\;  \WhileDo{c=1}{\PChoice{\Ass{c}{1}}{\nicefrac{1}{2}}{\Ass{c}{0}}}
\]
that simulates a geometric distribution by flipping a fair coin until
observing the first heads (represented by $0$) is almost-sure terminating: For all $n \geq
1$, the loop terminates after $n$ iterations with probability
$(\nicefrac{1}{2})^n$, thus the set of all its finite executions has
probability $\sum_{n \geq 1}  (\nicefrac{1}{2})^n = 1$.  Note that
besides these finite executions,  the
program also admits an infinite execution  where all coin flips return
tails (represented by $1$). However, as required for almost-sure termination, this execution has
probability $\lim_{n \To \infty}  (\nicefrac{1}{2})^n = 0$.

\medskip
\subsection{Proving Termination via Variants}
The traditional approach for establishing the total correctness of a
program, either deterministic or probabilistic, consists in combining
partial correctness with a termination argument. For example, for
the case of a probabilistic program $\prog$, if we know on the one hand
that it satisfies the partial correctness specification
$\pspec{\pree}{\prog}{\poste}$ and on the other hand, that it
terminates almost-surely from any state satisfying, say predicate $T$
(for termination), then we can conclude that it satisfies the total
correctness specification
$\tspec{\eval{T} \cdot \pree}{\prog}{\poste}$.

Since loops are the only possible source of divergence in our language,
let us focus on termination arguments for loops. For deterministic
programs, loop termination is established through the presence of a
so-called variant. Informally, a loop \emph{variant} is an integer expression
that decreases in each loop iteration and cannot decrease infinitely
many times without before leaving the loop. For a VCGen for
deterministic programs only, this would require adapting the
verification conditions generated by loops from
\begin{align}
  \vc{\IWhileDo{G}{I}{\prog}}{\post}  ~=~& \left\{ %
     G \land I \Rightarrow \wprec{\prog}{I}\, , \right. \notag \\* \notag %
    & \left. \lnot G \land I \Rightarrow \post  \right\} \;\; \cup  \\* \notag
    & \vc{\prog}{I}  \notag
      \intertext{to} \label{eq:vctd}
 \vct{\IWhileDo{G}{I \textcolor{BrickRed}{, v, \mathtt{l}}}{\prog}}{\post}  ~=~&  \left\{ %
     G \land I \textcolor{BrickRed}{\, \land \, v = v_0} \Rightarrow
                                           \wprec{\prog}{I
                                           \textcolor{BrickRed}{\:
                                           \land \: v < v_0} }\: ,
                                                                                   \right.\\ \notag %
    & \; \textcolor{BrickRed}{G \land I \Rightarrow v \geq
      \mathtt{l}\, ,} \\ \notag
    & \left. \lnot G \land I \Rightarrow \post  \right\} \;\; \cup  \\ \notag
    & \vct{\prog}{I \textcolor{BrickRed}{\: \land \: v <
      v_0}}   \notag
\end{align}
where $v$ denotes the loop variant, $\mathtt{l}$ a lower bound
thereof established by the loop invariant and guard, and $v_0$ a fresh
logical variable~\cite{Barros:2012}.\footnote{Without lost of generality, $\mathtt{l}$ can be
  considered to be 0. We prefer to leave it as an additional parameter in
  order to avoid (the otherwise required) adaptations of the variant
  $v$.} 

McIver and Morgan~\cite[Lemma 7.5.1]{McIver:2004} showed how to generalize this variant-based
termination argument to probabilistic loops. However, the
generalization deviates from the argument
for deterministic programs in two aspects. First, it does not require that the variant
decreases with probability $1$ in each loop iteration, but only with a
fixed positive probability $\epsilon > 0$. Second, besides being
bounded from below, the loop variant must be bounded also from above.

Even though adapting Equation~\ref{eq:vctd} to the probabilistic case by
accounting for these deviations is rather straightforward, another
change is also necessary. To see why, observe that the role of
invariant $I$ in Equation~\ref{eq:vctd}  is twofold: on the one hand, to
establish the desired partial correctness of the loop (in particular,
post-condition $\post$)
and, on the other hand, to encode a set of states from which the loop
is guaranteed to terminate (recall that ``partial correctness plus termination
implies total correctness''). However, for the case of probabilistic programs,
these two roles must be decoupled because the invariant
required to establish the partial correctness of the loop might be itself
probabilistic, \ie a \emph{proper} expectation, while the almost-sure termination
of the loop remains encoded by a set of states, that is, a \emph{predicate} over
states.

To reason about slices that preserve the total correctness of
probabilistic programs, we thus annotate loops as
\[
  \IWhileDo{G}{\inv, T, v, \mathtt{l}, \mathtt{u}, \epsilon}{c}~,
\]
where $\inv$ is an expectation representing the loop invariant (like
for the case of partial correctness), $T$ is a predicate representing
the sets of states from which the loop terminates almost surely, $v$
is an integer-valued function over program states representing the
loop variant, $\mathtt{l}$ and $\mathtt{u}$ are integers representing
a lower and upper bound for the variant, respectively, and $\epsilon$
is a probability in the interval $(0,1]$ with which the variant is
guaranteed to decrease in each iteration.

\medskip
\subsection{Verification Condition Generator}\label{sec:VCGentc}

In view of the above discussion, to define the VCGen for total
correctness specifications we adapt transformers $\wprecsymbol$ and
$\vcsymbol$ as follows, where for convenience we display adaptations
in red:
\begin{equation}
  \label{eq:wpt-loop}
\wprect{\IWhileDo{G}{\inv, \textcolor{BrickRed}{T, v, \mathtt{l}, \mathtt{u}, \epsilon}}{\prog}}{\poste}
~=~  \textcolor{BrickRed}{\charFun{T} \: \cdot\:} \inv
\end{equation}
\begin{align}
\MoveEqLeft[3]
   \vct{\IWhileDo{G}{\inv, \textcolor{BrickRed}{T, v, \mathtt{l}, \mathtt{u}, \epsilon}}{\prog}}{\poste} \label{eq:vct-loop} \\
&=~ \textcolor{BrickRed}{\left\{ %
    \charFun{G \land T} \;\pimplies\;
    \wprect{\prog}{\charFun{T}}~, \right.} \notag \\ %
& \hphantom{=~\left\{ \right.}  \textcolor{BrickRed}{\epsilon \, \charFun{G \land T \land  v = v_0}  \;\pimplies\;
    \wprect{\prog}{\charFun{v < v_0}}~,} \notag \\ 
&  \hphantom{=~\left\{ \right.}  \textcolor{BrickRed}{\charFun{G \land T} \;\pimplies\;
    \charFun{\mathtt{l} \leq v \leq \mathtt{u}} \left. \right\}  \;
    \cup} \notag \\ 
& \hphantom{=~\left\{ \right.}  \textcolor{BrickRed}{\vct{\prog}{\charFun{T}} \; \cup} \notag \\ 
& \hphantom{=~\left\{ \right.}  \textcolor{BrickRed}{\vct{\prog}{\charFun{v < v_0}} \;} \cup  \notag \\ 
& \hphantom{=~} ~\left\{ \charFun{G}  \cdot \inv \;\pimplies\;
    \wprec{\prog}{\inv}~,  \right. \notag \\
&  \hphantom{=~\left\{ \right.}   \charFun{\lnot G} \cdot \inv
    \;\pimplies\; \poste \left. \right\}  \;\cup \notag \\
& \hphantom{=~\left\{ \right.} \vc{\prog}{\inv} \notag
\end{align}
For the remaining language constructs, $\wprectsymbol$ and $\vctsymbol$
follow the same rules as their respective counterparts
for partial correctness $\wprecsymbol$ and $\vcsymbol$ (see
Figures~\ref{fig:wprec} and \ref{fig:VCGen}), $\vctsymbol$
making use of $\wprectsymbol$ instead of $\wprecsymbol$.

In Equation~\ref{eq:vct-loop}, the validity of the (added)
verification conditions in red entails that the loop terminates almost
surely from $T$ (and that $T$ is a standard, \ie non-probabilistic,
loop invariant)~\cite[Lemma 7.5.1]{McIver:2004}. The validity of the
remaining verification conditions (as generated also by
$\vcsymbol$) entails that $\inv$ is a valid \emph{partial correctness}
invariant, strong enough as to establish post-expectation
$\poste$. Combining these two results, we can conclude that the loop
satisfies the \emph{total correctness} specification given by pre-expectation
$\charFun{T} \cdot \inv$ (as reflected by
Equation~\ref{eq:wpt-loop}) and post-expectation $\poste$~\cite[Lemma
2.4.1-Case 2]{McIver:2004}.

With these adaptations in place, we can readily define the VCGen for
total correctness, mimicking the definition of the
VCGen for total correctness:
\begin{definition}[VCGen for total correctness]\label{def:VCGent}
  The set of verification conditions $\vcgt{\pree}{\prog}{\poste}$ for
  the total correctness of a \Lang program $\prog$ \wrt pre-expectation $\pree$ and
  post-expectation $\poste$ is defined as follows:
  \[
  \vcgt{\pree}{\prog}{\poste} ~~=~~ \{ \pree \pimplies
  \wprect{\prog}{\poste} \} \, \cup \, \vct{\prog}{\poste}~.
\]
\end{definition}

\noindent Having introduced $\vcgtsymbol$, we can restate the definition of $\vctsymbol$ more succinctly:
\begin{align}
\MoveEqLeft[3]
   \vct{\IWhileDo{G}{\inv, \textcolor{BrickRed}{T, v, \mathtt{l}, \mathtt{u}, \epsilon}}{\prog}}{\poste} \label{eq:vct-alt} \\
  =~&\textcolor{BrickRed}{\vcgt{\charFun{G \land
    T}}{\prog}{\charFun{T}} \; \cup}  \nonumber \\
  &\textcolor{BrickRed}{\vcgt{\epsilon \, \charFun{G
    \land T \land  v = v_0}}{\prog}{\charFun{v < v_0}} \; \cup} \nonumber \\
 &  \textcolor{BrickRed}{\left\{ \charFun{G \land T} \;\pimplies\;
     \charFun{\mathtt{l} \leq v \leq \mathtt{u}} \right\}  \;
     \cup} \nonumber \\ 
&\vcgt{\charFun{G}  \cdot \inv}{\prog}{\inv} \; \cup \nonumber \\
&  \left\{\charFun{\lnot G} \cdot \inv
    \;\pimplies\; \poste \right\} \nonumber
\end{align}

The VCGen for total correctness obeys the same properties of soundness
and monotonicity as the VCGen for partial correctness.
\begin{lemma}[Soundness of $\vcgtsymbol$]\label{thm:vcgtotal-sound}
For any $\Lang$ program $\prog$ and any two expectations $\pree, \poste \colon \Exp$,
\[
 \models \vcgt{\pree}{\prog}{\poste}  \quad \implies \quad \models \tspec{\pree}{\prog}{\poste}~.
\]
\end{lemma}
%

\begin{lemma}[Monotonicity of  $\vctsymbol$/$\vcgtsymbol$]\label{thm:vctotal-monot}
For any \Lang program $\prog$ and any four expectations $\pree,
\pree', \poste,
\poste' : \Exp$,
\[
  \begin{array}{c @{\qquad}c@{\qquad} c}
  \poste \pimplies \poste'   & \implies & \models
 \vct{\prog}{\poste} ~\Rightarrow~ \models \vct{\prog}{\poste'}~,\\[1ex]
\pree' \pimplies \pree \: \land \: \poste \pimplies \poste'   & \implies & \models
 \vcgt{\pree}{\prog}{\poste} ~\Rightarrow~ \models \vcgt{\pree'}{\prog}{\poste'}~.\\
  \end{array}
\]
\end{lemma}

As for the alternative characterization of $\vcgtsymbol$, the case of
loops requires the adaptations displayed in red. 

\begin{lemma}[Alt.~characterization of $\models \vcgtsymbol$]\label{thm:alt-vcgt}
For any \Lang program $\prog = \seqn{1}{n}$ and any two expectations $\pree,
\poste : \Exp$,
\begin{enumerate}
  \item If $\inst_j = \Cond{G}{\prog_1}{\prog_2}$ or  $\inst_j = \PChoice{\prog_1}{p}{\prog_2}$ for some $1 \leq j
    \leq n$, then 
    %
    \[
      \models \vcgt{\pree}{\prog}{\poste}
      \qquad \text{iff} \qquad
      \begin{array}[t]{l}
        \models \vcti{j+1}{\prog}{\poste}~~\land\\[1ex]
        \models \vct{\prog_1}{\wprecti{j+1}{\prog}{\poste}}~~\land\\[1ex]
        \models \vct{\prog_2}{\wprecti{j+1}{\prog}{\poste}}~~\land\\[1ex]
         \models \vcgti{j-1}{\pree}{\prog}{\wprecti{j}{\prog}{\poste}}~.
      \end{array}
    \]%
\item If $\inst_j = \IWhileDo{G}{\inv, \textcolor{BrickRed}{T, v, \mathtt{l}, \mathtt{u}, \epsilon}}{\prog'}$ for some $1 \leq j
    \leq n$, then%
    \[
      \models \vcgt{\pree}{\prog}{\poste}
      \qquad \text{iff} \qquad
      \begin{array}[t]{l}
        \models \vcgtii{j+1}{\charFun{\lnot G} \cdot \inv}{\prog}{\poste}~~\land\\[1ex]
        \models \vcg{\charFun{G} \cdot \inv}{\prog'}{\inv}~~\land\\[1ex]
         \models
        \vcgti{j-1}{\pree}{\prog}{\textcolor{BrickRed}{\charFun{T}
        \cdot}\inv}~~\land\\[1ex]
  \textcolor{BrickRed}{\models \vcgt{\charFun{G \land
    T}}{\prog'}{\charFun{T}}~~\land}  \\[1ex]
  \textcolor{BrickRed}{\models \vcgt{\epsilon \, \charFun{G
    \land T \land  v = v_0}}{\prog'}{\charFun{v < v_0}}~~\land}  \\[1ex]
 \textcolor{BrickRed}{\models \left\{ \charFun{G \land T} \;\pimplies\;
     \charFun{\mathtt{l} \leq v \leq \mathtt{u}} \right\}} 
\end{array}
\]%
  \end{enumerate}
\end{lemma}
%

\medskip
\subsection{Removing instructions}\label{sec:slicetc}

Armed with VCGen for total correctness, we can readily adapt the
notion of specification-based slice to preserve total correctness
properties: 

\begin{definition}[Program slicing based on total correctness specifications~\cite{Barros:2012}] We
  say that \Lang program $\prog'$ is a \emph{specification-based
    slice} of \Lang program $\prog$ with respect to the total
  correctness specification
  given by pre-expectation $\pree$ and post-expectation $\poste$,
  written $\tspecSlice{\prog'}{\pree}{g}{\prog}$, iff
    \begin{enumerate}
        \item $\prog' \preccurlyeq  \prog$, and
        \item $\models \vcgt{\pree}{\prog}{\poste} ~\Rightarrow~ \models \vcgt{\pree}{\prog'}{\poste}$
    \end{enumerate}
\end{definition}

The slicing criteria from previous section carry over to the case of
total correctness.

\begin{theorem}[Removing top-level instructions for total correctness]\label{thm:slice-top-total}
Let $\prog = \seqn{1}{n}$ be a \Lang program together with its respective pre- and
post-expectation $\pree$ and $\poste$. Moreover, let $1\leq j \leq k
\leq n$. If \[\wprecti{j}{\prog}{\poste} ~\pimplies~ 
\wprecti{k+1}{\prog}{\poste}\] then, \[\tspecSlice{\remove{j}{k}{\prog}}{\pree}{\poste}{\prog}~.\]
\end{theorem}

The criterion for removing nested instructions requires adapting the
notion of local specification induced by loops. To see why, observe that given a
program containing a loop, to generate valid slices of the program
from slices of the loop body, the latter must preserve not only the
invariant ($\inv$), but also the termination predicate ($T$) and the
probability of variant decrement ($\epsilon$). To account for these
simultaneous requirements, the \emph{total local specification}
relation associates total correctness specifications to \emph{sets}
of specifications rather than single specifications. We write
\[
\localSpect{\pree}{\prog}{\poste}{\prog'}{  \sspec{\pree_1}{\poste_1},
\ldots, \sspec{\pree_n}{\poste_n}}
\]
to denote that the total correctness specification
$\sspect{\pree}{\poste}$ of $\prog$ induces the local specifications $\sspec{\pree_1}{\poste_1},
\ldots, \sspec{\pree_n}{\poste_n}$ over subprogram $\prog'$.  Each local specification $\sspec{\pree_i}{\poste_i}$ can refer to
either partial ($\sspecp{\pree_i}{\poste_i}$) or total
($\sspect{\pree_i}{\poste_i}$) correctness. 

The rule for loops now reads:
\[
  \infrule{\inst_j = \IWhileDo{G}{\inv, T, v, \mathtt{l}, \mathtt{u}, \epsilon}{\prog'}}%
  {\localSpect{\pree}{\prog}{\poste}{\prog'}%
    {\begin{array}{l}
      \sspecp{\charFun{G} \cdot \inv}{\inv},\;%
      \sspect{\charFun{G \land T} }{\charFun{T}},\; \\ %
      \sspect{\epsilon \, \charFun{G \land T \land  v = v_0} } {\charFun{v < v_0}}\end{array}%
    }}~~\lrule{$\localSpecSymbol \texttt{while}^\downarrow$}
\]
The rule encoding transitivity also needs to be adjusted to account for the multiplicity
of induced local specifications: 
\begin{align*}
  \infrule{\begin{array}{c}%
\localSpect{\pree}{\prog}{\poste}{\prog'}{\sspec{\pree_i}{\poste_i}}_{i=1,\ldots,n}\\[1ex]
\localSpec{\pree_i}{\prog'}{\poste_i}{\prog''}{\sspec{\pree_{i,j}}{\poste_{i,j}}}_{j=1,\ldots,m_i}
             \quad \forall i=1,\ldots,n
           \end{array}}
  {\localSpect{\pree}{\prog}{\poste}{\prog''}{\sspec{\pree_{i,j}}{\poste_{i,j}}}_{\substack{i=1,\ldots,n
  \\ j=1,\ldots,m_i}}}~~\lrule{$\localSpecSymbol \texttt{trans}^\downarrow$}  
\end{align*}
The remaining rules mirror their counterpart from~Figure~\ref{fig:local-spec}.

\begin{theorem}[Removing nested instructions for total correctness]\label{thm:slice-nested-total}
 Let $\prog$ be a \Lang program together with its respective pre- and
post-expectation $\pree$ and $\poste$, and let $\prog'$ be a
subprogram of $\prog$ such that
$\localSpect{\pree}{\prog}{\poste}{\prog'}{\sspec{\pree_i}{\poste_i}}_{i=1,\ldots,n}$. If
$\prog''$ is a portion of $\prog'$ ($\prog'' \preccurlyeq \prog'$) such that 
for all $i=1,\ldots,n$,
\[
  \specSlice{\prog''}{\pree_i}{\poste_i}{\prog'}~,
\]
then
\[
  \tspecSlice{\prog\subst{\prog'}{\prog''}}{\pree}{\poste}{\prog}~,
\]
where $\prog\subst{\prog'}{\prog''}$ denotes the program that is
obtained from $\prog$ by replacing $\prog'$ with~$\prog''$. 
\end{theorem}

\noindent In the theorem statement, the correctness type (partial or total) of each premise
$\specSlice{\prog''}{\pree_i}{\poste_i}{\prog'}$ is inherited by the
type of the local specification $\sspec{\pree_i}{\poste_i}$ induced over
$\prog'$. For example, if for some $i$, the induced local specification refers to total
correctness ($\sspect{\pree_i}{\poste_i}$), then the corresponding
premise refers accordingly to total correctness ($\tspecSlice{\prog''}{\pree_i}{\poste_i}{\prog'}$).

We next illustrate the application of
Theorems~\ref{thm:slice-top-total} and \ref{thm:slice-nested-total} to
reason about program slicing that preserve total correctness specifications.

\begin{example}
Alice and Bob repeatedly flip each a fair coin until observing
a matching outcome, either both heads or both tails. However, Alice
decides to ``trick'' Bob and switches the outcome of her coin, before
comparing it to Bob's. The game can be encoded by program
\[
  \begin{array}{l}
    \codeComment{$\pree = \tfrac{1}{2^K} \charFun{K >0}$}\\
    \Ass{n}{0};\\
    \Ass{a,b}{0,1};\\
    \While~(a \neq b)~[\inv, T, v, \mathtt{l}, \mathtt{u}, \epsilon]~\Do\\
    \quad  \Ass{n}{n+1};\\
    \quad  \PChoice{\Ass{a}{0}}{\nicefrac{1}{2}}{\Ass{a}{1}};\\
    \quad  \Ass{a}{1-a};\\
    \quad  \PChoice{\Ass{b}{0}}{\nicefrac{1}{2}}{\Ass{b}{1}} \\
    \codeComment{$\poste = \charFun{n=K}$}
  \end{array}
\]
The program is instrumented with a variable $n$ that tracks the
required number of rounds until observing the first match. The program
terminates after $K$ loop iterations with probability
$\nicefrac{1}{2^K}$ provided $K>0$ and with probability 0
otherwise, satisfying the annotated specification. The specification
refers to total correctness. Indeed, we can prove that the loop
terminates almost-surely from any initial state, exploiting the fact that
$\charFun{a \neq b}$ is a loop variant, bounded by 0 and 1, which
decrements with probability $\nicefrac{1}{2}$ in each
iteration. Together with the invariant required to establish the
specification, these correspond to the following loop annotations:
\[
  \begin{array}{lcl}
    \inv &=& \charFun{a \neq b} \charFun{n < K} 2^{n-K} + \charFun{a =
      b} \charFun{n = K}\\
    T &=& \mathsf{true} \\
    v &=& \charFun{a \neq b} \\
    \mathtt{l} &=& 0\\
    \mathtt{u} &=& 1\\
    \epsilon &=& \nicefrac{1}{2}
  \end{array}
\]

By appealing to Theorem~\ref{thm:slice-nested-total}, we show that we
can remove the assignment $\Ass{a}{1-a}$ from the loop body, while preserving the
program total correctness specification. Said otherwise, switching the
outcome of Alice coin has no effect on the number of rounds required
until observing the first match. Formally, if we call
$\prog'$ the original loop body and $\prog''$ the slice of $\prog'$
obtained by removing the assignment, the application of
Theorem~\ref{thm:slice-nested-total} requires proving:
\begin{gather}
  \pspecSlice{\prog''}{\charFun{a \neq b} \cdot \inv}{\inv}{\prog'} \label{eq:ex4-1}\\
  \tspecSlice{\prog''}{\charFun{a \neq b \land \mathsf{true}}}{\charFun{\mathsf{true}}}{\prog'} \label{eq:ex4-2} \\
  \tspecSlice{\prog''}{\epsilon \, \charFun{a \neq b \land
      \mathsf{true} \land  \charFun{a \neq b} =
      v_0}}{\charFun{\charFun{a \neq b} < v_0}}{\prog'} \label{eq:ex4-3}
\end{gather}

\noindent To this end, we apply Theorem~\ref{thm:slice-top} (to 
\eqref{eq:ex4-1}) and Theorem~\ref{thm:slice-top-total} (to \eqref{eq:ex4-2} and \eqref{eq:ex4-3}), propagating the respective post-expectations
backward along $\prog'$, till traversing the assignment to be removed:
\[
  \begin{array}{l}
      \codeComment{$\tfrac{1}{2} \charFun{a \neq 1} \charFun{n < K} 2^{n-K} +
 \tfrac{1}{2}  \charFun{a = 1} \charFun{n = K} \; +$}\\[0.5ex]
    \codeComment{$\tfrac{1}{2} \charFun{a \neq 0} \charFun{n < K} 2^{n-K} +
 \tfrac{1}{2}  \charFun{a = 0} \charFun{n = K}$}\\[0.5ex]
   \Ass{a}{1-a};\\[0.5ex]
    \codeComment{$\tfrac{1}{2} \charFun{a \neq 0} \charFun{n < K} 2^{n-K} +
 \tfrac{1}{2}  \charFun{a = 0} \charFun{n = K} \; +$}\\[0.5ex]
    \codeComment{$\tfrac{1}{2} \charFun{a \neq 1} \charFun{n < K} 2^{n-K} +
 \tfrac{1}{2}  \charFun{a = 1} \charFun{n = K}$}\\[0.5ex]
    \PChoice{\Ass{b}{0}}{\nicefrac{1}{2}}{\Ass{b}{1}} \\[0.5ex]
    \codeComment{\inv}
  \end{array}
\]
\[
  \begin{array}[t]{l @{\qquad \qquad }l}
       \codeComment{$\charFun{\mathsf{true}}$}&\codeComment{$\tfrac{1}{2}\charFun{\charFun{a \neq 1} < v_0} + \tfrac{1}{2}\charFun{\charFun{a \neq 0} < v_0}$}\\[0.5ex]
    \Ass{a}{1-a}; & \Ass{a}{1-a};\\[0.5ex]
     \codeComment{$\charFun{\mathsf{true}}$} &  \codeComment{$\tfrac{1}{2}\charFun{\charFun{a \neq 0} < v_0} + \tfrac{1}{2}\charFun{\charFun{a \neq 1} < v_0}$}\\[0.5ex]
    \PChoice{\Ass{b}{0}}{\nicefrac{1}{2}}{\Ass{b}{1}} &  \PChoice{\Ass{b}{0}}{\nicefrac{1}{2}}{\Ass{b}{1}} \\[0.5ex]
    \codeComment{$\charFun{\mathsf{true}}$}&\codeComment{$\charFun{\charFun{a \neq b} < v_0}$}
  \end{array}
\]
\noindent In all three cases, we see that the expectation above the
assignment entails (in fact coincides with) the
expectation below it, allowing us to safely remove it while preserving
the (overall) program specification.

Finally, observe that, similarly to the previous examples, applying an analysis
based on data/control dependencies to produce a slice that preserves
the value of $n$ would be fruitless as the program admits no such
proper slice. \hfill $\triangle$
\end{example}


%% file: applications.tex
\section{Case Study}
\label{sec:applications}

We now showcase the applicability of our technique to the
field of probabilistic modelling, in particular, aiding in model understanding and
model simplification.

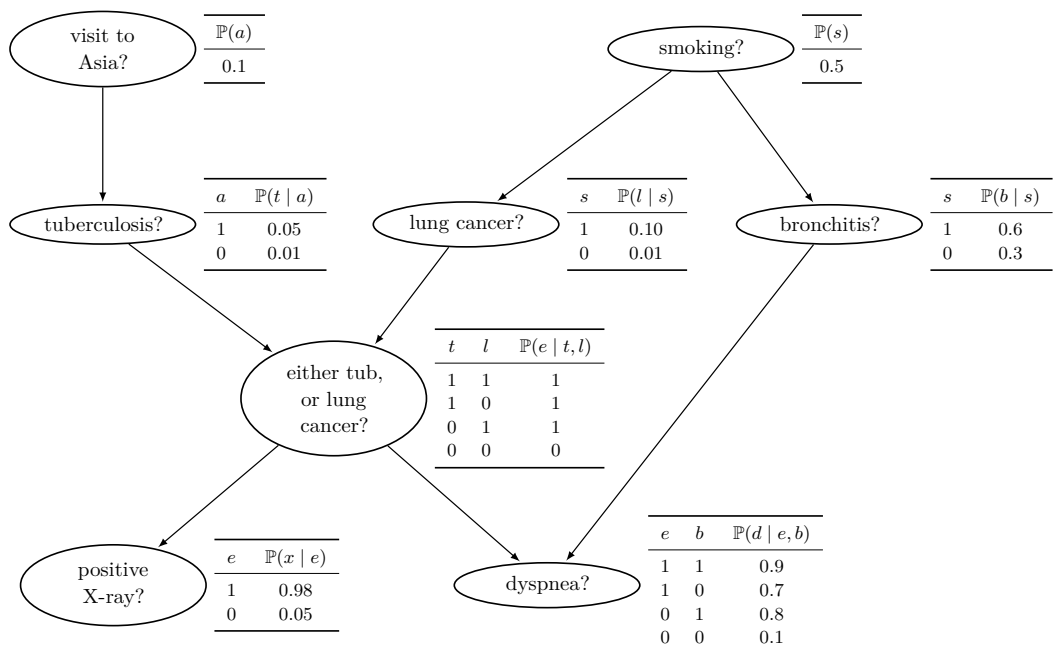
\begin{figure}[t]
  \centering
  \resizebox{\textwidth}{!}{%
     \input{cancerBN}
  }
  \caption{Bayesian network describing a fictional knowledge related
to different lung diseases and factors~\cite{Lauritzen:1988}.}
  \label{fig:BN}
\end{figure}

Since their introduction in the 80's, graphical models ---in
particular, Bayesian networks--- have been the \emph{de facto}
formalism for encoding probabilistic models due to their accessibility
and simplicity. For example, Figure~\ref{fig:BN} shows a Bayesian
network by Lauritzen and Spiegelhalter~\cite{Lauritzen:1988} modeling
(a quantitative version of) the following fictitious knowledge related
to different lung diseases (tuberculosis, lung cancer and bronchitis)
and factors (visit to Asia and smoking):

\begin{quote}
Shortness-of-breath (dyspnoea) may be due to tuberculosis, lung cancer
or bronchitis, or none of them, or more than one of them. A recent
visit to Asia increases the chances of tuberculosis, while smoking is
known to be a risk factor for both lung cancer and bronchitis. The
results of a single chest X-ray do not discriminate between lung
cancer and tuberculosis, as neither does the presence or absence of
dyspnoea.  
\end{quote}

\noindent The network topology encodes the dependencies among the
involved random variables. In particular, random
variables can be distributed either independently of the reminding
random variables, like $a$ (visit to Asia) or conditionally on a
subset of them, like $d$ (dyspnea). The probability distribution of the
random variables are specified by probability distribution
tables (in Figure~\ref{fig:BN} depicted on the right of the node
encoding the random variable). The network together with the
probability distribution tables uniquely determines the joint
distribution of all random variables. 

With the emergence of probabilistic programming systems in the past
years, probabilistic programs have become a more convenient formalism
for encoding such probabilistic models for several
reasons~\cite{DBLP:conf/icse/GordonHNR14}.  First, probabilistic
programs provide additional abstractions not provided by Bayesian
networks. For example, while Bayesian networks are inherently acyclic,
probabilistic programs allow encoding recursive models. Also,
probabilistic programs enable a more compositional approach to
modelling due to the presence of functional abstractions. Second,
modern probabilistic programming systems include state of the art
inference algorithms, which significantly improve inference
time. Altogether, these features allow a faster and more convenient prototyping than
Bayesian networks.

Through the use of traditional programming language abstractions,
probabilistic programs can encode intricate models, involving multiple
random variables which may be highly coupled. However, given such a
complex model, we might be interested in a \emph{partial view} thereof. For
example, while the program in Figure~\ref{fig:BNSlice1} encodes the
entire model described by the Bayesian network from
Figure~\ref{fig:BN}\footnote{The translation of Bayesian networks into
  probabilistic programs is rather straightforward; for a formal
  description, see, \eg, \cite{DBLP:conf/esop/BatzKKM18}.}, we may be
interested only in the probability that the X-ray of a patient turns
out positive, \eg, because we want to compute this probability or because we want to
identify the \emph{fragment} of the model that defines this probability.

\begin{figure}[t]
  \centering
\begin{multicols}{2}\noindent%
  {\small%
    \begin{align*}
      %
      \scalemath{0.8}{i_1: ~} & \PChoice{\Ass{a}{1}}{\nicefrac{1}{100}}{\Ass{a}{0}};             \\[-0.4ex]
      %
      \scalemath{0.8}{i_2: ~} & \PChoice{\Ass{s}{1}}{\nicefrac{1}{2}}{\Ass{s}{0}};               \\[-0.4ex]
      %
      \scalemath{0.8}{i_3: ~} & \If \, (a=1) \,\Then  ~ \{                                       \\[-0.4ex]
                              & \quad  \PChoice{\Ass{t}{1}}{\nicefrac{1}{2}}{\Ass{t}{0}}         \\[-0.4ex]
                              & \} ~\Else ~ \{                                                   \\[-0.4ex]
                              & \quad  \PChoice{\Ass{t}{1}}{\nicefrac{1}{100}}{\Ass{t}{0}}       \\[-0.4ex]
                              & \}                                                               \\[-0.4ex]
      %
      \scalemath{0.8}{i_4: ~} & \If \, (s=1) \,\Then  ~ \{                                       \\[-0.4ex]
                              & \quad  \PChoice{\Ass{l}{1}}{\nicefrac{1}{10}}{\Ass{l}{0}}        \\[-0.4ex]
                              & \} ~\Else ~ \{                                                   \\[-0.4ex]
                              & \quad  \PChoice{\Ass{l}{1}}{\nicefrac{1}{100}}{\Ass{l}{0}}       \\[-0.4ex]
                              & \}                                                               \\[-0.4ex]
      %
      \scalemath{0.8}{i_5: ~} & \SliceAway{\If \, (s=1) \,\Then  ~ \{}                                       \\[-0.4ex]
                              & \quad  \SliceAway{\PChoice{\Ass{b}{1}}{\nicefrac{6}{10}}{\Ass{b}{0}}}        \\[-0.4ex]
                              & \SliceAway{\} ~\Else ~ \{}                                                   \\[-0.4ex]
                              & \quad  \SliceAway{\PChoice{\Ass{b}{1}}{\nicefrac{3}{10}}{\Ass{b}{0}}}        \\[-0.4ex]
                              & \SliceAway{\}}                                                               \\[-0.4ex]
      %
      \scalemath{0.8}{i_6: ~} & \If \, (t=1 \land l=1) \,\Then  ~ \{                             \\[-0.4ex]
                              & \quad  \PChoice{\Ass{e}{1}}{1}{\Ass{e}{0}}                       \\[-0.4ex]
                              & \} ~\Else \, \If \, (t=1 \land l=0)  ~ \{                        \\[-0.4ex]
                              & \quad
                                \PChoice{\Ass{e}{1}}{1}{\Ass{e}{0}}
      \\[-0.4ex]
                              & \} ~\Else \, \If \, (t=0 \land l=1) ~ \{                         \\[-0.4ex]
                              & \quad  \PChoice{\Ass{e}{1}}{1}{\Ass{e}{0}}                       \\[-0.4ex]
                              & \} ~\Else ~ \{                                                   \\[-0.4ex]
                              & \quad \PChoice{\Ass{e}{1}}{0}{\Ass{e}{0}}                        \\[-0.4ex]
                              & \}                                                               \\[-0.4ex]
      \scalemath{0.8}{i_7: ~} & \If \, (e=1) \,\Then  ~ \{                                       \\[-0.4ex]
                              & \quad  \PChoice{\Ass{x}{1}}{\nicefrac{98}{100}}{ \SliceAway{\Ass{x}{0}}  }   \\[-0.4ex]
                              & \} ~\Else ~ \{                                                   \\[-0.4ex]
                              & \quad  \PChoice{ \Ass{x}{1}  }{\nicefrac{5}{100}}{  \SliceAway{\Ass{x}{0}} } \\[-0.4ex]
                              & \}                                                               \\[-0.4ex]
      %
      \scalemath{0.8}{i_8: ~} & \SliceAway{\If \, (e=1 \land b=1) \,\Then   ~ \{}                            \\[-0.4ex]
                              & \quad  \SliceAway{\PChoice{\Ass{d}{1}}{\nicefrac{9}{10}}{\Ass{d}{0}}}        \\[-0.4ex]
                              & \SliceAway{\} ~\Else \, \If \, (e=1 \land b=0) ~ \{}                         \\[-0.4ex]
                              & \quad  \SliceAway{\PChoice{\Ass{d}{1}}{\nicefrac{7}{10}}{\Ass{d}{0}}}        \\[-0.4ex]
                              & \SliceAway{\} ~\Else \, \If \,  (e=0 \land b=1) ~ \{}                       \\[-0.4ex]
                              & \quad    \SliceAway{\PChoice{\Ass{d}{1}}{\nicefrac{8}{10}}{\Ass{d}{0}}}      \\[-0.4ex]
                              & \SliceAway{\} ~\Else ~ \{}                                                   \\[-0.4ex]
                              & \quad  \SliceAway{\PChoice{\Ass{d}{1}}{\nicefrac{1}{10}}{\Ass{d}{0}}}        \\[-0.4ex]
                              & \SliceAway{\}}%
    \end{align*}%
  }%
\end{multicols}
\caption{Probabilistic program describing a model that relates
  different lung diseases 
  and factors. Code fragments in \SliceAway{red} can be sliced away when
  considering post-expectation $\charFun{x=1}$.}
  \label{fig:BNSlice1}
\end{figure}

Applying the slicing technique from Section~\ref{sec:extension}, we
can conclude that the program fragments from Figure~\ref{fig:BNSlice1}
colored in \SliceAway{red} are extraneous to this probability and can
thus be sliced away of the program. 
%
%
Formally, the resulting program represents a slice of the original
program \wrt post-expectation $\charFun{x=1}$, and any
pre-expectation. The detailed derivation can be found in Appendix~B.

\begin{figure}[t]
  \centering
\begin{multicols}{2}\noindent%
  {\small%
    \begin{align*}
      %
      \scalemath{0.8}{i_1: ~} & \PChoice{\Ass{a}{1}}{\nicefrac{1}{100}}{\Ass{a}{0}};             \\[-0.4ex]
      %
      \scalemath{0.8}{i_2: ~} & \PChoice{\Ass{s}{1}}{\nicefrac{1}{2}}{\Ass{s}{0}};               \\[-0.4ex]
      %
      \scalemath{0.8}{i_3: ~} & \If \, (a=1) \,\Then  ~ \{                                       \\[-0.4ex]
                              & \quad  \PChoice{\Ass{t}{1}}{\nicefrac{1}{2}}{\SliceAway{\Ass{t}{0}}}         \\[-0.4ex]
                              & \} ~\Else ~ \{                                                   \\[-0.4ex]
                              & \quad  \PChoice{\Ass{t}{1}}{\nicefrac{1}{100}}{\SliceAway{\Ass{t}{0}}}       \\[-0.4ex]
                              & \}                                                               \\[-0.4ex]
      %
      \scalemath{0.8}{i_4: ~} & \If \, (s=1) \,\Then  ~ \{                                       \\[-0.4ex]
                              & \quad  \PChoice{\Ass{l}{1}}{\nicefrac{1}{10}}{\SliceAway{\Ass{l}{0}}}        \\[-0.4ex]
                              & \} ~\Else ~ \{                                                   \\[-0.4ex]
                              & \quad  \PChoice{\Ass{l}{1}}{\nicefrac{1}{100}}{\SliceAway{\Ass{l}{0}}}       \\[-0.4ex]
                              & \}                                                               \\[-0.4ex]
      %
      \scalemath{0.8}{i_5: ~} & \SliceAway{\If \, (s=1) \,\Then  ~ \{}                                       \\[-0.4ex]
                              & \quad  \SliceAway{\PChoice{\Ass{b}{1}}{\nicefrac{6}{10}}{\Ass{b}{0}}}        \\[-0.4ex]
                              & \SliceAway{\} ~\Else ~ \{}                                                   \\[-0.4ex]
                              & \quad  \SliceAway{\PChoice{\Ass{b}{1}}{\nicefrac{3}{10}}{\Ass{b}{0}}}        \\[-0.4ex]
                              & \SliceAway{\}}                                                               \\[-0.4ex]
      %
      \scalemath{0.8}{i_6: ~} & \SliceAway{\If \, (t=1 \land l=1) \,\Then  ~ \{}                             \\[-0.4ex]
                              & \quad  \SliceAway{\PChoice{\Ass{e}{1}}{1}{\Ass{e}{0}}}                       \\[-0.4ex]
                              & \SliceAway{\} ~\Else \, \If \, (t=1 \land l=0)  ~ \{}                        \\[-0.4ex]
                              & \quad
                                \SliceAway{\PChoice{\Ass{e}{1}}{1}{\Ass{e}{0}}}
      \\[-0.4ex]
                              & \SliceAway{\} ~\Else \, \If \, (t=0 \land l=1) ~ \{}                         \\[-0.4ex]
                              & \quad  \SliceAway{\PChoice{\Ass{e}{1}}{1}{\Ass{e}{0}}}                       \\[-0.4ex]
                              & \SliceAway{\} ~\Else ~ \{ }                                                  \\[-0.4ex]
                              & \quad \SliceAway{\PChoice{\Ass{e}{1}}{0}{\Ass{e}{0}}  }                      \\[-0.4ex]
                              & \SliceAway{\}}                                                               \\[-0.4ex]
      \scalemath{0.8}{i_7: ~} & \SliceAway{\If \, (e=1) \,\Then  ~ \{}                                       \\[-0.4ex]
                              & \quad  \SliceAway{\PChoice{\Ass{x}{1}}{\nicefrac{98}{100}}{ \Ass{x}{0} }}   \\[-0.4ex]
                              & \SliceAway{\} ~\Else ~ \{}                                                   \\[-0.4ex]
                              & \quad  \SliceAway{\PChoice{ \Ass{x}{1}  }{\nicefrac{5}{100}}{\Ass{x}{0}}} \\[-0.4ex]
                              & \}                                                               \\[-0.4ex]
      %
      \scalemath{0.8}{i_8: ~} & \SliceAway{\If \, (e=1 \land b=1) \,\Then   ~ \{}                            \\[-0.4ex]
                              & \quad  \SliceAway{\PChoice{\Ass{d}{1}}{\nicefrac{9}{10}}{\Ass{d}{0}}}        \\[-0.4ex]
                              & \SliceAway{\} ~\Else \, \If \, (e=1 \land b=0) ~ \{}                         \\[-0.4ex]
                              & \quad  \SliceAway{\PChoice{\Ass{d}{1}}{\nicefrac{7}{10}}{\Ass{d}{0}}}        \\[-0.4ex]
                              & \SliceAway{\} ~\Else \, \If \,  (e=0 \land b=1) ~ \{}                       \\[-0.4ex]
                              & \quad    \SliceAway{\PChoice{\Ass{d}{1}}{\nicefrac{8}{10}}{\Ass{d}{0}}}      \\[-0.4ex]
                              & \SliceAway{\} ~\Else ~ \{}                                                   \\[-0.4ex]
                              & \quad  \SliceAway{\PChoice{\Ass{d}{1}}{\nicefrac{1}{10}}{\Ass{d}{0}}}        \\[-0.4ex]
                              & \SliceAway{\}}%
    \end{align*}%
  }%
\end{multicols}
\caption{Probabilistic program describing a model that relates
  different lung diseases 
  and factors. Code fragments in \SliceAway{red} can be sliced away when
  considering post-expectation $\charFun{t=1 \land l=1}$.}
  \label{fig:BNSlice2}
\end{figure}

In a similar way, if we are interested only in the probability that a
patient suffers from both tuberculosis and lung cancer at the same
time, we can slice the program \wrt post-expectation
$\charFun{t=1 \land l=1}$. This allows a more aggressive slicing, as
depicted in~Figure~\ref{fig:BNSlice2}. The detailed derivation is also
found in Appendix~B.

Notoriously, in both cases we obtain more precise slices than the one
yield by traditional slicing techniques based on data and control
dependencies. Concretely, in the first case these techniques fail to
identify 
assignment $\Ass{x}{0}$ as a removable piece of code, and in the
second case they fail to identify assignments $\Ass{l}{0}$ and
$\Ass{b}{0}$ as such.

%% file: cancerBN.tex
\begin{tikzpicture}[
  ->,>=stealth',shorten >=1pt,auto,node distance=2cm,
		semithick,
  state/.style={draw,ellipse,text width=2cm,align=center,thick}
]
\node[state] (Asia) {visit to Asia?};
\node[state] (Smok) [right=7cm of Asia] {smoking?};

\node[state] (Tub) [below =2cm of Asia] {tuberculosis?};
\node[state] (Lcancer) [right=3cm of Tub] {lung cancer?};
\node[state] (Bron) [right=3cm  of Lcancer] {bronchitis?};

\node[state] (Either) [below left =2cm and  0mm of Lcancer] {either tub, or lung cancer?};

\node[state] (Xray) [below left =2cm and 1.5cm of Either] {positive X-ray?};
\node[state] (Dysp) [right=4.2cm of Xray]{dyspnea?};


\path (Asia) edge[-latex] (Tub);
\path (Tub) edge[-latex] (Either);
\path (Either) edge[-latex] (Xray);
\path (Either) edge[-latex] (Dysp);
\path (Lcancer) edge[-latex] (Either);
\path (Smok) edge[-latex] (Bron);
\path (Smok) edge[-latex] (Lcancer);
\path (Bron) edge[-latex] (Dysp);

\node[right=0.001cm of Asia]
{\small
  \begin{tabular}{c}
    \toprule
    $\prob{a}$\\
    \cmidrule{1-1}
    0.1\\
    \bottomrule
\end{tabular}
};

\node[right=0.001cm of Smok]
{\small
  \begin{tabular}{c}
    \toprule
    $\prob{s}$\\
    \cmidrule{1-1}
    0.5\\
    \bottomrule
\end{tabular}
};

\node[right=0.001cm of Tub]
{\small
  \begin{tabular}{c @{\hspace{1.5em}} c}
    \toprule
    $a$ & $\prob{t\mid a}$\\
    \cmidrule{1-2}
    1 & 0.05\\
    0 & 0.01\\
    \bottomrule
\end{tabular}
};

\node[right=0.001cm of Lcancer]
{\small
  \begin{tabular}{c @{\hspace{1.5em}} c}
    \toprule
    $s$ & $\prob{l\mid s}$\\
    \cmidrule{1-2}
    1 & 0.10\\
    0 & 0.01\\
    \bottomrule
\end{tabular}
};

\node[right=0.001cm of Bron]
{\small
  \begin{tabular}{c @{\hspace{1.5em}} c}
    \toprule
    $s$ & $\prob{b\mid s}$\\
    \cmidrule{1-2}
    1 & 0.6\\
    0 & 0.3\\
   \bottomrule
\end{tabular}
};

\node[right=0.001cm of Either]
{\small
  \begin{tabular}{cc@{\hspace{1.5em}}c}
    \toprule
    $t$ & $l$ & $\prob{e\mid t,l}$\\
    \cmidrule{1-3} 
    1 & 1 & 1\\
    1 & 0 & 1\\
    0 & 1 & 1\\
    0 & 0 & 0\\
    \bottomrule
\end{tabular}
};

\node[right=0.001cm of Dysp]
{\small
  \begin{tabular}{c c @{\hspace{1.5em}} c}
    \toprule
    $e$ & $b$ & $\prob{d\mid e,b}$\\
    \cmidrule{1-3}
    1 & 1 & 0.9\\
    1 & 0 & 0.7\\
    0 & 1 & 0.8\\
    0 & 0 & 0.1\\
    \bottomrule
\end{tabular}
};

\node[right=0.001cm of Xray]
{\small
  \begin{tabular}{c @{\hspace{1.5em}} c}
    \toprule
    $e$ & $\prob{x\mid e}$\\
    \cmidrule{1-2}
    1 & 0.98\\
    0 & 0.05\\
    \bottomrule
\end{tabular}
};

\end{tikzpicture}

%% file: graphs.tex
\section{Slicing Algorithm}
\label{sec:graphs}

In this section we present an algorithm for computing program slices. The algorithm is based on
the construction of a \emph{slice graph}
(Definition~\ref{def:slicegraph}) and returns the \emph{least} slice (in a
sense to be defined later) that can be derived by the application of
Theorems~\ref{thm:slice-top}/\ref{thm:slice-top-total} and
\ref{thm:slice-nested}/\ref{thm:slice-nested-total}. This algorithm
provides a starting point for a (semi-)automated application of the
slicing technique developed in Sections~\ref{sec:slicing} and \ref{sec:extension}.

Despite being a mild adaptation of the algorithm introduced by
Barros~\etal~\cite{Barros:2012} for slicing deterministic programs---while we use
only a backward propagation of post-conditions,
Barros~\etal combine backward propagation of
post-conditions with forward propagation of pre-conditions (see
Section~\ref{sec:discussion} for a further discussion)---we prefer
to include a full description of the adaptation here to make the
presentation more self-contained.

The slice graph of a program is obtained by extending its control flow
graph first with semantic information (assertions) in the labels,
yielding an intermediate \emph{labelled control flow graph}, and then with
additional edges that ``short-circuit'' removable
instructions. Determining the least program slice is then cast as a
(generalization of a) weighted shortest path problem on the slice graph. 

For convenience, we develop the algorithm for computing program slices
that preserve partial correctness, and then discuss the necessary
adaptations for total correctness.

\begin{figure}[t!]
 \centering
  \resizebox{0.8\textwidth}{!}{%
  \input{cfg}
  }
       \caption{Excerpt of the slice graph associated to the program from
         Example~\ref{ex:randint}.}
    \label{fig:cfg}
  \end{figure}
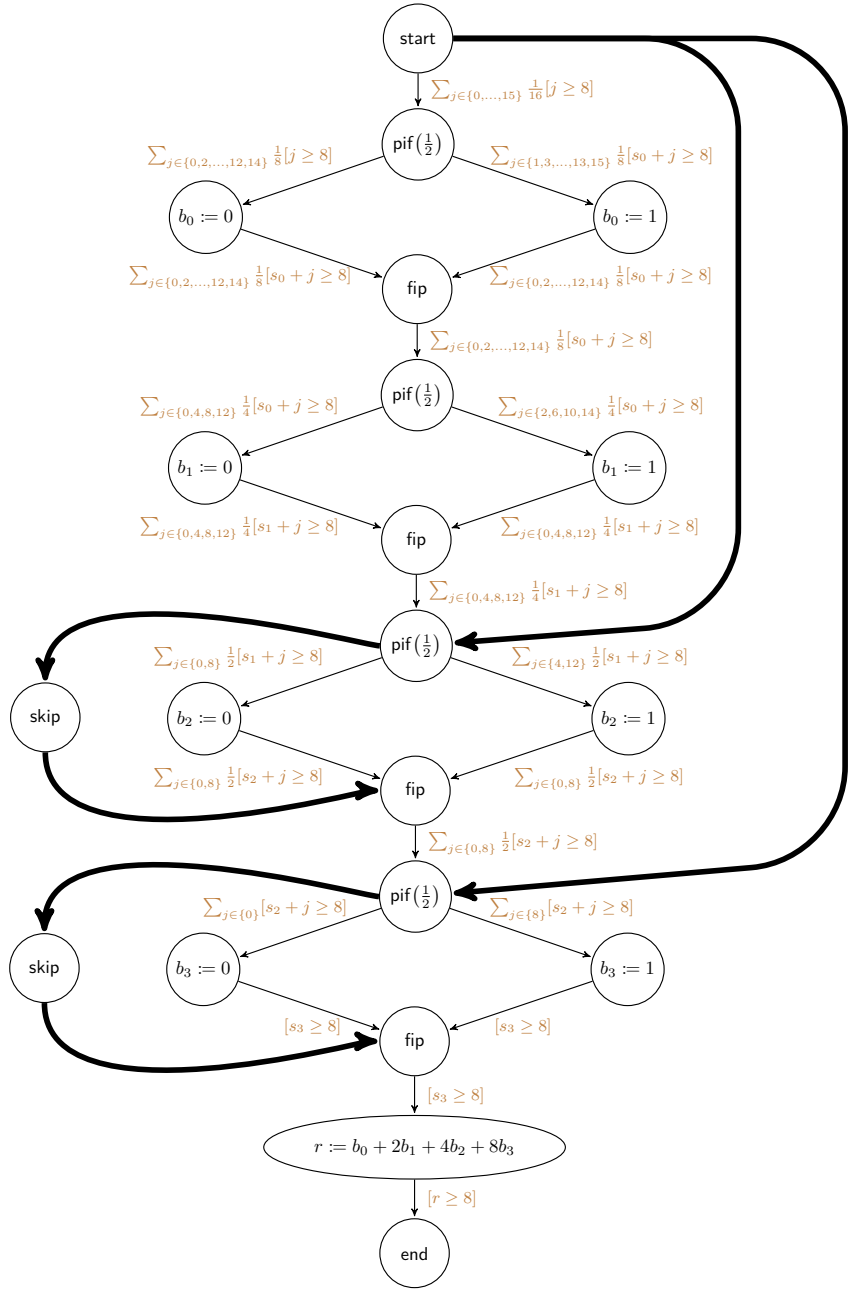

Intuitively, the labelled control flow graph of a program
$\prog = \seqn{1}{n}$ with respect to a post-expectation $\poste$
associates to each edge $(\inst_j, \inst_{j+1})$ the expectation that
is obtained by propagating (via transformer $\wprecsymbol$) the
post-expectation $\poste$ backward, till traversing $\inst_{j+1}$. For
example, the labelled control flow graph of the program from
Example~\ref{ex:randint} \wrt post-expectation
$\charFun{r \geq 8}$ is depicted in Figure~\ref{fig:cfg} (thick edges
are not part of the labelled control flow graph, but of the slice graph).

Formally, it is defined as follows:
\begin{definition}[Labelled control flow graph; adapted
  from~\cite{Barros:2012}]
 Given a \Lang program $\prog = \seqn{1}{n}$ and a (post-)
 expectation $\poste \colon \Exp$, the \emph{labelled control
   flow graph} \lcfg{\prog}{\poste} is a directed acyclic
 graph, whose edges are labelled with expectations. To construct it,
 we make use of the auxiliary functions $\ing{\inst_j}$ and
 $\outg{\inst_j}$ that associate each instruction $\inst_j$ of
 $\prog$ with a respective \emph{input} and \emph{output node} in
 \lcfg{\prog}{\poste}. The graph \lcfg{\prog}{\poste}  is then constructed as follows:

 \begin{enumerate}
   \setlength\itemsep{2ex}
 \item Each instruction $\inst_j$ induces one (\Skip or assignments) or
   two (conditional branches, probabilistic choices or loops) nodes in
   \lcfg{\prog}{\poste}:
   \begin{itemize}
   \item If  $\inst_j = \Skip$ or  $\inst_j = \Ass{x}{E}$, then
     $\inst_j$ is a node of  \lcfg{\prog}{\poste}. Moreover, we
     define $\ing{\inst_j} = \outg{\inst_j } = \inst_j$.
   \item If  $\inst_j = \Cond{G}{\prog_1}{\prog_2}$, then $\bif{G}$
     and $\fib$ are nodes of  \lcfg{\prog}{\poste}. Moreover,
     we define $\ing{\inst_j} = \bif{G}$ and $\outg{\inst_j } = \fib$.
   \item If  $\inst_j = \PChoice{\prog_1}{p}{\prog_2}$, then $\pif{p}$
     and $\fip$ are nodes of  \lcfg{\prog}{\poste}. Moreover,
     we define $\ing{\inst_j} = \pif{p}$ and $\outg{\inst_j } = \fip$.
    \item If  $\inst_j = \IWhileDo{G}{\inv}{\prog'}$, then $\don{G}$
     and $\odn$ are nodes of  \lcfg{\prog}{\poste}. Moreover,
     we define $\ing{\inst_j} = \don{G}$ and $\outg{\inst_j } = \odn$.
   \end{itemize}
   \item \start and \dne are two distinguished nodes of \lcfg{\prog}{\poste}.
   \item $(\start, \ing{\inst_1})$,
     $(\ing{\inst_j},\outg{\inst_{j+1}})$ and $(\outg{\inst_{n}},\dne)$
         are edges of \lcfg{\prog}{\poste} for each
         $j=1,\ldots,n-1$. The labels of these edges are defined as
         follows:
         \[
           \begin{array}[t]{lcl}
             \glabel{\start, \ing{\inst_1}}  &=& \wpreci{1}{\prog}{\poste} \\
             \glabel{\outg{\inst_j},\ing{\inst_{j+1}}} &=&
                                                           \wpreci{j+1}{\prog}{\poste}\quad 
                                                           \forall\; j=1,\ldots,n-1 \\
             \glabel{\outg{\inst_{n}},\dne} &=& \poste
           \end{array}
         \]
   \item If  $\inst_j = \Cond{G}{\prog_1}{\prog_2}$ or  $\inst_j = \PChoice{\prog_1}{p}{\prog_2}$  for some
     $j=1,\ldots,n$, we recursively construct the labelled control
     flow graphs
     \[
       \lcfg{\prog_1}{ \wpreci{j+1}{\prog}{\poste}}%
       \qquad\text{and}\qquad
       \lcfg{\prog_2}{ \wpreci{j+1}{\prog}{\poste}}~.
     \]
     This pair of graphs are incorporated into  \lcfg{\prog}{\poste}
     by removing their \start nodes and setting $\ing{\inst_j}$ as the
     origin of the dangling edges, and similarly removing their \dne
     nodes and setting $\outg{\inst_j}$ as the destination of the
     dangling edges.
     \item If $\inst_j = \IWhileDo{G}{\inv}{\prog'}$ for some
     $j=1,\ldots,n$, we recursively construct the labelled control
     flow graph
     \[
       \lcfg{\prog'}{\inv}~.
     \]
     This graph is incorporated into  \lcfg{\prog}{\poste}
     by removing its \start node and setting $\ing{\inst_j}$ as the
     origin of the dangling edge, and similarly removing its \dne
     node and setting $\outg{\inst_j}$ as the destination of the
     dangling edge.     
 \end{enumerate}
\end{definition}

Observe that the labelled control flow graph of a program can be
constructed by first building the traditional control flow graph, and
then traversing it backward to propagate the post-expectation.

The \emph{slice graph} of a program is obtained by extending
its labelled control flow graph with edges that short-circuit
removable instructions, as identified by Theorems~\ref{thm:slice-top}
and \ref{thm:slice-nested}.

\begin{definition}[Slice graph; adapted from~\cite{Barros:2012}]\label{def:slicegraph}
The \emph{slice graph}
\slg{\prog}{\pree}{\poste} of a \Lang program  $\prog$ \wrt pre-expectation $\pree$ and post-expectation $\poste$ is obtained by extending
its labelled control flow graph \lcfg{\prog}{\poste} with 
additional edges and $\Skip$ nodes. Concretely, for each subprogram $\prog'={\inst'_1},\ldots, {\inst'_n}$ of $\prog$ such
that $\localSpecp{\pree}{\prog}{\poste}{\pree'}{\prog'}{\poste'}$ we
proceed as follows:
\begin{enumerate}
\item If $\pree' \pimplies' \poste'$, we add a new $\Skip$ node,
  together with the pair of edges $(\ing{\prog'}, \Skip)$ and
  $(\Skip,\outg{\prog'})$;

\item For all $j=1,\ldots,n$, if  $\pree' \pimplies
  \wpreci{j+1}{\prog'}{\poste'}$, we add edge $(\ing{\prog'},
  \ing{\inst'_{j+1}})$;
  
\item For all $j=1,\ldots,n$, if  $\wpreci{j}{\prog'}{\poste'} \pimplies
  \poste'$, we add edge $(\outg{\inst'_{j-1}}, \outg{\prog'})$; 

 \item For all $j,k=1,\ldots,n$ such that $j <k$, if
  $\wpreci{j}{\prog'}{\poste'} \pimplies \wpreci{k}{\prog'}{\poste'}$,
  we add edge $(\outg{\inst'_{j-1}}, \ing{\inst'_{k}} )$.
\end{enumerate}
\end{definition}

Returning to the program from
Example~\ref{ex:randint}, the thick edges in Figure~\ref{fig:cfg}
represent a subset of the edges incorporated by the slice graph. An edge
in the slice graph that is not depicted in the figure is, for
example, the one short-circuiting the probabilistic choice assigning a
value to $b_0$, only.

It is not hard to see that, by construction, all slices of a program
that can be derived by (the repeated application of) Theorems~\ref{thm:slice-top} and
\ref{thm:slice-nested} are represented in the slice graph. We are thus
left to choose the \emph{minimal} slice. In this regard, we define the
size of a slice to be the number of atomic instructions
in the subgraph representing the slice.

\medskip
\noindent\textbf{Slicing algorithm.} For straight-line programs, \ie programs free of conditional branches
and probabilistic choices, the minimal slice can be computed by
calculating the shortest path between the \start and \dne
vertices of the slice graph. However, for branching program, this will
select a single branch. To address this problem, Barros~\etal~\cite{Barros:2012} suggests
combining a weighted shortest path algorithm with graph rewriting as
follows:

\begin{enumerate}
\item Assign weight 1 to every edge of the slice graph $G$.
\item For all branching instructions that do not contain any other
  branching instruction as subprogram,
  \begin{enumerate}
  \item run a shortest path algorithm on each of the two branches and
    let $s = 1 + l + r$, where $l$ and $r$ are the lengths of the
    shortest paths of each of branch;
   \item replace the pair of branches with a single edge joining the
     origin ($\bif{G}/\pif{p}$) and the destination ($\fib/\fip$) of
     the branching instruction, and assign it weight $s$.
       \end{enumerate}
   \item Go to step 2 if the resulting graph still contains any
     branching instruction with straight-line branches (observe that the
     step 2 above could have created new such instructions).
\end{enumerate}

When applied to the slice graph of Figure~\ref{fig:cfg}, this
algorithm can choose to keep the assignment of $0$ to $b_3$
instead of replacing them with a $\Skip$ (observe that this is
consistent with our notion of slice with the least number of atomic
instructions). However, this is not what one would expect in
practice. To address this issue, in step 1 we can assign weight, \eg,
$\nicefrac{1}{2}$ (instead of 1) to all edges incident to \Skip vertices.

Finally, to compute slices that preserve the total correctness of programs,
recall that programs containing while loops induce three local
specifications on the loop body (number which can grow larger in the
presence of nested loops). Therefore, the labels of control flow graphs must consist in tuples of
expectations rather than single expectations. The entailment relation
between expectations is naturally extended to tuples by
taking the canonical lifting, \eg, $(\pree_1, \pree_2, \pree_3) \pimplies (\pree'_1, \pree'_2,
\pree'_3)$ iff $\pree_1 \pimplies \pree'_1$, $\pree_2 \pimplies
\pree'_2$ and $\pree_3 \pimplies \pree'_3$. Observe that the labelling
of the graph will not necessarily be uniform, since different
subprograms may be labeled with tuples of different sizes.

%% file: cfg.tex
\begin{tikzpicture}[->,>=stealth',shorten >=1pt,auto,node distance=2cm,
		semithick,
		state/.style={circle, draw, minimum size=1.3cm} ]
	\node[state]                                (A)                    {$\start$};
	\node[state]                                (Bif) [below of=A]                      	{$\pif{\frac{1}{2}}$};
	\node[state]                                (Bp1) [below left=0.4cm and 3cm of Bif]		{$\Ass{b_0}{0}$};
	\node[state]                                (Bp2) [below right=0.4cm and 3cm of Bif] 	{$\Ass{b_0}{1}$};
	\node[state]                                (Bfi) [below right=0.4cm and 3cm of Bp1] 	{$\fip$};
	\node[state]                                (Cif) [below of=Bfi] 											{$\pif{\frac{1}{2}}$};
	\node[state]                                (Cp1) [below left=0.4cm and 3cm of Cif] 	{$\Ass{b_1}{0}$};
	\node[state]                                (Cp2) [below right=0.4cm and 3cm of Cif] 	{$\Ass{b_1}{1}$};
	\node[state]                                (Cfi) [below right=0.4cm and 3cm of Cp1] 	{$\fip$};
	\node[state]                                (Dif) [below of=Cfi] 											{$\pif{\frac{1}{2}}$};
	\node[state]                                (Dp1) [below left=0.4cm and 3cm of Dif] 	{$\Ass{b_2}{0}$};
	\node[state]                                (Dp2) [below right=0.4cm and 3cm of Dif] 	{$\Ass{b_2}{1}$};
	\node[state]                                (skip2) [below left=0.4cm and 6cm of Dif] 	{$\Skip$};
	\node[state]                                (Dfi) [below right=0.4cm and 3cm of Dp1] 	{$\fip$};
	\node[state]                                (Eif) [below of=Dfi] 											{$\pif{\frac{1}{2}}$};
	\node[state]                                (Ep1) [below left=0.4cm and 3cm of Eif] 	{$\Ass{b_3}{0}$};
	\node[state]                                (Ep2) [below right=0.4cm and 3cm of Eif] 	{$\Ass{b_3}{1}$};
	\node[state]                                (skip3) [below left=0.4cm and 6cm of Eif] 	{$\Skip$};
	\node[state]                                (Efi) [below right=0.4cm and 3cm of Ep1] 	{$\fip$};
	\node[ellipse,draw, minimum height = 1.2cm] (r) 	[below of=Efi] 											{$\Ass{r}{b_0+2b_1+4b_2+8b_3}$};
	\node[state] (end) [below of=r] {$\dne$};

	\path (A) 	edge	node {\codeGraph{$\sum_{j \in \{0,\ldots,15\}} \tfrac{1}{16}  \charFun{j \geq 8}$}} (Bif);
	\path (Bif) edge  node[above,xshift=-40pt,yshift=3pt] {\codeGraph{$\sum_{j \in \{0,2,\ldots,12,14\}} \tfrac{1}{8}  \charFun{j \geq 8}$}} (Bp1);
	\path (Bif) edge  node[above,xshift=40pt,yshift=3pt] {\codeGraph{$\sum_{j \in \{1,3,\ldots,13,15\}} \tfrac{1}{8}  \charFun{s_0+j  \geq 8}$}} (Bp2);
	\path (Bp1) edge  node[below,xshift=-40pt, yshift=-3pt] {\codeGraph{$\sum_{j \in \{0,2,\ldots,12,14\}} \tfrac{1}{8}  \charFun{s_0+j  \geq 8}$}} (Bfi);
	\path (Bp2) edge  node[below,xshift=40pt, yshift=-3pt] {\codeGraph{$\sum_{j \in \{0,2,\ldots,12,14\}} \tfrac{1}{8}  \charFun{s_0+j  \geq 8}$}} (Bfi);
	\path (Bfi) edge  node {\codeGraph{$\sum_{j \in \{0,2,\ldots,12,14\}} \tfrac{1}{8}  \charFun{s_0+j  \geq 8}$}} (Cif);
	\path (Cif) edge  node[above,xshift=-40pt,yshift=3pt] {\codeGraph{$\sum_{j \in \{0,4,8,12\}} \tfrac{1}{4}  \charFun{s_0+j  \geq 8}$}} (Cp1);
	\path (Cif) edge  node[above,xshift=40pt,yshift=3pt] {\codeGraph{$\sum_{j \in \{2,6,10,14\}} \tfrac{1}{4}  \charFun{s_0+j  \geq 8}$}} (Cp2);
	\path (Cp1) edge  node[below,xshift=-40pt, yshift=-3pt] {\codeGraph{$\sum_{j \in \{0,4,8,12\}} \tfrac{1}{4}  \charFun{s_1+j  \geq 8}$}} (Cfi);
	\path (Cp2) edge  node[below,xshift=40pt, yshift=-3pt] {\codeGraph{$\sum_{j \in \{0,4,8,12\}} \tfrac{1}{4}  \charFun{s_1+j  \geq 8}$}} (Cfi);
	\path (Cfi) edge  node {\codeGraph{$\sum_{j \in \{0,4,8,12\}} \tfrac{1}{4}  \charFun{s_1+j  \geq 8}$}} (Dif);
	\path (Dif) edge  node[above,xshift=-40pt,yshift=3pt] {\codeGraph{$\sum_{j \in \{0,8\}} \tfrac{1}{2}  \charFun{s_1+j  \geq 8}$}} (Dp1);
	\path (Dif) edge  node[above,xshift=40pt,yshift=3pt] {\codeGraph{$\sum_{j \in \{4,12\}} \tfrac{1}{2}  \charFun{s_1+j  \geq 8}$}} (Dp2);
	\path (Dp1) edge  node[below,xshift=-40pt, yshift=-3pt] {\codeGraph{$\sum_{j \in \{0,8\}} \tfrac{1}{2}  \charFun{s_2+j  \geq 8}$}} (Dfi);
	\path (Dp2) edge  node[below,xshift=40pt, yshift=-3pt] {\codeGraph{$\sum_{j \in \{0,8\}} \tfrac{1}{2}  \charFun{s_2+j  \geq 8}$}} (Dfi);
	\path (Dfi) edge  node {\codeGraph{$\sum_{j \in \{0,8\}} \tfrac{1}{2}  \charFun{s_2+j  \geq 8}$}} (Eif);
	\path (Eif) edge  node[above,xshift=-20pt,yshift=3pt] {\codeGraph{$\sum_{j \in \{0\}}  \charFun{s_2+j \geq 8}$}} (Ep1);
	\path (Eif) edge  node[above,xshift=20pt,yshift=3pt] {\codeGraph{$\sum_{j \in \{8\}}  \charFun{s_2+j \geq 8}$}} (Ep2);
	\path (Ep1) edge  node[below,xshift=0pt, yshift=-3pt] {\codeGraph{$\charFun{s_3 \geq 8}$}} (Efi);
	\path (Ep2) edge  node[below,xshift=0pt, yshift=-3pt] {\codeGraph{$\charFun{s_3 \geq 8}$}} (Efi);
	\path (Efi) edge  node {\codeGraph{$\charFun{s_3 \geq 8}$}} (r);
	\path (r)   edge  node {\codeGraph{$\charFun{r \geq 8}$}} (end);
	\draw[->,line width=3pt] (Dif) .. controls +(left:1cm) and +(up:3cm) .. node[above,sloped] {} (skip2);
	\draw[->,line width=3pt] (skip2) .. controls +(down:3cm) and +(left:1cm) .. node[above,sloped] {} (Dfi);
	\draw[->,line width=3pt] (Eif) .. controls +(left:1cm) and +(up:3cm) .. node[above,sloped] {} (skip3);
	\draw[->,line width=3pt] (skip3) .. controls +(down:3cm) and +(left:1cm) .. node[above,sloped] {} (Efi);
        \draw[->,line width=3pt] (A) [rounded corners=50pt] -- (8,0) -- (8,-15.5) -- (Eif);
        \draw[->,line width=3pt] (A) [rounded corners=50pt] -- (6,0) -- (6,-11) -- (Dif);
                                                                                                                                                

\end{tikzpicture}

%% file: discussion.tex
\section{Discussion}
\label{sec:discussion}

In this section we discuss some design decisions, extensions and
limitations behind the developed slicing approach, pointing out some 
relevant directions of future work. 

\subparagraph*{Termination of probabilistic programs.}
The termination problem for probabilistic programs is significantly
more challenging than for deterministic programs. For example, at the
computational hardness level, while determining whether a
deterministic program terminates on a given input is a semi-decidable
problem (lying in the $\Sigma_1^0$-complete class of the arithmetical
hierarchy), determining whether a probabilistic program almost-surely
terminates on a given input is not (lying in the $\Pi_2^0$-complete
class)~\cite{Kaminski:AI,KaKa2015}. Moreover, deciding almost-sure
termination of a probabilistic program \emph{on a single input} is as
hard as deciding termination of an ordinary (deterministic) program
\emph{on all inputs}.

Matching this intuition, while the variant-based termination argument
for deterministic programs overviewed in Section~\ref{sec:extension}
is complete, the probabilistic version by McIver and Morgan~\cite{McIver:2004} internalized by our VCGen is not. For example, it
is unable to establish the almost-sure termination of the program
below, representing a 1-dimensional (one-side bounded) random walk:
\[
\WhileDo{x \neq 0}{\PChoice{\Ass{x}{x-1}}{\nicefrac{1}{2}}{\Ass{x}{x+1}}}~.
\]

In a recent work~\cite{McIver:POPL18},  McIver~\etal generalized the
termination argument internalized by our VCGen, incrementing its
expressivity so as to establish the termination, for example, of the
above program. The new rule strengthens the original rule in three
aspects: \emph{i}) the variant need not be upper-bounded, \emph{ii})
the variant may be real-valued, and \emph{iii}) the variant decrement
probability may vary across iterations. 

Even though for the sake of presentation accessibility, in
Section~\ref{sec:extension} we designed our VCGen for total
correctness internalizing the original rule, the more recent version
can be internalized following a similar approach. Nevertheless, note
that even though the more recent rule is (strictly) more expressive
than its original version, completeness remains an open problem.

An interesting direction for future research is investigating how our
VCGen can internalize other classes of termination argument, in
particular, those based on the notion of super-martingales, as 
developed in several recent works~\cite{DBLP:conf/cav/ChakarovS13,DBLP:conf/popl/FioritiH15,DBLP:conf/popl/ChatterjeeNZ17,DBLP:conf/cav/ChatterjeeFG17,DBLP:journals/toplas/ChatterjeeFNH18,DBLP:conf/aplas/HuangFC18,DBLP:journals/pacmpl/AgrawalC018,DBLP:conf/vmcai/FuC19,DBLP:journals/pacmpl/Huang0CG19}.

\subparagraph*{Forward propagation of pre-conditions.}
Specification-based slicing techniques
require the combination of both backward propagation of
post-conditions and forward propagation of pre-conditions to yield
minimal slices~\cite{Lee:2001,Chung:2001}. For example, consider program
\[
  \begin{array}{l@{}l}
    \If~(y>0)& \ \Then~\{\Ass{x}{100};\; \Ass{x}{x+50};\;
                 \Ass{x}{x-100} \}\\
    &\ \Else~\{\Ass{x}{x-150};\; \Ass{x}{x-100};\;
                 \Ass{x}{x+100} \}~,
  \end{array}
\]
together with the specification given by pre-condition $y>0$ and
post-condition $x \geq 0$. The minimal slice that preserves the
specification is:
\[
  \begin{array}{l@{}l}
    \If~(y>0) &\ \Then~\{\Ass{x}{100} \}\\
    &\ \Else~\{\Skip\}
  \end{array}
\]
To obtain it, we can, for example, first propagate the pre-condition
$y>0$ forward (via the \emph{strongest post-condition transformer}~\cite{Dijkstra:90}), removing this
way all the (dead) instructions in the $\Else$ branch, and in the
resulting program then propagate the post-condition
$x \geq 0$  backward (via the \emph{weakest pre-condition
  transformer}~\cite{Dijkstra:90}) removing this way the last two instructions of the $\Then$ branch. Any slicing
based only on either kind of propagation will produce less precise slices.

In her PhD thesis~\cite{Jones:1992}, Jones proved that it is unfortunately not
possible to define an analogue of the strongest post-condition transformer
for expectations.  To see why, consider program
\[
  \PChoice{\Ass{x}{0}}{\nicefrac{1}{2}}{\Ass{x}{1}}~.
\]
With respect to pre-expectation $\nicefrac{1}{2}$, two valid
post-expectations are $\charFun{x=0}$ and $\charFun{x=1}$. Thus, the
strongest post-expectation must bound both
$\charFun{x=0}$ and $\charFun{x=1}$  from below (recall Definition~\ref{def:exp-entailment} of
expectation entailment). The only common lower bound is the
constantly null expectation  
$\charFun{\mathsf{false}}$, which is clearly not a valid post-expectation of
the program. 

An important line of future work is then to investigate the design of
a slicing approach that allows both the propagation of pre-conditions
forward and of post-conditions backward. A promising starting point
here is to consider program logics that instead of representing pre- and
post-conditions as real-valued functions over 
states (like expectations), represent them as Boolean predicates over
state distributions. While many such logics already
exist~\cite{Hartog:1999,
  DBLP:journals/tcs/ChadhaCMS07,DBLP:journals/entcs/RandZ15,Barthe:2018:ESOPa},
to the best of our knowledge, none provides an analogue of a strongest
post-condition transformer.

\subparagraph*{Efficiency vs precision tradeoff.}
When designing our slicing technique, we privileged computation
efficiency over slice precision. This is particularly reflected by
Theorems~\ref{thm:slice-nested} and \ref{thm:slice-nested-total},
which embody a modular approach to slicing based on \emph{local
  reasoning principles}: We can slice a
program that contains, \eg, a probabilistic choice by slicing either
of its branches, and this slicing requires only the local
specification induced on the branch---no other contextual information
(like the local specification of the other branch) is required.

This design decision trades computational efficiency for slice precision. To illustrate this,
consider the program below, together with its specification:
\[
\pspec[\quad]{\tfrac{1}{2}}{\PChoice{\Ass{x}{1}}{\nicefrac{3}{4}}{\Ass{x}{x+1}}}{\charFun{x
  = 1}}~.
 \]
The right branch of the probabilistic choice can be removed yielding a
valid specification-preserving slice. However,
Theorem~\ref{thm:slice-nested}  fails to
identify it as a removable fragment: The local specification induced on the right
branch is
\[
\pspec[\quad]{\charFun{x=0}}{\Ass{x}{x+1}}{\charFun{x
  = 1}}~,
 \]
and clearly, $\charFun{x=0} \not \pimplies \charFun{x =
  1}$. Intuitively, the problem is that the information available to
slice the right branch (its local specification) does not account for the fact that the left
branch can by itself already establish the post-condition.

To improve precision, the slicing approach should incorporate a mutual
dependence analysis between the two branches, which might become highly
expensive as nesting level increases. We leave as future work exploring
a better compromise between efficiency and precision.


%% file: related.tex
\section{Related Work}
\label{sec:related}

There is a vast body of work on program slicing; we refere the reader
to \cite{Tip:JFP95} for an overview of different slicing techniques
and to \cite{Xu:2005} for an overview of different applications. Here we will focus only on
specification-based slicing, slicing approaches for
probabilistic programs and VCGens for establishing probabilistic
program specifications.

\subparagraph*{Specification-based slicing.} 
The notion of slicing with respect to a pre- and post-condition of
programs, \ie specification--based slicing, was introduced by Comuzzi
and Hart~\cite{Comuzzi:1996}. Since then, the approach has been
extended and refined by several
authors~\cite{Lee:2001,Chung:2001,DaCruz:2010,Barros:2012}. While the
original approach of Comuzzi and Hart~\cite{Comuzzi:1996} uses a
backward reasoning (\ie weakest pre-conditions) for constructing
slices, Lee~\etal~\cite{Lee:2001,Chung:2001} combine a backward with a
forward reasoning (the latter through strongest postconditions),
sequentially. Da Cruz~\etal~\cite{DaCruz:2010} extend
specification--based slicing to a contract--based setting, where
slicing is simultaneously performed over a set of procedures. Finally,
Barros~\etal~\cite{Barros:2012} show that a \emph{simultaneous}
(rather than a sequential) combination of forward and backward
reasoning is necessary (and sufficient) to deliver optimal slices. All
of these approaches are restricted to deterministic programs. Our
approach for probabilistic programs is along the lines of
Barros~\etal~\cite{Barros:2012} approach, but restricted to backward
reasoning, only, due to the limitations laid out in
Section~\ref{sec:discussion}.

\subparagraph*{Slicing of probabilistic programs.}
Hur~\etal~\cite{Hur:2014} were the first to explore the problem of
slicing for probabilistic programs. They observed that the classical
slicing approach based on data and control dependences becomes unsound
for probabilistic programs with conditioning, and showed how to
extended it with a new class of dependence to recover soundness. In a
later work, Amtoft and Banerjee~\cite{Amtoft:2016,Amtoft:TOPLAS:20}
introduce the notion of \emph{probabilistic control-flow graphs},
which allows a direct adaptation of conventional slicing machinery
such as data dependence, postdominators and relevant variables to the
case of probabilistic programs. Both the approaches by
Hur~\etal~\cite{Hur:2014} and Amtoft and
Banerjee~\cite{Amtoft:2016,Amtoft:TOPLAS:20} perform a conventional
slicing with respect to a set of (output) variables of interest.

\subparagraph*{VCGen for probabilistic programs.}
Hur~\etal~\cite{DBLP:journals/tcs/HurdMM05} present a formalization of the
$\wllpsymbol$ expectation transformers in the \textsf{HOL4} proof assistant for
\textsf{pGCL}, an extension of our language \Lang with (demonic)
non-determinism and failure. They define a VCGen for establishing the
partial correctness of programs annotated with loops invariants in the
same line as our VCGen from Section~\ref{sec:VCGen}. However, since the VCGen is
implemented in a Prolog interpreter instead of in the same proof
assistant, they are unable to provide a mechanized proof of the VCGen
soundness. On the contrary, we provide VCGens for establishing both the total and
partial correctness of programs, together with their respective (paper-and-pencil)
soundness proofs. 

Cock~\cite{Cock:AFP14,DBLP:journals/corr/abs-1211-6197} develops another
formalization of the $\wllpsymbol$ expectation transformers for
\textsf{pGCL} in the \textsf{Isabelle/HOL} proof assistant. In
contrast to  Hur~\etal~\cite{DBLP:journals/tcs/HurdMM05} who adopt a deep
embedding, Cock~\cite{Cock:AFP14} adopts a shallow embedding to take
advantage of the proof assistant mechanization. He also implements a
VCGen, but it is limited to loop-free programs. 

In contrast to ours, Hur~\etal's~\cite{DBLP:journals/tcs/HurdMM05} and
Cock's~\cite{Cock:AFP14} approaches, which are based on $\ZO$-valued
assertions over
states---expectations---Barthe~\etal~\cite{Barthe:2018:ESOPa} present
a Hoare logic based on Boolean assertions over state
distributions. Even though they do not introduce a VCGen itself, they
provide all the ingredients to do so: a weakest pre-condition
transformer for non-looping programs and syntactic conditions for
discharging the premises of (a subset of) the loop proof
rules. However, the problem of assertion entailments in this logic (as
required for slice computation) seems to be harder than that of
expectation entailment. Finally, Chadha~\etal~\cite{DBLP:journals/tcs/ChadhaCMS07}
provide a decidable Hoare logic also based on Boolean assertions over
state distributions. However, the logic is limited to straight-line
programs, only.

%% file: concl.tex
\section{Conclusion}
\label{sec:conclusion}

We have developed the first slicing approach for
probabilistic programs based on specifications. The slicing approach
is based on the backward
propagation of post-conditions and features appealing properties such
as termination-sensitivity and modularity via local reasoning
principles.

By applying our approach to a set of examples, we have shown that the
main benefit of specification-based program slicing---increased
precision---carries over the class of probabilistic
programs. This is particularly interesting due to the recent
resurgence of probabilistic programing and its intrinsic
complexity---any tool aiding
program understanding becomes vital.

We have identified several directions of future work. These include
improving slice precision by either incorporating the forward
propagation of pre-conditions or dispensing with the local reasoning
principle underlying modular slicing. Another interesting research avenues
comprise extending the language with conditioning---a key ingredient
of probabilistic modelling---and incorporating termination arguments
based on martingales.\vfill

%% file: appendixA.tex
\begin{proof}[Proof of Lemma~\ref{thm:vcgpartial-sound}]
	We prove that
	\[
		\models \vc{\prog}{\poste} %
		\quad \implies \quad %
		\wprec{\prog}{\poste} \pimplies \wlp{\prog}{\poste}~,
	\]
	by induction on the structure of $\prog$.

	\subparagraph*{Case $\prog = \Skip$}
	\begin{align*}
		  & \wprec{\Skip}{\poste}                                       \\
		= & \qquad \by{def of $\wprecsymbol$ for no-op}\displaybreak[0] \\
		  & \poste                                                      \\
		= & \qquad \by{def of $\wlpsymbol$ for no-op}\displaybreak[0]   \\
		  & \wlp{\Skip}{\poste}
	\end{align*}
	\subparagraph*{Case $\prog = \Ass{x}{E}$}
	\begin{align*}
		  & \wprec{\Ass{x}{E}}{\poste}                                       \\
		= & \qquad \by{def of $\wprecsymbol$ for assignment}\displaybreak[0] \\
		  & \poste\subst{x}{E}                                               \\
		= & \qquad \by{def of $\wlpsymbol$ for assignment}\displaybreak[0]   \\
		  & \wlp{\Ass{x}{E}}{\poste}
	\end{align*}
	\subparagraph*{Case $\prog = \Cond{G}{\prog_1}{\prog_2}$}
	From the hypothesis and the definition of $\vcsymbol$ we have
	$\models \vc{\prog_1}{\poste} $ and $\models
		\vc{\prog_2}{\poste}$. Thus
	\begin{align*}
		          & \wprec{\Cond{G}{\prog_1}{\prog_2}}{\poste}                                             \\
		=         & \qquad \by{def of $\wprecsymbol$ for conditional branching}\displaybreak[0]            \\
		          & \eval{G} \cdot\wprec{\prog_1}{\poste} +  \eval{ \lnot G} \cdot \wprec{\prog_2}{\poste} \\
		\pimplies & \qquad \by{inductive hypothesis}\displaybreak[0]                                       \\
		          & \eval{G} \cdot\wlp{\prog_1}{\poste} +  \eval{ \lnot G} \cdot \wlp{\prog_2}{\poste}     \\
		=         & \qquad \by{def of $\wlpsymbol$ for conditional branching}\displaybreak[0]              \\
		          & \wlp{\Cond{G}{\prog_1}{\prog_2}}{\poste}
	\end{align*}
	\subparagraph*{Case $\prog = \PChoice{\prog_1}{p}{\prog_2}$}
	From the hypothesis and the definition of $\vcsymbol$ we have
	$\models \vc{\prog_1}{\poste} $ and $\models
		\vc{\prog_2}{\poste}$. Thus
	\begin{align*}
		          & \wprec{\PChoice{\prog_1}{p}{\prog_2}}{\poste}                               \\
		=         & \qquad \by{def of $\wprecsymbol$ for probabilisitic choice}\displaybreak[0] \\
		          & p\cdot \wprec{\prog_1}{\poste} + (1-p)\cdot \wprec{\prog_2}{\poste}         \\
		\pimplies & \qquad \by{inductive hypothesis}\displaybreak[0]                            \\
		          & p\cdot \wlp{\prog_1}{\poste} + (1-p)\cdot \wlp{\prog_2}{\poste}             \\
		=         & \qquad \by{def of $\wlpsymbol$ for probabilisitic choice}\displaybreak[0]   \\
		          & \wlp{\Cond{G}{\prog_1}{\prog_2}}{\poste}
	\end{align*}
	\subparagraph*{Case $\prog =\prog_1\,;\,\prog_2$}
	From the hypothesis and the definition of $\vcsymbol$ we have
	$\models \vc{\prog_1}{\wprec{\prog_2}{g}}$ and $\models
		\vc{\prog_2}{\poste}$. Thus
	\begin{align*}
		          & \wprec{\prog_1\,;\,\prog_2}{\poste}                                          \\
		=         & \qquad \by{def of $\wprecsymbol$ for sequential composition}\displaybreak[0] \\
		          & \wprec{\prog_1}{\wprec{\prog_2}{\poste}}                                     \\
		\pimplies & \qquad \by{inductive hypothesis}\displaybreak[0]                             \\
		          & \wlp{\prog_1}{\wprec{\prog_2}{\poste}}                                       \\
		\pimplies & \qquad \by{inductive hypothesis}\displaybreak[0]                             \\
		          & \wlp{\prog_1}{\wlp{\prog_2}{\poste}}                                         \\
		=         & \qquad \by{def of $\wlpsymbol$ for sequential composition}\displaybreak[0]   \\
		          & \wlp{\prog_1\,;\,\prog_2}{\poste}
	\end{align*}

	\subparagraph*{Case $\prog = \IWhileDo{G}{\inv}{c}$}
	\renewcommand{\theequation}{\arabic{equation}}
	From the hypothesis and the definition of $\vcsymbol$  we have
	\begin{align}
		 & \models \left\{ \charFun{G} \cdot \inv
		\pimplies \wprec{\prog}{\inv} \right
		\}   ~\land              \label{eqn:hyp-wpre-fixpoint}        \\
		 & \models \left\{   \charFun{\lnot G} \cdot \inv
		\pimplies \poste \right\}  ~\land  \label{eqn:hyp-g-fixpoint} \\
		 & \models \vc{\prog}{\inv} \label{eqn:hyp-inv-hi}
	\end{align}
	Thus
	\begin{align*}
		          & \wprec{\IWhileDo{G}{\inv}{c}}{\poste}                                                          \\
		=         & \qquad \by{def of $\wprecsymbol$ for guarded loop}\displaybreak[0]                             \\
		          & \inv                                                                                           \\
		=         & \qquad \by{algebra}\displaybreak[0]                                                            \\
		          & \charFun{\lnot G}\cdot\inv + \charFun{G}\cdot\inv                                              \\
		=         & \qquad \by{idempotency}\displaybreak[0]                                                        \\
		          & \charFun{\lnot G}\cdot
		\left(\charFun{\lnot G}\cdot\inv\right) +
		\charFun{G}\cdot\left(\charFun{G} \cdot \inv\right)                                                        \\
		\pimplies & \qquad \by{using \eqref{eqn:hyp-wpre-fixpoint} and \eqref{eqn:hyp-g-fixpoint}}\displaybreak[0] \\
		          & \charFun{\lnot G}\cdot \poste + \charFun{G} \cdot \wprec{\prog}{\inv}                          \\
		\pimplies & \qquad \by{inductive hypothesis, using \eqref{eqn:hyp-inv-hi}}\displaybreak[0]                                               \\
		          & \charFun{\lnot G}\cdot \poste  + \charFun{G} \cdot \wlp{\prog}{\inv}
	\end{align*}
	Let us define $H(h)=\charFun{\lnot G}\cdot \poste + \charFun{G} \cdot
		\wlp{\prog}{h}$. From the above derivation we can conclude that $$\inv
		\pimplies \charFun{\lnot G}\cdot \poste + \charFun{G} \cdot
		\wlp{\prog}{\inv} = H(\inv)~.$$
	Now by Park's Theorem~\cite{Wechler:2012},
	\begin{align*}
		          & \inv                                                                   \\
		\pimplies & \qquad \by{Park's Theorem}\displaybreak[0]                             \\
		          & \nu h.\: H(h)                                                          \\
		=         & \qquad \by{def of $H$ and \wlpsymbol for guarded loop}\displaybreak[0] \\
		          & \wlp{\IWhileDo{G}{\inv}{c}}{\poste}
	\end{align*}
	Combining the results, we obtain
	\[
		\wprec{\IWhileDo{G}{\inv}{c}}{\poste} ~=~ \inv ~\pimplies
		\wlp{\IWhileDo{G}{\inv}{c}}{\poste}\qedhere
		~\]
\end{proof}

\bigskip
\begin{proof}[Proof of Lemma~\ref{thm:vc-monot}]
	Let $\poste \pimplies \poste'$. We prove the monotonicity of $\vcd{\prog}$ by induction on the structure of $c$ (the monotonicity of $\vcgd{\prog}$ follows as immediate corollary).

	\subparagraph*{Case $\prog = \Skip$}
	Vacuously true.

	\subparagraph*{Case $\prog =\Ass{x}{E}$}
	Vacuously true.
	%
	%
	\subparagraph*{Case $\prog =\Cond{G}{c_1}{c_2}$}
	\begin{align*}
		%
		                & \models \vc{\Cond{G}{c_1}{c_2}}{\poste}           \\
		\Leftrightarrow & \qquad \by{def of $\vcsymbol$}\displaybreak[0]  \\
		                & \models  \vc{c_1}{\poste} \cup \vc{c_2}{\poste}   \\
		\Rightarrow     & \qquad \by{inductive hypothesis} \displaybreak[0] \\
		                & \models \vc{c_1}{\poste'} \cup \vc{c_2}{\poste'}   \\
		\Leftrightarrow & \qquad \by{def of $\vcsymbol$} \displaybreak[0]   \\
		                & \models \vc{\Cond{G}{c_1}{c_2}}{\poste'}
	\end{align*}

	\subparagraph*{Case $\prog =\PChoice{c_1}{p}{c_2}$}
	Analogous to the previous case.

	\subparagraph*{Case $\prog =c_1;c_2$}
	\begin{align*}
		                & \models \vc{c_1;c_2}{\poste}                                                      \\
		\Leftrightarrow & \qquad \by{def of $\vcsymbol$}\displaybreak[0]                                   \\
		                & \models  \vc{c_1}{\wprec{c_2}{\poste}} \cup \vc{c_2}{\poste}                      \\
		\Rightarrow     & \qquad \by{inductive hypothesis and $\wprecsymbol$ monotonicity} \displaybreak[0] \\
		                & \models \vc{c_1}{\wprec{c_2}{\poste'}} \cup \vc{c_2}{\poste'}                     \\
		\Leftrightarrow & \qquad \by{def of $\vcsymbol$} \displaybreak[0]                                   \\
		                & \models \vc{c_1;c_2}{\poste'}
	\end{align*}

	\subparagraph*{Case $\prog =\IWhileDo{G}{\inv}{c}$}

	\begin{align*}
		%
		                & \models \vc{\IWhileDo{G}{\inv}{c}}{\poste}                                                                                 \\
		\Leftrightarrow & \qquad \by{def of $\vcsymbol$}\displaybreak[0]                                                                             \\
		                & \models  \{ \eval{G} \cdot \inv \pimplies \wprec{c}{\inv}, \eval{\lnot G} \cdot \inv \pimplies \poste \} \cup \vc{c}{\inv} \\
		\Rightarrow     & \qquad \by{transitivity of $\pimplies $} \displaybreak[0]                                                                  \\
		                & \models \{ \eval{G} \cdot \inv \pimplies \wprec{c}{\inv}, \eval{\lnot G} \cdot \inv \pimplies \poste' \} \cup \vc{c}{\inv} \\
		\Leftrightarrow & \qquad \by{def of $\vcsymbol$} \displaybreak[0]                                                                           \\
		                & \models
		\vc{\IWhileDo{G}{\inv}{c}}{\poste'} \qedhere
	\end{align*}

\end{proof}

\bigskip
\begin{proof}[Proof of Lemma~\ref{thm:alt-vcg}]~

	\begin{enumerate}
		\setlength\itemsep{2ex}

		\item Let $\inst_j = \Cond{G}{\prog_1}{\prog_2}$ for some $1\leq j \leq n$. Then
		      \begin{align*}
			                      & \models \vcg{\pree}{\prog}{\poste}                                                                                                                              \\
			      \Leftrightarrow & \qquad \by{def of $\vcgsymbol$}\displaybreak[0]                                                                                                                 \\
			                      & \models \{ \pree \pimplies \wprec{\prog}{\poste} \} \, \cup \, \vc{\prog}{\poste}                                                                               \\
			      \Leftrightarrow & \qquad \by{def of \vcsymbol for sequential composition}\displaybreak[0]                                                                                         \\
			                      & \models \{ \pree \pimplies \wprec{\prog}{\poste} \} \, \cup \, \vc{\seqn{1}{j}}{\wpreci{j+1}{\prog}{\poste}} \, \cup \, \vci{j+1}{\prog}{\poste}                \\
			      \Leftrightarrow & \qquad \by{def of \vcsymbol for sequential composition}\displaybreak[0]                                                                                         \\
			                      & \models \{ \pree \pimplies \wprec{\prog}{\poste} \} \, \cup \, \vc{\seqn{1}{j-1}}{\wprec{\inst_j}{\wpreci{j+1}{\prog}{\poste}}}                                 \\
			                      & \phantom{\models} \, \cup \, \vc{\inst_j}{\wpreci{j+1}{\prog}{\poste}} \, \cup \, \vci{j+1}{\prog}{\poste}                                                      \\
			      \Leftrightarrow & \qquad \by{def of \vcsymbol for conditional branching}\displaybreak[0]                                                                                          \\
			                      & \models \{ \pree \pimplies \wprec{\prog}{\poste} \} \, \cup \, \vc{\seqn{1}{j-1}}{\wprec{\inst_j}{\wpreci{j+1}{\prog}{\poste}}}                                 \\
			                      & \phantom{\models} \, \cup \, \vc{\prog_1}{\wpreci{j+1}{\prog}{\poste}} \, \cup \, \vc{\prog_2}{\wpreci{j+1}{\prog}{\poste}} \, \cup \, \vci{j+1}{\prog}{\poste} \\
			      \Leftrightarrow & \qquad \by{def of \wprecsymbol for sequential composition}\displaybreak[0]                                                                                      \\
			                      & \models \{ \pree \pimplies \wprec{\seqn{1}{j-1}}{\wpreci{j}{\prog}{\poste}} \}                                                                                  \\
			                      & \phantom{\models} \, \cup \, \vc{\seqn{1}{j-1}}{\wpreci{j}{\prog}{\poste}} \, \cup \, \vc{\prog_1}{\wpreci{j+1}{\prog}{\poste}}                                 \\
			                      & \phantom{\models} \, \cup \, \vc{\prog_2}{\wpreci{j+1}{\prog}{\poste}} \, \cup \, \vci{j+1}{\prog}{\poste}                                                      \\
			      \Leftrightarrow & \qquad \by{def of $\vcgid{k}$ and rearrange terms}\displaybreak[0]                                                                                           \\
			                      & \models \vci{j+1}{\prog}{\poste}                                                                                                                                \\
			                      & \phantom{\models} \, \cup \, \vc{\prog_1}{\wpreci{j+1}{\prog}{\poste}}                                                                                          \\
			                      & \phantom{\models} \, \cup \, \vc{\prog_2}{\wpreci{j+1}{\prog}{\poste}}                                                                                          \\
			                      & \phantom{\models} \, \cup \,\vcgi{j-1}{\pree}{\prog}{\wpreci{j}{\prog}{\poste}}
		      \end{align*}

		      The case where $\inst_j = \PChoice{\prog_1}{p}{\prog_2}$ follows the same argument.

		\item $\inst_j = \IWhileDo{G}{\inv}{\prog'}$ for some $1 \leq j \leq n$. Then
		      \begin{align*}
			                      & \models \vcg{\pree}{\prog}{\poste}                                                                                                                                    \\
			      \Leftrightarrow & \qquad \by{def of $\vcgsymbol$}\displaybreak[0]                                                                                                                       \\
			                      & \models \{ \pree \pimplies \wprec{\prog}{\poste} \} \, \cup \, \vc{\prog}{\poste}                                                                                     \\
			      \Leftrightarrow & \qquad \by{def of $\wprecsymbol$ and $\vcsymbol$ for sequential composition}\displaybreak[0]                                                                          \\
			                      & \models \{ \pree \pimplies \wprec{\seqn{1}{j-1}}{\wprec{\inst_j}{\wpreci{j+1}{\prog}{\poste}}} \}                                                                     \\
			                      & \phantom{\models} \, \cup \, \vc{\seqn{1}{j}}{\wpreci{j+1}{\prog}{\poste}} \, \cup \, \vci{j+1}{\prog}{\poste}                                                        \\
			      \Leftrightarrow & \qquad \by{def of $\wprecsymbol$ for guarded loop and def of $\vcsymbol$ for sequential composition}\displaybreak[0]                                                  \\
			                      & \models \{ \pree \pimplies \wprec{\seqn{1}{j-1}}{\inv} \}                                                                                                             \\
			                      & \phantom{\models} \, \cup \, \vc{\seqn{1}{j-1}}{\wprec{\inst_j}{\wpreci{j+1}{\prog}{\poste}}}                                                                         \\
			                      & \phantom{\models} \, \cup \, \vc{\inst_j}{\wpreci{j+1}{\prog}{\poste}} \, \cup \, \vci{j+1}{\prog}{\poste}                                                            \\
			      \Leftrightarrow & \qquad \by{def of $\wprecsymbol$ and $\vcsymbol$ for guarded loop}\displaybreak[0]                                                                                    \\
			                      & \models \{ \pree \pimplies \wprec{\seqn{1}{j-1}}{\inv} \} \, \cup \, \vc{\seqn{1}{j-1}}{\inv}                                                                         \\
			                      & \phantom{\models} \, \cup \, \left\{\charFun{G} \cdot \inv \pimplies \wprec{\prog'}{\inv},\charFun{\lnot G} \cdot \inv \pimplies \wpreci{j+1}{\prog}{\poste} \right\} \\
			                      & \phantom{\models} \, \cup \, \vc{\prog'}{\inv} \, \cup \, \vci{j+1}{\prog}{\poste}                                                                                    \\
			      \Leftrightarrow & \qquad \by{def of $\vcgid{k}$ and algebra}\displaybreak[0]                                                                                                            \\
			                      & \models \vcgi{j-1}{\pree}{\prog}{\inv}                                                                                                                                \\
			                      & \phantom{\models} \, \cup \, \left\{\charFun{G} \cdot \inv \pimplies \wprec{\prog'}{\inv}\right\}  \, \cup \, \vc{\prog'}{\inv}                                       \\
			                      & \phantom{\models} \, \cup \, \left\{\charFun{\lnot G} \cdot \inv \pimplies \wpreci{j+1}{\prog}{\poste} \right\} \, \cup \, \vci{j+1}{\prog}{\poste}                   \\
			      \Leftrightarrow & \qquad \by{def of \vcgsymbol and rearrange terms}\displaybreak[0]                                                                                                    \\
			                      & \models \vci{j+1}{\prog}{\poste}                                                                                                                                      \\
			                      & \phantom{\models} \, \cup \, \left\{\charFun{\lnot G} \cdot \inv \pimplies \wpreci{j+1}{\prog}{\poste} \right\}                                                       \\
			                      & \phantom{\models} \, \cup \, \vcg{\charFun{G}\cdot \inv}{\prog'}{\inv}                                                                                                \\
			                      & \phantom{\models} \, \cup \, \vcgi{j-1}{\pree}{\prog}{\inv} \qedhere
		      \end{align*}
	\end{enumerate}
\end{proof}

\bigskip
\begin{proof}[Proof of Theorem~\ref{thm:slice-nested}]
	\renewcommand{\theequation}{\arabic{equation}}

	By induction on the derivation of the relation
	$\localSpecp{\pree}{\prog}{\poste}{\pree'}{\prog'}{\poste'}$
	(see Figure \ref{fig:local-spec}).

	\subparagraph*{Case $\lrule{\localSpecSymbol\texttt{ift}}$} In this case we have
	\[
		\begin{array}{lcl}
			\prog   & = &
			\seqn{1}{j-1};\Cond{G}{\prog'}{\prog_{f}};\seqn{j+1}{n}                     \\
			\pree'  & = & \charFun{G} \cdot \wprec{\prog'}{\wpreci{j+1}{\prog}{\poste}} \\
			\poste' & = & \wpreci{j+1}{\prog}{\poste}
		\end{array}
	\]
	We must show that $\prog\subst{\prog'}{\prog''} \preccurlyeq \prog$ and
	$\models \vcg{\pree}{\prog}{\poste} \implies
		\vcg{\pree}{\prog\subst{\prog'}{\prog''}}{\poste}$. The
	first proof obligation is straightforward  (see
	Figure~\ref{fig:is-portion-of}). To establish the
	second proof obligation, we exploit
	Lemma~\ref{thm:alt-vcg} and the fact that
	$\prog\subst{\prog'}{\prog''}$ coincides with
	$\prog$ in all but the $j$-th instruction. Therefore, assuming
	\begin{align}
		\quad     & \models \vci{j+1}{\prog}{\poste}  \nonumber                                                 \\
		\land ~~~ & \models \vc{\prog'}{\wpreci{j+1}{\prog}{\poste}} \label{eq:hyp-speclocal}                   \\
		\land ~~~ & \models \vc{\prog_f}{\wpreci{j+1}{\prog}{\poste}} \nonumber                                 \\
		\land ~~~ & \models \vcgi{j-1}{\pree}{\prog}{\wpreci{j}{\prog}{\poste}} \label{eq:vcgti-transformwprec}
	\end{align}
	we have to conclude that
	\begin{align}
		\quad     & \models \vci{j+1}{\prog}{\poste} \nonumber                                                                              \\
		\land ~~~ & \models \vc{\prog''}{\wpreci{j+1}{\prog}{\poste}} \label{eq:hyp-speclocal-goal}                                         \\
		\land ~~~ & \models \vc{\prog_f}{\wpreci{j+1}{\prog}{\poste}} \nonumber                                                             \\
		\land ~~~ & \models \vcgi{j-1}{\pree}{\prog}{\wpreci{j}{\prog\subst{\prog'}{\prog''}}{\poste}} \label{eq:vcgti-transformwprec-goal}
	\end{align}
	We need to prove only
	Equations~\eqref{eq:hyp-speclocal-goal}  and
	\eqref{eq:vcgti-transformwprec-goal} since the
	other two are already part of the premises. Let us start with
	Equation~\eqref{eq:hyp-speclocal-goal}. From hypothesis
	$\pspecSlice{\prog''}{\pree'}{\poste'}{\prog'}$, we have:
	\begin{equation}\label{eqn:ift-spec-local}
		\begin{array}{l@{}l}
			                & \models \left\{ \charFun{G} \cdot \wprec{\prog'}{\wpreci{j+1}{\prog}{\poste}} \pimplies \wprec{\prog'}{\wpreci{j+1}{\prog}{\poste}} \right\}  \\
			                & ~~~ \, \cup \, \vc{\prog'}{\wpreci{j+1}{\prog}{\poste}}                                                                                       \\
			\Rightarrow ~~~ & \models \left\{ \charFun{G} \cdot \wprec{\prog'}{\wpreci{j+1}{\prog}{\poste}} \pimplies \wprec{\prog''}{\wpreci{j+1}{\prog}{\poste}} \right\} \\
			                & ~~~ \, \cup \, \vc{\prog''}{\wpreci{j+1}{\prog}{\poste}}
		\end{array}
	\end{equation}
	\noindent The premise of the above implication holds true from the
	fact that $\charFun{G} \cdot h \pimplies h$ for any expectation $h$,
	and from Equation \eqref{eq:hyp-speclocal}. Thus we can conclude, in
	particular,
	\[
		\vc{\prog''}{\wpreci{j+1}{\prog}{\poste}}~,
	\]
	which amounts to Equation~\eqref{eq:hyp-speclocal-goal}.
	\noindent Finally, we show that Equation~\eqref{eq:vcgti-transformwprec-goal}
	follows from Equation~\eqref{eq:vcgti-transformwprec}:
	\begin{align*}
		                & \models \vcgi{j-1}{\pree}{\prog}{\wpreci{j}{\prog}{\poste}}                                                                                                              \\
		\Leftrightarrow & \qquad \by{def of \wprecsymbol for sequential composition}\displaybreak[0]                                                                                               \\
		                & \models \vcgi{j-1}{\pree}{\prog}{\wprec{\inst_j}{\wpreci{j+1}{\prog}{\poste}}}                                                                                           \\
		\Leftrightarrow & \qquad \by{def of \wprecsymbol for conditional branching}\displaybreak[0]                                                                                                \\
		                & \models \vcgi{j-1}{\pree}{\prog}{
							\begin{array}{l}
								\charFun{G}\cdot \wprec{\prog'}{\wpreci{j+1}{\prog}{\poste}} + \\
								\charFun{\lnot G}\cdot \wprec{\prog_f}{\wpreci{j+1}{\prog}{\poste}}                 
								\end{array}}	\\
		\Leftrightarrow & \qquad \by{idempotency}\displaybreak[0]                                                                                                                                  \\
		                & \models \vcgi{j-1}{\pree}{\prog}{
							\begin{array}{l}
								\charFun{G}\cdot \charFun{G}\cdot \wprec{\prog'}{\wpreci{j+1}{\prog}{\poste}} + \\
								 \charFun{\lnot G} \cdot \wprec{\prog_f}{\wpreci{j+1}{\prog}{\poste}}
								\end{array}	}\\
		\Rightarrow     & \qquad \by{by \eqref{eqn:ift-spec-local} and monotonicity of \vcgsymbol}\displaybreak[0]                                                                                 \\
		                & \models \vcgi{j-1}{\pree}{\prog}{
								\begin{array}{l}
									\charFun{G}\cdot \wprec{\prog''}{\wpreci{j+1}{\prog}{\poste}} +\\
									 \charFun{\lnot G} \cdot \wprec{\prog_f}{\wpreci{j+1}{\prog}{\poste}}
									\end{array}}\\
		\Leftrightarrow & \qquad \by{def of \wprecsymbol for conditional branching}\displaybreak[0]                                                                                                \\
		                & \models \vcgi{j-1}{\pree}{\prog}{\wprec{\Cond{G}{\prog''}{\prog_f}}{\wpreci{j+1}{\prog}{\poste}}}                                                                        \\
		\Leftrightarrow & \qquad \by{def of $\prog\subst{\prog'}{\prog''}$}\displaybreak[0]                                                                                                        \\
		                & \models \vcgi{j-1}{\pree}{\prog}{\wpreci{j}{\prog\subst{\prog'}{\prog''}}{\poste}}
	\end{align*}

	\subparagraph*{Case $\lrule{\localSpecSymbol\texttt{iff}}$} Analogous to
	case $\lrule{\localSpecSymbol\texttt{ift}}$.

	\subparagraph*{Case $\lrule{\localSpecSymbol\texttt{pl}}$} We
	proceed analogously to the case of rule $\lrule{\localSpecSymbol\texttt{ift}}$. Now we have
	\[
		\begin{array}{lcl}
			\prog   & = &
			\seqn{1}{j-1};\PChoice{\prog'}{p}{\prog_r};\seqn{j+1}{n}  \\
			\pree'  & = & \wprec{\prog'}{\wpreci{j+1}{\prog}{\poste}} \\
			\poste' & = & \wpreci{j+1}{\prog}{\poste}
		\end{array}
	\]
	Again, we must show that $\prog\subst{\prog'}{\prog''} \preccurlyeq \prog$ and
	$\models \vcg{\pree}{\prog}{\poste} \implies
		\vcg{\pree}{\prog\subst{\prog'}{\prog''}}{\poste}$. The
	first proof obligation is straightforward  (see
	Figure~\ref{fig:is-portion-of}). To establish the
	second proof obligation, we exploit
	Lemma~\ref{thm:alt-vcg} and the fact that
	$\prog\subst{\prog'}{\prog''}$ coincides with
	$\prog$ in all but the $j$-th instruction. Therefore, assuming
	\begin{align}
		\quad     & \models \vci{j+1}{\prog}{\poste}  \nonumber                                                    \\
		\land ~~~ & \models \vc{\prog'}{\wpreci{j+1}{\prog}{\poste}} \label{eq:pl-hyp-speclocal}                   \\
		\land ~~~ & \models \vc{\prog_r}{\wpreci{j+1}{\prog}{\poste}} \nonumber                                    \\
		\land ~~~ & \models \vcgi{j-1}{\pree}{\prog}{\wpreci{j}{\prog}{\poste}} \label{eq:pl-vcgti-transformwprec}
	\end{align}
	we have to conclude that
	\begin{align}
		\quad     & \models \vci{j+1}{\prog}{\poste} \nonumber                                                                                 \\
		\land ~~~ & \models \vc{\prog''}{\wpreci{j+1}{\prog}{\poste}} \label{eq:pl-hyp-speclocal-goal}                                         \\
		\land ~~~ & \models \vc{\prog_r}{\wpreci{j+1}{\prog}{\poste}} \nonumber                                                                \\
		\land ~~~ & \models \vcgi{j-1}{\pree}{\prog}{\wpreci{j}{\prog\subst{\prog'}{\prog''}}{\poste}} \label{eq:pl-vcgti-transformwprec-goal}
	\end{align}
	We need to prove only
	Equations~\eqref{eq:pl-hyp-speclocal-goal}  and
	\eqref{eq:pl-vcgti-transformwprec-goal} since the
	other two are already part of the premises. Let us start with
	Equation~\eqref{eq:pl-hyp-speclocal-goal}. From hypothesis
	$\pspecSlice{\prog''}{\pree'}{\poste'}{\prog'}$, we have:
	\begin{equation}\label{eqn:pl-spec-local}
		\begin{array}{l@{}l}
			                & \models \left\{ \wprec{\prog'}{\wpreci{j+1}{\prog}{\poste}} \pimplies \wprec{\prog'}{\wpreci{j+1}{\prog}{\poste}} \right\}  \\
			                & ~~~ \, \cup \, \vc{\prog'}{\wpreci{j+1}{\prog}{\poste}}                                                                     \\
			\Rightarrow ~~~ & \models \left\{ \wprec{\prog'}{\wpreci{j+1}{\prog}{\poste}} \pimplies \wprec{\prog''}{\wpreci{j+1}{\prog}{\poste}} \right\} \\
			                & ~~~ \, \cup \, \vc{\prog''}{\wpreci{j+1}{\prog}{\poste}}
		\end{array}
	\end{equation}
	\noindent The premise of the above implication holds true from the
	fact that $h \pimplies h$ for any expectation $h$,
	and from Equation \eqref{eq:pl-hyp-speclocal}. Thus we can conclude, in
	particular,
	\[
		\vc{\prog''}{\wpreci{j+1}{\prog}{\poste}}~,
	\]
	which amounts to Equation~\eqref{eq:pl-hyp-speclocal-goal}.
	\noindent Finally, we show that Equation~\eqref{eq:pl-vcgti-transformwprec-goal}
	follows from Equation~\eqref{eq:pl-vcgti-transformwprec}:
	\begin{align*}
		                & \models \vcgi{j-1}{\pree}{\prog}{\wpreci{j}{\prog}{\poste}}                                                                                   \\
		\Leftrightarrow & \qquad \by{def of \wprecsymbol for sequential composition}\displaybreak[0]                                                                    \\
		                & \models \vcgi{j-1}{\pree}{\prog}{\wprec{\inst_j}{\wpreci{j+1}{\prog}{\poste}}}                                                                \\
		\Leftrightarrow & \qquad \by{def of \wprecsymbol for probabilistic choice}\displaybreak[0]                                                                      \\
		                & \models \vcgi{j-1}{\pree}{\prog}{\begin{array}{l}
							p\cdot \wprec{\prog'}{\wpreci{j+1}{\prog}{\poste}} + \;\\
							(1-p)\cdot\wprec{\prog_r}{\wpreci{j+1}{\prog}{\poste}}
						\end{array}}\\
							%
		%
		\Rightarrow     & \qquad \by{using \eqref{eqn:pl-spec-local} and monotonicity of \vcgsymbol}\displaybreak[0]                                                    \\
		                & \models \vcgi{j-1}{\pree}{\prog}{\begin{array}{l}
							p\cdot \wprec{\prog''}{\wpreci{j+1}{\prog}{\poste}} + \;\\
							(1-p)\cdot\wprec{\prog_r}{\wpreci{j+1}{\prog}{\poste}}
						\end{array}
						} \\
		\Leftrightarrow & \qquad \by{def of \wprecsymbol for probabilistic choice}\displaybreak[0]                                                                      \\
		                & \models \vcgi{j-1}{\pree}{\prog}{\wprec{\PChoice{\prog''}{p}{\prog_r}}{\wpreci{j+1}{\prog}{\poste}}}                                          \\
		\Leftrightarrow & \qquad \by{Use def of $\prog\subst{\prog'}{\prog''}$}\displaybreak[0]                                                                        \\
		                & \models \vcgi{j-1}{\pree}{\prog}{\wpreci{j}{\prog\subst{\prog'}{\prog''}}{\poste}}
	\end{align*}

	\subparagraph*{Case $\lrule{\localSpecSymbol\texttt{pr}}$}  Analogous to
	case  $\lrule{\localSpecSymbol\texttt{pl}}$
	\subparagraph*{Case $\lrule{\localSpecSymbol\texttt{while}}$} In this case we have
	\[
		\begin{array}{lcl}
			\prog   & = &
			\seqn{1}{j-1};\IWhileDo{G}{inv}{\prog'};\seqn{j+1}{n} \\
			\pree'  & = & \charFun{G}\cdot\inv                    \\
			\poste' & = & \inv
		\end{array}
	\]
	We must show that $\prog\subst{\prog'}{\prog''} \preccurlyeq \prog$ and
	$\models \vcg{\pree}{\prog}{\poste} \implies \allowbreak
		\vcg{\pree}{\prog\subst{\prog'}{\prog''}}{\poste}$. The
	first proof obligation is straightforward  (see
	Figure~\ref{fig:is-portion-of}). To establish the
	second proof obligation, we exploit
	Lemma~\ref{thm:alt-vcg} and the fact that
	$\prog\subst{\prog'}{\prog''}$ coincides with
	$\prog$ in all but the $j$-th instruction. Therefore, assuming
	\begin{align}
		\quad     & \models \vci{j+1}{\prog}{\poste}  \nonumber                                                  \\
		\land ~~~ & \models \{ \charFun{\lnot G} \cdot  \inv  \pimplies \wpreci{j+1}{\prog}{\poste} \} \nonumber \\
		\land ~~~ & \models \vcg{\charFun{G} \cdot \inv }{\prog'}{\inv} \label{eq:while-vcg-spec}                \\
		\land ~~~ & \models \vcgi{j-1}{\pree}{\prog}{\inv} \nonumber
	\end{align}
	we have to conclude that
	\begin{align}
		\quad     & \models \vci{j+1}{\prog}{\poste} \nonumber                                                   \\
		\land ~~~ & \models \{ \charFun{\lnot G} \cdot  \inv  \pimplies \wpreci{j+1}{\prog}{\poste} \} \nonumber \\
		\land ~~~ & \models \vcg{\charFun{G} \cdot \inv }{\prog''}{\inv} \label{eq:while-vcg-spec-goal}          \\
		\land ~~~ & \models \vcgi{j-1}{\pree}{\prog}{\inv} \nonumber
	\end{align}
	We need to prove only
	Equation~\eqref{eq:while-vcg-spec} since the
	other three are already part of the premises. From hypothesis
	$\pspecSlice{\prog''}{\pree'}{\poste'}{\prog'}$, we have:
	$$\models \vcg{\charFun{G} \cdot \inv}{\prog'}{\inv} ~\Rightarrow~ \vcg{\charFun{G} \cdot \inv}{\prog''}{\inv}$$
	The premise of the above implication holds true from Equation~\eqref{eq:while-vcg-spec}. Then we can conclude $\vcg{\charFun{G} \cdot \inv}{\prog''}{\inv}$ and \eqref{eq:while-vcg-spec-goal} is proved.
	\subparagraph*{Case $\lrule{\localSpecSymbol\texttt{refl}}$} This case is inmediate because $\pree' = \pree$ and $\poste' = \poste$ and $\prog =\prog'$.

	\subparagraph*{Case $\lrule{\localSpecSymbol\texttt{trans}}$} Let $\prog^{*}$ be a \Lang with its respective pre- and
	post-expectation $\pree^{*}$ and $\poste^{*}$ such that $\prog'$ is a subprogram of $\prog^{*}$, $\localSpecp{\pree}{\prog}{\poste}{\pree^{*}}{\prog^{*}}{\poste^{*}}$ and $\localSpecp{\pree^{*}}{\prog^{*}}{\poste^{*}}{\pree'}{\prog'}{\poste'}$.
	Let us apply the induction hypothesis to $\prog^{*}$ and $\prog'$, this gives $\pspecSlice{\prog''}{\pree'}{\poste'}{\prog'} \Rightarrow \pspecSlice{\prog^{*}\subst{\prog'}{\prog''}}{\pree^{*}}{\poste^{*}}{\prog^{*}}$.
	We now apply this argument again $\pspecSlice{\prog^{*}\subst{\prog'}{\prog''}}{\pree^{*}}{\poste^{*}}{\prog^{*}} \Rightarrow \pspecSlice{\prog\subst{\prog^{*}}{\prog^{*}\subst{\prog'}{\prog''}}}{\pree}{\poste}{\prog}$.
	\noindent This completes the proof.
\end{proof}
\bigskip
\begin{proof}[Proof of Lemma~\ref{thm:vcgtotal-sound}]
	\renewcommand{\theequation}{\arabic{equation}}
	We proceed to show that
	\[
		\models \vct{\prog}{\poste} %
		\quad \implies \quad %
		\wprect{\prog}{\poste} \pimplies \wp{\prog}{\poste}~
	\]
	by induction on the structure of $\prog$. We only provide the case of loops as the remaining
	cases follows the same argument as for the counterpart  $\vcgsymbol$
	for partial correctness (see proof of Lemma~\ref{thm:vcgpartial-sound}). The proof of lemma follows as an immediate corollary of previos property.

	\subparagraph*{Case $\prog = \IWhileDo{G}{\inv, T, v, \mathtt{l}, \mathtt{u}, \epsilon}{\prog'}$}

	From the hypothesis \\ $\models \vct{\IWhileDo{G}{\inv, T, v, \mathtt{l}, \mathtt{u}, \epsilon}{\prog'}}{\poste}$ we have
	\begin{align}
		 & \models \left\{ \charFun{G\land T} \pimplies \wprect{\prog}{\charFun{T}} \right\}      \; \land                                 \label{eqn:variant-rule-2}   \\
		 & \models \left\{ \epsilon \cdot \charFun{G \land T \land v=v_0} \pimplies \wprect{\prog}{\charFun{v<v_0}} \right\} \; \land  \label{eqn:variant-rule-3}       \\
		 & \models \left\{ \charFun{G \land T} \pimplies \charFun{\mathtt{l} \leq v \leq \mathtt{u}} \right\}     \; \land         \nonumber                            \\
		 & \models \vct{\prog}{\charFun{T}}                                                                               \; \land           \label{eqn:variant-rule-5} \\
		 & \models \vct{\prog}{\charFun{v<v_0}}                                                                  \; \land           \label{eqn:variant-rule-6}          \\
		 & \models \left\{ \charFun{G}\cdot \inv \pimplies \wprec{\prog}{\inv} \right\}                           \; \land       \label{eqn:variant-rule-9}             \\
		 & \models \left\{ \charFun{\lnot G}\cdot \inv \pimplies \poste \right\}                                 \; \land         \label{eqn:variant-rule-8}            \\
		 & \models \vc{\prog}{\inv} \label{eqn:variant-rule-10}
	\end{align}
	%
	%
	First, we begin proving that the loop terminates almost-surely
	from any state in $T$. To this end, we apply \cite[Lemma
		7.5.1]{McIver:2004}, which requires proving that:
	\begin{align}
		 & \charFun{G \land T} \pimplies \charFun{\mathtt{l} \leq v \leq \mathtt{u}} \; \land \nonumber                     \\
		 & \charFun{G \land T} \pimplies \wp{\prog}{\charFun{T}} \; \land \label{eqn:lemma7-hyp2}                           \\
		 & \epsilon \cdot \charFun{G \land T \land v=v_0} \pimplies \wprect{\prog}{\charFun{v<v_0}} \label{eqn:lemma7-hyp3}
	\end{align}
	We need to prove only Equations~\eqref{eqn:lemma7-hyp2} and \eqref{eqn:lemma7-hyp3} since the first equeation is already part of the premises.
	To establish Equation~\eqref{eqn:lemma7-hyp2}, we apply inductive
	hypothesis on $\prog$ and from premise \eqref{eqn:variant-rule-5}, we
	conclude that $\wprect{\prog}{\charFun{T}} \pimplies \wp{\prog}{\charFun{T}}$, which
	together with premise \eqref{eqn:variant-rule-2} readily establishes
	Equation~\eqref{eqn:lemma7-hyp2}.  To establish
	Equation~\eqref{eqn:lemma7-hyp3}, we follow the same argument (exploiting premises
	\eqref{eqn:variant-rule-6} and \eqref{eqn:variant-rule-3}).

	\noindent In the proof of Lemma~\ref{thm:vcgpartial-sound}, we showed that
	\[
		\models \vc{\prog}{\poste} %
		\quad \implies \quad %
		\wprec{\prog}{\poste} \pimplies \wlp{\prog}{\poste}~.
	\]
	Following the same argument as above, and exploiting premises
	\eqref{eqn:variant-rule-10} and \eqref{eqn:variant-rule-9}, we
	conclude that
	\[
		\charFun{G}\cdot \inv \pimplies \wlp{\prog}{\inv}~,
	\]
	which says that $\inv$ is a (weak) loop invariant. From this, and the
	fact that the loop terminates almost-surely from $T$, we can conclude
	the proof appealing to \cite[Lemma 2.4.1-Case 2]{McIver:2004} as
	follows:
	\begin{align*}
		          & \wprect{\prog}{\poste}                                                                               \\
		=         & \qquad \by{def of $\wprectsymbol$ for guarded loop}\displaybreak[0]                                  \\
		          & \charFun{T}\cdot\inv                                                                                 \\
		\pimplies & \qquad \by{\cite[Lemma 2.4.1-Case 2]{McIver:2004}}\displaybreak[0]                                   \\
		          & \wp{\IWhileDo{G}{\inv, T, v, \mathtt{l}, \mathtt{u}, \epsilon}{\prog'}}{\charFun{\lnot G}\cdot \inv} \\
		\pimplies & \qquad \by{use \eqref{eqn:variant-rule-8} and monotonicity of \wpsymbol}\displaybreak[0]             \\
		          & \wp{\IWhileDo{G}{\inv, T, v, \mathtt{l}, \mathtt{u}, \epsilon}{\prog'}}{\poste}                      \\
		\pimplies & \qquad \by{def of $\prog$}\displaybreak[0]                                                           \\
		          & \wp{c}{\poste} \qedhere
	\end{align*}
\end{proof}

\bigskip
\begin{proof}[Proof of Lemma~\ref{thm:vctotal-monot}]
	The monotonicity proof of $\vctsymbol$ proceeds by induction on the
	program structure. We only provide the case of loops as the remaining
	cases follows the same argument as for the counterpart  $\vcsymbol$
	for partial correctness (see proof of Lemma~\ref{thm:vc-monot}).

	\subparagraph*{Case $\prog = \IWhileDo{G}{\inv, T, v, \mathtt{l}, \mathtt{u}, \epsilon}{\prog'}$}

	\begin{align*}
		%
		                & \models \vct{\IWhileDo{G}{\inv, T, v, \mathtt{l}, \mathtt{u}, \epsilon}{\prog'}}{\poste}                                            \\
		\Leftrightarrow & \qquad \by{def of $\vctsymbol$ using \eqref{eq:vct-alt} from section \ref{sec:extension}}\displaybreak[0]                           \\
		                & \models \vcgt{\charFun{G \land T}}{\prog'}{\charFun{T}} \; \cup                                                                     \\
		                & \hphantom{=~\left\{ \right.}  \vcgt{\epsilon \, \charFun{G \land T \land  v = v_0}}{\prog'}{\charFun{v < v_0}} \; \cup              \\
		                & \hphantom{=~\left\{ \right.}  \left\{ \charFun{G \land T} \;\pimplies\; \charFun{\mathtt{l} \leq v \leq \mathtt{u}} \right\} \;\cup \\
		                & \hphantom{=~\left\{ \right.} \vcgt{\charFun{G}  \cdot \inv}{\prog'}{\inv} \; \cup \nonumber                                         \\
		                & \hphantom{=~\left\{ \right.}  \left\{\charFun{\lnot G} \cdot \inv \;\pimplies\; \poste \right\} \nonumber                           \\
		\Rightarrow     & \qquad \by{hypothesis and transitivity of $\pimplies $} \displaybreak[0]                                                            \\		%
		                & \models \vcgt{\charFun{G \land T}}{\prog'}{\charFun{T}} \; \cup                                                                     \\
		                & \hphantom{=~\left\{ \right.}  \vcgt{\epsilon \, \charFun{G \land T \land  v = v_0}}{\prog'}{\charFun{v < v_0}} \; \cup              \\
		                & \hphantom{=~\left\{ \right.}  \left\{ \charFun{G \land T} \;\pimplies\; \charFun{\mathtt{l} \leq v \leq \mathtt{u}} \right\} \;\cup \\
		                & \hphantom{=~\left\{ \right.} \vcgt{\charFun{G}  \cdot \inv}{\prog'}{\inv} \; \cup \nonumber                                         \\
		                & \hphantom{=~\left\{ \right.}  \left\{\charFun{\lnot G} \cdot \inv \;\pimplies\; \poste' \right\} \nonumber                           \\
		\Leftrightarrow & \qquad \by{def of $\vctsymbol$} \displaybreak[0]                                                                                    \\		%
		                & \models \vct{\IWhileDo{G}{\inv, T, v, \mathtt{l}, \mathtt{u}, \epsilon}{\prog'}}{\poste'}
	\end{align*}

	The monotonicity proof of $\vcgtsymbol$ follows as immediate corollary.
\end{proof}

\bigskip
\begin{proof}[Proof of Lemma~\ref{thm:alt-vcgt}]
	We give the proof only for the case where $\inst_j$ is a loop since
	the other cases (where $\inst_j$ is a conditional branching or a
	probabilistic choice) follow the same argument as the counterpart result
	for partial correctness (see proof of Lemma~\ref{thm:alt-vcg}). Let $\inst_j =
		\IWhileDo{G}{\inv, T, v, \mathtt{l}, \mathtt{u}, \epsilon}{\prog'}$. Then

	\begin{align*}
		                & \models \vcgt{\pree}{\prog}{\poste}                                                                                                 \\
		\Leftrightarrow & \qquad \by{def of $\vcgtsymbol$ }\displaybreak[0]                                                                                   \\
		                & \models \{ \pree \pimplies \wprect{\prog}{\poste} \} \, \cup \, \vct{\prog}{\poste}                                                 \\
		\Leftrightarrow & \qquad \by{def of $\vctsymbol$ and $\wprectsymbol$ for sequential composition}\displaybreak[0]                                      \\
		                & \models \{ \pree \pimplies \wprect{\seqn{1}{j-1}}{\wprect{\inst_j}{\wprecti{j+1}{\prog}{\poste}}} \}                                \\
		                & \, \cup \, \vct{\seqn{1}{j}}{\wprecti{j+1}{\prog}{\poste}} \, \cup \, \vcti{j+1}{\prog}{\poste}                                     \\
		\Leftrightarrow & \qquad \by{def of $\wprectsymbol$ for guarded loop and def of $\vctsymbol$ for sequential composition}\displaybreak[0]              \\
		                & \models \{ \pree \pimplies \wprect{\seqn{1}{j-1}}{\charFun{T}\cdot \inv} \} \, \cup \,                                              \\
		                & \phantom{\models} \vct{\seqn{1}{j-1}}{\wprect{\inst_j}{\wprecti{j+1}{\prog}{\poste}}} \, \cup \,                                    \\
		                & \phantom{\models}  \vct{\inst_j}{\wprecti{j+1}{\prog}{\poste}} \, \cup \, \vcti{j+1}{\prog}{\poste}                                 \\
		\Leftrightarrow & \qquad \by{def of $\wprectsymbol$ for guarded loop and}\displaybreak[0]                                                             \\
		                & \models \{ \pree \pimplies \wprect{\seqn{1}{j-1}}{\charFun{T}\cdot \inv} \} \, \cup \,                                              \\
		                & \phantom{\models} \vct{\seqn{1}{j-1}}{\charFun{T}\cdot \inv} \, \cup \,                                                             \\
		                & \phantom{\models}  \vct{\inst_j}{\wprecti{j+1}{\prog}{\poste}} \, \cup \, \vcti{j+1}{\prog}{\poste}                                 \\
		\Leftrightarrow & \qquad \by{def of $\vcgtsymbol$ and def of $\vctsymbol$ for guarded loop}\displaybreak[0]                                           \\
		                & \models \vcgti{j-1}{\pree}{\prog}{\charFun{T} \cdot \inv} \; \cup                                                                      \\
		                & \hphantom{=~\left\{ \right.} \vcgt{\charFun{G \land T}}{\prog'}{\charFun{T}} \; \cup                                                \\
		                & \hphantom{=~\left\{ \right.}  \vcgt{\epsilon \, \charFun{G \land T \land  v = v_0}}{\prog'}{\charFun{v < v_0}} \; \cup              \\
		                & \hphantom{=~\left\{ \right.}  \left\{ \charFun{G \land T} \;\pimplies\; \charFun{\mathtt{l} \leq v \leq \mathtt{u}} \right\} \;\cup \\
		                & \hphantom{=~\left\{ \right.} \vcgt{\charFun{G}  \cdot \inv}{\prog'}{\inv} \; \cup                                                   \\
		                & \hphantom{=~\left\{ \right.}  \left\{\charFun{\lnot G} \cdot \inv \;\pimplies\; \wprecti{j+1}{\prog}{\poste} \right\} \; \cup          \\
		                & \hphantom{=~\left\{ \right.}  \vcti{j+1}{\prog}{\poste}                                                                             \\
		\Leftrightarrow & \qquad \by{def of $\vcgtsymbol$ and rearrange terms}\displaybreak[0]                                                                \\
		                & \models \vcgti{j+1}{\charFun{\lnot G} \cdot \inv}{\prog}{\poste} \; \cup                                                            \\
		                & \hphantom{=~\left\{ \right.} \vcgt{\charFun{G}  \cdot \inv}{\prog'}{\inv} \; \cup                                                   \\
		                & \hphantom{=~\left\{ \right.} \vcgti{j-1}{\pree}{\prog}{\charFun{T} \cdot \inv} \; \cup                                              \\
		                & \hphantom{=~\left\{ \right.} \vcgt{\charFun{G \land T}}{\prog'}{\charFun{T}} \; \cup                                                \\
		                & \hphantom{=~\left\{ \right.}  \vcgt{\epsilon \, \charFun{G \land T \land  v = v_0}}{\prog'}{\charFun{v < v_0}} \; \cup              \\
		                & \hphantom{=~\left\{ \right.}  \left\{ \charFun{G \land T} \;\pimplies\; \charFun{\mathtt{l} \leq v \leq \mathtt{u}} \right\}\qedhere
	\end{align*}
\end{proof}

\bigskip

\begin{proof}[Proof of Theorem~\ref{thm:slice-top-total}]
	It follows the same argument as the counterpart result  for partial
	correctness (Theorem~\ref{thm:slice-top}).
\end{proof}

\bigskip

\begin{proof}[Proof of Theorem~\ref{thm:slice-nested-total}]
	\renewcommand{\theequation}{\arabic{equation}}
	By induction on the derivation of the relation
	$\localSpect{\pree}{\prog}{\poste}{\prog'}{\sspec{\pree_i}{\poste_i}}_{i=1,\ldots,n}$	(see Figure \ref{fig:local-spec}).
	We only provide the case of loops and transitivity as the remaining
	cases follows the same argument of the counterpart removing nested instructions
	for partial correctness (see proof of Theorem~\ref{thm:slice-nested}).

	\subparagraph*{Case \lrule{$\localSpecSymbol \texttt{while}^\downarrow$}} In this case we have
	\begin{align*}
		 & \prog   =
		\seqn{1}{j-1};\Cond{G}{\prog'}{\prog_{f}};\seqn{j+1}{n}
	\end{align*}
	\noindent also we have
	\begin{align}
		 & \pspecSlice{\prog''}{\charFun{G}\cdot \inv}{\inv}{\prog'}    \label{eq:hyp-tspeclocal-1}                                    \\
		 & \tspecSlice{\prog''}{\charFun{G \land T}}{\charFun{T}}{\prog'}                      \label{eq:hyp-tspeclocal-2}            \\
		 & \tspecSlice{\prog''}{\epsilon \cdot \charFun{G\land T \land v=v_{0}}}{\charFun{v< v_0}}{\prog'} \label{eq:hyp-tspeclocal-3}
	\end{align}

	We must show that $\prog\subst{\prog'}{\prog''} \preccurlyeq \prog$ and
	$\models \vcg{\pree}{\prog}{\poste} \implies
		\vcg{\pree}{\prog\subst{\prog'}{\prog''}}{\poste}$. The
	first proof obligation is straightforward  (see
	Figure~\ref{fig:is-portion-of}). To establish the
	second proof obligation, we exploit
	Lemma~\ref{thm:alt-vcgt} and the fact that
	$\prog\subst{\prog'}{\prog''}$ coincides with
	$\prog$ in all but the $j$-th instruction. Therefore, assuming

	\begin{align}
		\quad     & \models \vcgtii{j+1}{\charFun{\lnot G} \cdot \inv}{\prog}{\poste}  \nonumber                                                  \\
		\land ~~~ & \models \vcg{\charFun{G} \cdot \inv}{\prog'}{\inv} \label{eq:total-nested-hyp-2} \\
		\land ~~~ &  \models \vcgti{j-1}{\pree}{\prog}{\charFun{T} \cdot \inv} \nonumber             \\
		\land ~~~ & \models \vcgt{\charFun{G \land T}}{\prog'}{\charFun{T}} \label{eq:total-nested-hyp-4}  \\
		\land ~~~ & \models \vcgt{\epsilon \, \charFun{G  \land T \land  v = v_0}}{\prog'}{\charFun{v < v_0}} \label{eq:total-nested-hyp-5} \\
		\land ~~~ & \models \left\{ \charFun{G \land T} \;\pimplies\;  \charFun{\mathtt{l} \leq v \leq \mathtt{u}} \right\} \nonumber
	\end{align}
	we have to conclude that
	\begin{align}
		\quad     & \models \vcgtii{j+1}{\charFun{\lnot G} \cdot \inv}{\prog}{\poste}  \nonumber                                                  \\
		\land ~~~ & \models \vcg{\charFun{G} \cdot \inv}{\prog''}{\inv} \label{eq:total-nested-goal-2} \\
		\land ~~~ &  \models \vcgti{j-1}{\pree}{\prog}{\charFun{T} \cdot \inv} \nonumber               \\
		\land ~~~ & \models \vcgt{\charFun{G \land T}}{\prog''}{\charFun{T}} \label{eq:total-nested-goal-4} \\
		\land ~~~ & \models \vcgt{\epsilon \, \charFun{G  \land T \land  v = v_0}}{\prog''}{\charFun{v < v_0}} \label{eq:total-nested-goal-5}\\
		\land ~~~ & \models \left\{ \charFun{G \land T} \;\pimplies\;  \charFun{\mathtt{l} \leq v \leq \mathtt{u}} \right\} \nonumber
	\end{align}
	We need to prove only
	Equations~\eqref{eq:total-nested-goal-2}, \eqref{eq:total-nested-goal-4} and \eqref{eq:total-nested-goal-5} since the
	others are already part of the premises. But each equation is straightforward since
	from Equations~\eqref{eq:hyp-tspeclocal-1} and \eqref{eq:total-nested-hyp-2} we can conclude
	\[
		\models \vcg{\charFun{G} \cdot \inv}{\prog''}{\inv},
	\]
	\noindent Also, from Equations~\eqref{eq:hyp-tspeclocal-2} and \eqref{eq:total-nested-hyp-4} we get
	\[
		\models \vcgt{\charFun{G \land T}}{\prog''}{\charFun{T}}.
	\]
	%
	\noindent Finally, from Equations~\eqref{eq:hyp-tspeclocal-3} and \eqref{eq:total-nested-hyp-5} we obtain
	\[
		\models \vcgt{\epsilon \cdot \charFun{G \land T \land v=v_0}}{\prog''}{\charFun{v<v_0}}.
	\]

	\subparagraph*{Case \lrule{$\localSpecSymbol \texttt{trans}^\downarrow$}} Let $\prog^{*}$ be a $\Lang$ with its respective
	sets of specifications $\{\sspec{f_i}{g_i}\}_{i=1,\dots,n}$ such that
	$\prog' \preccurlyeq \prog^{*}$ and
	\begin{align}
		& \localSpect{\pree}{\prog}{\poste}{\prog^{*}}{\sspec{\pree_i}{\poste_i}}_{i=1,\ldots,n} \label{eq:hi-nested-total}\\
		& \localSpec{\pree_i}{\prog^{*}}{\poste_i}{\prog'}{\sspec{\pree_{i,j}}{\poste_{i,j}}}_{j=1,\ldots,m_i} \quad \forall i=1,\ldots,n  \label{eq:hyp-weak-hi}
	\end{align}
	We must show that  $\tspecSlice{\prog\subst{\prog^{*}}{\prog^{*}\subst{\prog'}{\prog''}}}{\pree}{\poste}{\prog}$.
	Let us consider a pair of arbitrary $i \in [1,\dots,n]$
        and $j \in [1,\dots,m_i]$ and do a case analysis on the
        kind of local specification $\sspec{\pree_{i}}{\poste_{i}}$
        refers to. If it refers to a total correctness
        specification, \ie  $\prog^{*} \vdash
        \sspect{\pree_{i}}{\poste_{i}}$, by inductive hypothesis, Equation~\eqref{eq:hyp-weak-hi} and the fact that $\specSlice{\prog''}{\pree_{i,j}}{\poste_{i,j}}{\prog'}$ we can conclude that
	\[
	\tspecSlice{\prog^{*}\subst{\prog'}{\prog''}}{\pree_{i}}{\poste_{i}}{\prog^{*}}. 
	\]
	If on the other hand, $\sspec{\pree_{i}}{\poste_{i}}$ refers
        to a partial correctness specification,  \ie $\prog^{*} \vdash
        \sspecp{\pree_{i}}{\poste_{i}}$, 
	by Theorem~\ref{thm:slice-nested}, Equation~\eqref{eq:hyp-weak-hi} and the fact that $\specSlice{\prog''}{\pree_{i,j}}{\poste_{i,j}}{\prog'}$ we get
	\[
	\pspecSlice{\prog^{*}\subst{\prog'}{\prog''}}{\pree_{i}}{\poste_{i}}{\prog^{*}}~.
	\] 

	\noindent In either case, we obtain a specitifacion-based slice of $\prog^{*}$.
	It follows that 
	\begin{equation}\label{eq:hyp-ind-nested-total}
		\forall i = 1,\dots,n, ~~~ \specSlice{\prog^{*}\subst{\prog'}{\prog''}}{\pree_{i}}{\poste_{i}}{\prog^{*}}~.
	\end{equation}

	Finally, from Equations~\eqref{eq:hi-nested-total} and
        \eqref{eq:hyp-ind-nested-total} and by inductive hypothesis,
        we can conclude that
        \begin{equation*}
         \tspecSlice{\prog\subst{\prog^{*}}{\prog^{*}\subst{\prog'}{\prog''}}}{\pree}{\poste}{\prog}\qedhere
        \end{equation*}
\end{proof}\vfill

%% file: appendixB.tex
\subsection*{Detailed analysis from Section~\ref{sec:applications}}

We slice the program from Figure~\ref{fig:BNSlice1} \wrt post-expectation $\poste = \charFun{x=1}$
and an arbitrary pre-expectation $\pree$. To this end, we start by propagating post-expectation $\poste$
backward along the program, as shown in Figure~\ref{fig:BNSlicing1Full}. Observe that 
$\wprecid{5}  = \wprecid{6}$ and $\wprecid{8} = \wprecid{9}$. Thus, in view of Theorem \ref{thm:slice-top-total} we can remove instructions $i_5$ and $i_7$.

Furthermore, let us consider the local specification induced over the right branch $\{ \Ass{x}{0} \}$ of the probabilistic choice in the \emph{true} branch of the conditional branching $\inst_7$. To compute it, we first compute the local specification induced over the  \emph{true} branch of the conditional branching, obtaining post-expectation $ \charFun{x=1}$  and pre-expectation $\tfrac{98}{100}\charFun{e=1}$. This induces itself local specification given by post-expectation $\poste_{7} = \charFun{x=1}$ and pre-expectation $\pree_{7} = 0$ on the right branch of the probabilistic choice (here, we use $0$ to denote the constant expectation $\lambda \state. 0$). Since $\pree_{7} = 0$, it trivially holds that $\pree_{7} \pimplies \poste_{7}$ and an application of Theorem~\ref{thm:slice-top-total} together with a double application of Theorem~\ref{thm:slice-nested-total} allows slicing away the whole content of right branch of the probabilistic choice, namely assignment $\Ass{x}{0}$.

With a similar reasoning, we can also slice away the right branch of the probabilistic choice in the \emph{false} branch of the conditional branching $\inst_7$. All the removable code above identified is colored in red in Figure~\ref{fig:BNSlicing1Full}.



\begin{figure}[t]
  \centering
  \begin{multicols}{2}\noindent%
    {\small
      \begin{align*}
                                & \codeComment{$\pree$}                                                                                                      \\[-0.4ex]
                                & \codeComment{$\wprecid{1}=\tfrac{21623}{4\cdot 10^6}+\tfrac{99\cdot 219877}{2\cdot 10^8}$}                                 \\[-0.4ex]
        \scalemath{0.8}{\inst_1: ~} & \PChoice{\Ass{a}{1}}{\nicefrac{1}{100}}{\Ass{a}{0}};                                                                       \\[-0.4ex]
                                & \codeComment{$\wprecid{2}=\tfrac{21623}{4\cdot 10^4}\charFun{a{\,=\,}1}+\tfrac{219877}{2\cdot 10^6}\charFun{a{\,\neq\,} 1}$}               \\[-0.4ex]
        %
        \scalemath{0.8}{\inst_2: ~} & \PChoice{\Ass{s}{1}}{\nicefrac{1}{2}}{\Ass{s}{0}};                                                                         \\[-0.4ex]
                                & \codeComment{$\wprecid{3} = \pree_{3,1}\charFun{a=1} + \pree_{3,2}\charFun{a\neq 1} $}                                         \\[-0.4ex]
                                & \codeComment{$\pree_{3,1} = \tfrac{1123}{2\cdot 10^3}\charFun{s=1}+\tfrac{10393}{2\cdot 10^4}\charFun{s\neq 1}$}               \\[-0.4ex]
                                & \codeComment{$\pree_{3,2} = \tfrac{15137}{10^5}\charFun{s=1} + \tfrac{68507}{10^6}\charFun{s\neq 1}$}                          \\[-0.4ex]
        \scalemath{0.8}{\inst_3: ~} & \If \, (a=1) \,\Then  ~ \{                                                                                                 \\[-0.4ex]
                                & \quad  \PChoice{\Ass{t}{1}}{\nicefrac{1}{2}}{\Ass{t}{0}}                                                                   \\[-0.4ex]
                                & \} ~\Else ~ \{                                                                                                             \\[-0.4ex]
                                & \quad  \PChoice{\Ass{t}{1}}{\nicefrac{1}{100}}{\Ass{t}{0}}                                                                 \\[-0.4ex]
                                & \}                                                                                                                         \\[-0.4ex]
                                & \codeComment{$\wprecid{4} = \pree_{4,1}\charFun{s=1} + \pree_{4,2}\charFun{s\neq 1}$}                                          \\[-0.4ex]
                                & \codeComment{$\pree_{4,1} = \tfrac{98}{10^3} + \tfrac{9\cdot 98}{10^3}\charFun{t=1}+\tfrac{9\cdot 5}{10^3}\charFun{t\neq 1}$}  \\[-0.4ex]
                                & \codeComment{$\pree_{4,2} =\tfrac{98}{10^4} + \tfrac{99\cdot 98}{10^4}\charFun{t=1}+\tfrac{99\cdot 5}{10^4}\charFun{t\neq 1}$} \\[-0.4ex]
        \scalemath{0.8}{\inst_4: ~} & \If \, (s=1) \,\Then  ~ \{                                                                                                 \\[-0.4ex]
                                & \quad  \PChoice{\Ass{l}{1}}{\nicefrac{1}{10}}{\Ass{l}{0}}                                                                  \\[-0.4ex]
                                & \} ~\Else ~ \{                                                                                                             \\[-0.4ex]
                                & \quad  \PChoice{\Ass{l}{1}}{\nicefrac{1}{100}}{\Ass{l}{0}}                                                                 \\[-0.4ex]
                                & \}                                                                                                                         \\[-0.4ex]
                                & \codeComment{$\wprecid{5} = \wprecid{6}$}                                                                                  \\[-0.4ex]
        \scalemath{0.8}{\inst_5: ~} & \SliceAway{\If \, (s=1) \,\Then  ~ \{                                   }                                                              \\[-0.4ex]
                                & \SliceAway{\quad  \PChoice{\Ass{b}{1}}{\nicefrac{6}{10}}{\Ass{b}{0}}  }                                                                \\[-0.4ex]
                                & \SliceAway{\} ~\Else ~ \{                                              }                                                               \\[-0.4ex]
                                & \SliceAway{\quad  \PChoice{\Ass{b}{1}}{\nicefrac{3}{10}}{\Ass{b}{0}} }                                                                  \\[-0.4ex]
                                & \SliceAway{\}  }                                                                                                                       \\[-0.4ex]
                                & \codeComment{$\wprecid{6} = \pree_{6,1} + $}                                                   \\[-0.4ex]
                                & \codeComment{$ \charFun{\neg(t=1 \land l=1)}\Big(\pree_{6,2} +$}                                                               \\[-0.4ex]
                                & \codeComment{$  \charFun{\neg(t=1 \land l\neq 1)}( \pree_{6,3}  + \pree_{6,4})\Big)$;}                                             \\[-0.4ex]
                              \end{align*}
                              \vfill\null
                                \columnbreak\noindent
                                \begin{align*}
                                & \codeComment{$\pree_{6,1} =  \tfrac{98}{100} \charFun{t=1 \land l=1}$}                                                      \\[-0.4ex]
                                & \codeComment{$\pree_{6,2} =  \tfrac{98}{100}\charFun{t=1 \land l\neq 1}$}                                                      \\[-0.4ex]
                                & \codeComment{$\pree_{6,3} =  \tfrac{98}{100} \charFun{t\neq 1 \land l=1}$}                                                     \\[-0.4ex]
                                & \codeComment{$\pree_{6,4} =  \tfrac{5}{100}\charFun{\neg(t\neq 1 \land l=1)}$}                                                 \\[-0.4ex]
        \scalemath{0.8}{\inst_6: ~} & \If \, (t=1 \land l=1) \,\Then  ~ \{                                                                                       \\[-0.4ex]
                                & \quad  \PChoice{\Ass{e}{1}}{1}{\Ass{e}{0}}                                                                                 \\[-0.4ex]
                                & \} ~\Else \, \If \, (t=1 \land l\neq 1 )  ~ \{                                                                                  \\[-0.4ex]
                                & \quad  \PChoice{\Ass{e}{1}}{1}{\Ass{e}{0}}                                                                                 \\[-0.4ex]
                                & \} ~\Else \, \If \, (t\neq 1 \land l=1) ~ \{                                                                                   \\[-0.4ex]
                                & \quad  \PChoice{\Ass{e}{1}}{1}{\Ass{e}{0}}                                                                                 \\[-0.4ex]
                                & \} ~\Else ~ \{                                                                                                             \\[-0.4ex]
                                & \quad \PChoice{\Ass{e}{1}}{0}{\Ass{e}{0}}                                                                                  \\[-0.4ex]
                                & \}                                                                                                                         \\[-0.4ex]
                                & \codeComment{$\wprecid{7} = \tfrac{98}{100}\charFun{e=1} + \tfrac{5}{100}\charFun{e\neq 1}$}                                   \\[-0.4ex]
        \scalemath{0.8}{\inst_7: ~} & \If \, (e=1) \,\Then  ~ \{                                                                                                 \\[-0.4ex]
                                & \quad  \PChoice{\Ass{x}{1}}{\nicefrac{98}{100}}{\codeComment{$\!\pree_7$} ~  \SliceAway{\Ass{x}{0}} ~ \codeComment{$\!\poste_7$}  }      \\[-0.4ex]
                                & \} ~\Else ~ \{                                                                                                             \\[-0.4ex]
                                & \quad  \PChoice{ \Ass{x}{1}  }{\nicefrac{5}{100}}{  \codeComment{$\! \pree_7$} ~  \SliceAway{\Ass{x}{0}} ~ \codeComment{$\!\poste_7$} } \\[-0.4ex]
                                & \}                                                                                                                         \\[-0.4ex]
                                & \codeComment{$\wprecid{8} = \charFun{x=1}$}                                                                                \\[-0.4ex]
        \scalemath{0.8}{\inst_8: ~} & \SliceAway{\If \, (e=1 \land b=1) \,\Then   ~ \{}                                                                                      \\[-0.4ex]
                                & \SliceAway{\quad  \PChoice{\Ass{d}{1}}{\nicefrac{9}{10}}{\Ass{d}{0}} }                                                                  \\[-0.4ex]
                                & \SliceAway{\} ~\Else \, \If \, (e=1 \land b\neq 1 ) ~ \{}                                                                                   \\[-0.4ex]
                                & \SliceAway{\quad  \PChoice{\Ass{d}{1}}{\nicefrac{7}{10}}{\Ass{d}{0}} }                                                                 \\[-0.4ex]
                                & \SliceAway{\} ~\Else \, \If \,  (e\neq 1 \land b=1) ~ \{}                                                                                  \\[-0.4ex]
                                & \SliceAway{\quad    \PChoice{\Ass{d}{1}}{\nicefrac{8}{10}}{\Ass{d}{0}}}                                                                \\[-0.4ex]
                                & \SliceAway{\} ~\Else ~ \{}                                                                                                             \\[-0.4ex]
                                & \SliceAway{\quad  \PChoice{\Ass{d}{1}}{\nicefrac{1}{10}}{\Ass{d}{0}}}                                                                  \\[-0.4ex]
                                & \SliceAway{\}}                                                                                                                        \\[-0.4ex]
                                & \codeComment{$\wprecid{9} = \poste = \charFun{x=1}$}                                                                            \\[-0.4ex]
      \end{align*}
    }
  \end{multicols}
  \caption{Program from Figure~\ref{fig:BNSlice1}, annotated with the backward propagation of post-expectation $\poste = \charFun{x=1}$ and the local specification induced over the right branches of the probabilistic choices in $\inst_7$. Code in \SliceAway{red} can be sliced away when considering post-expectation~$\poste$.}
  \label{fig:BNSlicing1Full}
\end{figure}

Now we slice the same program, but this time \wrt post-expectation $\poste' = \charFun{t=1 \land l=1}$ (and an arbitrary pre-expectation $\pree$). Similarly, we propagate the post-expectation backward along the program, obtaining the result in Figure~\ref{fig:BNSlicing2Full}. Since $\wprecid{5} = \wprecid{9}$, Theorem~\ref{thm:slice-top-total} allows deleting the sequence of instructions 
$\inst_5; \inst_6; \inst_7; \inst_8$. Computing the local specifications of the right branches of the probabilistic choices in $\inst_3$ and $\inst_4$ yields:
\begin{align*}
  \pree_{3} &~=~  \tfrac{1}{10}\charFun{0=1}\charFun{s=1} + \tfrac{1}{100}\charFun{0=1}\charFun{s\neq 1} ~=~ 0\\
  \poste_{3} &~=~  \tfrac{1}{10}\charFun{t=1}\charFun{s=1} + \tfrac{1}{100}\charFun{t=1}\charFun{s\neq 1} \\
  \pree_{4} &~=~ \charFun{t=1 \land 0=1} ~=~ 0\\
  \poste_{4} &~=~\charFun{t=1 \land l=1}
\end{align*}
Since $\pree_{3} \pimplies \poste_{3}$ and $\pree_{4} \pimplies \poste_{4}$, following a similar reasoning as before we can remove the right branches of the probabilistic choices in $\inst_3$ and $\inst_4$.




%
\begin{figure}[t]
  \centering
  \noindent
  \begin{multicols}{2}\noindent
    {\small
      \begin{align*}
        %
                                & \codeComment{$\wprecid{1}=\tfrac{11}{4\cdot 10^4} + \tfrac{99\cdot 11}{2\cdot 10^6}$}                                    \\[-0.4ex]
        \scalemath{0.8}{\inst_1: ~} & \PChoice{\Ass{a}{1}}{\nicefrac{1}{100}}{\Ass{a}{0}};                                                                     \\[-0.4ex]
                                & \codeComment{$\wprecid{2}=\tfrac{11}{4\cdot 10^2}\charFun{a=1}+\tfrac{11}{2\cdot 10^4}\charFun{a\neq 1}$}                    \\[-0.4ex]
        %
        \scalemath{0.8}{\inst_2: ~} & \PChoice{\Ass{s}{1}}{\nicefrac{1}{2}}{\Ass{s}{0}};                                                                       \\[-0.4ex]
                                & \codeComment{$\wprecid{3} = \pree_{3,1}\charFun{a=1} + \pree_{3,2}\charFun{a\neq 1} $}                                       \\[-0.4ex]
                                & \codeComment{$\pree_{3,1} = \tfrac{1}{2}(\tfrac{1}{10}\charFun{s=1} + \tfrac{1}{100}\charFun{s\neq 1}) $}                    \\[-0.4ex]
                                & \codeComment{$\pree_{3,2} = \tfrac{1}{100}(\tfrac{1}{10}\charFun{s=1} + \tfrac{1}{100}\charFun{s\neq 1})$}                   \\[-0.4ex]
        \scalemath{0.8}{\inst_3: ~} & \If \, (a=1) \,\Then  ~ \{                                                                                               \\[-0.4ex]
                                & \quad  \PChoice{\Ass{t}{1}}{\nicefrac{1}{2}}{\codeComment{$\pree_{3}$} ~ \SliceAway{\Ass{t}{0}} ~ \codeComment{$\poste_{3}$}}      \\[-0.4ex]
                                & \} ~\Else ~ \{                                                                                                           \\[-0.4ex]
                                & \quad  \PChoice{\Ass{t}{1}}{\nicefrac{1}{100}}{\codeComment{$\pree_{3}$} ~ \SliceAway{\Ass{t}{0}} ~ \codeComment{$\poste_{3}$}}  \\[-0.4ex]
                                & \}                                                                                                                       \\[-0.4ex]
                                & \codeComment{$\wprecid{4} = \pree_{4,1}\charFun{s=1} + \pree_{4,2}\charFun{s\neq 1}$}                                        \\[-0.4ex]
                                & \codeComment{$ \pree_{4,1}  = \tfrac{1}{10}\charFun{t=1}$}                                                               \\[-0.4ex]
                                & \codeComment{$ \pree_{4,2}  = \tfrac{1}{100}\charFun{t=1}$}                                                              \\[-0.4ex]
        \scalemath{0.8}{\inst_4: ~} & \If \, (s=1) \,\Then  ~ \{                                                                                               \\[-0.4ex]
                                & \quad  \PChoice{\Ass{l}{1}}{\nicefrac{1}{10}}{\codeComment{$\pree_{4}$} ~ \SliceAway{\Ass{l}{0}} ~ \codeComment{$\poste_{4}$}}     \\[-0.4ex]
                                & \} ~\Else ~ \{                                                                                                           \\[-0.4ex]
                                & \quad  \PChoice{\Ass{l}{1}}{\nicefrac{1}{100}}{ \codeComment{$\pree_{4}$} ~ \SliceAway{\Ass{l}{0}} ~ \codeComment{$\poste_{4}$}} \\[-0.4ex]
                                & \}                                                                                                                       \\[-0.4ex]
                                & \codeComment{$\wprecid{5} = \charFun{t=1 \land l=1}$}                                                                    \\[-0.4ex]
        \scalemath{0.8}{\inst_5: ~} & \SliceAway{\If \, (s=1) \,\Then  ~ \{   }                                                                                            \\[-0.4ex]
                                & \SliceAway{\quad  \PChoice{\Ass{b}{1}}{\nicefrac{6}{10}}{\Ass{b}{0}} }                                                                \\[-0.4ex]
                                & \SliceAway{\} ~\Else ~ \{  }                                                                                                         \\[-0.4ex]
                                & \SliceAway{\quad  \PChoice{\Ass{b}{1}}{\nicefrac{3}{10}}{\Ass{b}{0}}}                                                                \\[-0.4ex]
                                & \SliceAway{\}}                                                                                                                       \\[-0.4ex]
      \end{align*}
      \vfill\null
        \columnbreak\noindent
        \begin{align*}
                                & \codeComment{$\wprecid{6} = \charFun{t=1 \land l=1} $}                                                                   \\[-0.4ex]
        \scalemath{0.8}{\inst_6: ~} & \SliceAway{\If \, (t=1 \land l=1) \,\Then  ~ \{ }                                                                                    \\[-0.4ex]
                                & \SliceAway{\quad  \PChoice{\Ass{e}{1}}{1}{\Ass{e}{0}}}                                                                               \\[-0.4ex]
                                & \SliceAway{\} ~\Else \, \If \, (t=1 \land l\neq 1 )  ~ \{ }                                                                               \\[-0.4ex]
                                & \SliceAway{\quad  \PChoice{\Ass{e}{1}}{1}{\Ass{e}{0}} }                                                                              \\[-0.4ex]
                                & \SliceAway{\} ~\Else \, \If \, (t\neq 1 \land l=1) ~ \{     }                                                                            \\[-0.4ex]
                                & \SliceAway{\quad  \PChoice{\Ass{e}{1}}{1}{\Ass{e}{0}}  }                                                                             \\[-0.4ex]
                                & \SliceAway{\} ~\Else ~ \{ }                                                                                                          \\[-0.4ex]
                                & \SliceAway{\quad \PChoice{\Ass{e}{1}}{0}{\Ass{e}{0}}}                                                                                \\[-0.4ex]
                                & \SliceAway{\}}                                                                                                                      \\[-0.4ex]
                                & \codeComment{$\wprecid{7} = \charFun{t=1 \land l=1}$}                                                                    \\[-0.4ex]
        \scalemath{0.8}{\inst_7: ~} & \SliceAway{\If \, (e=1) \,\Then  ~ \{}                                                                                               \\[-0.4ex]
                                & \SliceAway{\quad  \PChoice{\Ass{x}{1}}{\nicefrac{98}{100}}{  \Ass{x}{0}  }}                                                          \\[-0.4ex]
                                & \SliceAway{\} ~\Else ~ \{}                                                                                                           \\[-0.4ex]
                                & \SliceAway{\quad  \PChoice{ \Ass{x}{1}  }{\nicefrac{5}{100}}{    \Ass{x}{0}}  }                                                      \\[-0.4ex]
                                & \SliceAway{\}}                                                                                                                       \\[-0.4ex]
                                & \codeComment{$\wprecid{8} = \charFun{t=1 \land l=1}$}                                                                    \\[-0.4ex]
        \scalemath{0.8}{\inst_8: ~} & \SliceAway{\If \, (e=1 \land b=1) \,\Then   ~ \{ }                                                                       \\[-0.4ex]
                                & \SliceAway{\quad  \PChoice{\Ass{d}{1}}{\nicefrac{9}{10}}{\Ass{d}{0}} }                                                   \\[-0.4ex]
                                & \SliceAway{\} ~\Else \, \If \, (e=1 \land b\neq 1 ) ~ \{ }                                                                    \\[-0.4ex]
                                & \SliceAway{\quad  \PChoice{\Ass{d}{1}}{\nicefrac{7}{10}}{\Ass{d}{0}}}                                                    \\[-0.4ex]
                                & \SliceAway{\} ~\Else \, \If \,  (e\neq 1 \land b=1) ~ \{ }                                                                   \\[-0.4ex]
                                & \SliceAway{\quad    \PChoice{\Ass{d}{1}}{\nicefrac{8}{10}}{\Ass{d}{0}} }                                                 \\[-0.4ex]
                                & \SliceAway{\} ~\Else ~ \{ }                                                                                              \\[-0.4ex]
                                & \SliceAway{\quad  \PChoice{\Ass{d}{1}}{\nicefrac{1}{10}}{\Ass{d}{0}} }                                                   \\[-0.4ex]
                                & \SliceAway{\}}                                                                                                                       \\[-0.4ex]
                                & \codeComment{$\wprecid{9} = \poste' = \charFun{t=1 \land l=1}$}                                                                \\[-0.4ex]
      \end{align*}
    }
  \end{multicols}
\caption{Program from Figure~\ref{fig:BNSlice2}, annotated with the backward propagation of post-expectation $\poste' = \charFun{t=1 \land l=1}$ and the local specification induced over the right branches of the probabilistic choices in $\inst_3$ and $\inst_4$. Code in \SliceAway{red} can be sliced away when considering post-expectation~$\poste'$.}
  \label{fig:BNSlicing2Full}
\end{figure}
\noindent

\noindent